\documentclass[aps,preprint,showkeys,superscriptaddress,nofootinbib,floatfix]{revtex4-1}
\usepackage[pdftex]{graphicx}  %%  for arXiv
\usepackage{bbm}
\usepackage{amsmath}
\usepackage{simplewick}

\newcommand{\Slash}[1]{{\ooalign{\hfil/\hfil\crcr$#1$}}}
\DeclareMathOperator{\re}{Re}

\DeclareMathOperator{\Li}{Li}

\def\simge{\mathrel{%
       \rlap{\raise 0.511ex \hbox{$>$}}{\lower 0.511ex \hbox{$\sim$}}}}
\def\simle{\mathrel{
       \rlap{\raise 0.511ex \hbox{$<$}}{\lower 0.511ex \hbox{$\sim$}}}}
\def \unitmatrix{\textbf{1}}

\begin{document} 

\title{\boldmath $N_f=2+1$ QCD thermodynamics with gradient flow using two-loop matching coefficients}

\preprint{UTHEP-748, UTCCS-P-132, J-PARC-TH-0216, KYUSHU-HET-211}

\author{Yusuke Taniguchi}
\email{tanigchi@het.ph.tsukuba.ac.jp}
\affiliation{Center for Computational Sciences, University of Tsukuba,
Tsukuba, Ibaraki 305-8571, Japan}
\author{Shinji Ejiri}
\email{ejiri@muse.sc.niigata-u.ac.jp}
\affiliation{Department of Physics, Niigata University, Niigata 950-2181,
Japan}
\author{Kazuyuki Kanaya}
\email{kanaya@ccs.tsukuba.ac.jp}
\affiliation{Tomonaga Center for the History of the Universe, University of Tsukuba, Tsukuba, Ibaraki 305-8571, Japan}
\author{Masakiyo Kitazawa}
\email{kitazawa@phys.sci.osaka-u.ac.jp}
\affiliation{Department of Physics, Osaka University, Osaka 560-0043, Japan}
\affiliation{J-PARC Branch, KEK Theory Center, Institute of Particle and Nuclear Studies,
KEK, 203-1, Shirakata, Tokai, Ibaraki, 319-1106, Japan}
\author{Hiroshi Suzuki}
\email{hsuzuki@phys.kyushu-u.ac.jp}
\affiliation{Department of Physics, Kyushu University, 744 Motooka, Nishi-ku, Fukuoka
819-0395, Japan}
\author{Takashi Umeda}
\email{tumeda@hiroshima-u.ac.jp}
\affiliation{Graduate School of Education, Hiroshima University,
Higashihiroshima, Hiroshima 739-8524, Japan}
\collaboration{WHOT-QCD Collaboration}
\noaffiliation

\date{\today}

\begin{abstract}
We study thermodynamic properties of $N_f=2+1$ QCD on the lattice 
adopting a nonperturbatively $O(a)$-improved Wilson quark action and the renormalization group-improved Iwasaki gauge action. 
To cope with the problems due to explicit violation of the Poincar\'e and chiral symmetries, we apply the \textbf{S}mall  \textbf{F}low-\textbf{\textit{t}}ime e\textbf{X}pansion (\textbf{SF\textit{t}X}) method based on the gradient flow, which is a general method  to correctly calculate any renormalized observables on the lattice.
In this method, the matching coefficients in front of operators in the small flow-time expansion are calculated by perturbation theory thanks to the asymptotic freedom around the small flow-time limit. 
In a previous study using one-loop matching coefficients, we found that the SF\textit{t}X method works well for the equation of state extracted from diagonal components of the energy-momentum tensor and for the chiral condensates and susceptibilities. 
In this paper, we study the effect of two-loop matching coefficients which have been calculated by Harlander \textit{et al.}\  recently.
We also test the influence of the renormalization scale in the SF\textit{t}X method. 
We find that, by adopting the $\mu_0$ renormalization scale of Harlander \textit{et al.}\ instead of the conventional $\mu_d=1/\sqrt{8t}$ scale, the linear behavior at large flow-times is improved so that we can perform the $t\to0$ extrapolation of the SF\textit{t}X method more confidently.
In the calculation of the two-loop matching coefficients by Harlander \textit{et al.}, the equation of motion for quark fields was used. 
For the entropy density in which the equation of motion has no effects, we find that the results using the two-loop coefficients agree well with those using one-loop coefficients. 
On the other hand, for the trace anomaly which is affected by the equation of motion, we find discrepancies between the one- and two-loop results at high temperatures.
By comparing the results of one-loop coefficients with and without using the equation of motion, the main origin of the discrepancies is suggested to be attributed to contamination of $O\!\left((aT)^2\right)=O\!\left(1/N_t^2\right)$ discretization errors in the equation of motion at~$N_t \simle 10$. 
\end{abstract}

\maketitle

%%%%%%%%%%%%%%%%%%%%%%%%%%%%%%%%%%%%%%%%%%%%%%%%%%%%
\section{Introduction}
\label{sec:intro}

The gradient flow (GF) opened us a variety of new methods to significantly simplify the calculation of physical observables on the lattice~\cite{Narayanan:2006rf,Luscher:2009eq,Luscher:2010iy,Luscher:2011bx,Luscher:2013cpa}.
For reviews, see Refs.~\cite{Luscher:2013vga,Ramos:2015dla,Lat16suzuki}.
In this paper, we study finite-temperature QCD with $2+1$ flavors of dynamical quarks by applying the \textbf{Small Flow-\textit{t}ime eXpansion} (\textbf{SF\textit{t}X}) \textbf{method} based on the GF~\cite{Suzuki:2013gza,Makino:2014taa,Endo:2015iea,Hieda:2016lly}.

The GF modifies the fields according to a flow equation, which is given by the gradient of the action in the case of pure Yang-Mills theory and is a kind of diffusion equation in term of a fictitious time (flow-time)~$t$.
Fields at positive flow-time $t>0$ can be viewed as smeared fields averaged over a mean-square physical radius of $\sqrt{8t}$ in four dimensions. 
Salient features of the GF are the UV-finiteness and the absence of short-distance
singularities in the expectation values of operators constructed by flowed fields at $t>0$. 
This finiteness enables us to identify these expectation values and corresponding operators as renormalized ones. 
We call this renormalization scheme as GF-scheme. 

At small flow-times, operators in the GF-scheme (``flowed operators'') can be expanded in terms of operators at $t=0$ in a conventional renormalization scheme, say the $\overline{\textrm{MS}}$-scheme~\cite{Luscher:2011bx}. 
By inverting the relation, we can also expand correctly renormalized physical observables in conventional schemes in terms of flowed operators at small $t$. 
Thanks to the asymptotic freedom of QCD around the small flow-time limit,  we can calculate the matching coefficients relating the operators in two schemes by perturbation theory. 
Basic idea of the SF\textit{t}X method is that, because the flowed operators are finite, we can evaluate their expectation values directly on the lattice without further renormalization~\cite{Suzuki:2013gza}.
We can thus extract the values of correctly renormalized physical observables by extrapolating proper combinations of expectation values in the GF-scheme to the small flow-time limit $t\to0$. 
In these extrapolations, the matching coefficients act not only to match the renormalization schemes but also to make the $t$-dependence milder by removing calculable part of operator mixings and $t$-dependences around the small flow-time limit. 
Note that the method is applicable also to observables whose founding symmetry is violated explicitly on the lattice, provided that the lattice theory has the correct continuum limit in which the symmetry is restored.

The SF\textit{t}X method was first tested in quenched QCD to calculate the energy-momentum tensor (EMT)~\cite{Asakawa:2013laa,FlowQCD1,Iritani2019,Yanagihara2018}, which has not been easy to evaluate on the lattice due to the explicit violation of the Poincar\'e invariance by the lattice regularization.%
\footnote{For a recent development in lattice determination of the EMT, see Refs.~\cite{DBrida-Giusti-Pepe,DallaBrida:2020gux}.}  
It was found that the equation of state (EoS) calculated from the diagonal components of the EMT correctly reproduces previous estimation using the conventional integral methods~\cite{Boyd:1996bx,Okamoto:1999hi,Umeda:2008bd,Borsanyi:2012ve}.
The SF\textit{t}X method was tested successfully also in solvable models~\cite{Makino:2014cxa,Suzuki:2015fka}.

We note that the SF\textit{t}X method is applicable also to chiral observables~\cite{Endo:2015iea,Hieda:2016lly}. 
We thus apply the method to QCD with dynamical Wilson-type quarks, with which the correct continuum limit is guaranteed, to cope with the problems due to explicit violation of the chiral symmetry on the lattice~\cite{Taniguchi:2016ofw,Taniguchi:2016tjc,Lat2017-kanaya,Taniguchi:2017ibr}.
To reduce the finite lattice spacing effects, we adopt the renormalization-group improved Iwasaki gauge action~\cite{Iwasaki:2011np,Iwasaki:1985we} and the $O(a)$-improved Wilson quark action~\cite{Sheikholeslami:1985ij} with a nonperturbatively estimated clover coefficient using the Schr\"odinger functional method~\cite{SAoki2006}.

In our previous study of $(2+1)$-flavor QCD with slightly heavy $u$ and $d$ quarks ($m_\pi/m_\rho\simeq0.63$) and approximately physical $s$ quark ($m_{\eta_{ss}}/m_\phi\simeq0.74$), we calculated the EMT as well as chiral condensates and disconnected chiral susceptibilities in the temperature range of $T\simeq174$--697 MeV ($N_t=16$--4, where $N_t$ is the lattice size in the temporal direction)~\cite{Taniguchi:2016ofw}.
The lattices are relatively fine with the lattice spacing $a\simeq0.07$ fm.
Adopting one-loop matching coefficients calculated in Refs.~\cite{Makino:2014taa,Hieda:2016lly}, 
we found that the EoS extracted from diagonal components of EMT by the SF\textit{t}X method is well consistent with that estimated with the conventional $T$-integration method at $T\simle280\,\mathrm{MeV}$ ($N_t \simge 10$)~\cite{Umeda:2012er}.
At the same time, the two estimates of EoS deviate at $T\simge348\,\mathrm{MeV}$, suggesting contamination of $a$-independent lattice artifacts of $O\!\left((aT)^2=1/N_t^2\right)$ at $N_t \simle 8$ in the EMT.
We also found that the chiral condensates bend sharply and the disconnected chiral susceptibilities show peak at $T\simeq190$ MeV which was suggested as the pseudocritical temperature from other observables~\cite{Umeda:2012er}. 
We have further studied topological properties of QCD by the SF\textit{t}X method on these lattices~\cite{Taniguchi:2016tjc}.
We found that the topological susceptibilities estimated with the gluonic and fermionic definitions agree well with each other at $T\simle 279$ MeV even at finite lattice spacing of $a=0.07$ fm.
This is in clear contrast to their conventional lattice estimations: 
For example, a study with improved staggered quarks reports more than hundred times larger gluonic susceptibility than fermionic one at similar lattice spacings~\cite{Petreczky:2016vrs}%
\footnote{In Ref.~\cite{Petreczky:2016vrs}, unlike the study of Ref.~\cite{Taniguchi:2016tjc} with the SF\textit{t}X method, the topological susceptibilities with the fermionic definition were measured by approximating the disconnected pseudoscalar susceptibility by the disconnected scalar susceptibility. 
This approximation should be valid in the continuum limit at high temperatures where the chiral symmetry is well restored. 
However, significant cutoff effects were observed in the results of topological susceptibilities up to $N_t=12$ they studied. 
}.
These suggest that the SF\textit{t}X method is powerful in calculating observables from lattice simulations.

Recently, Harlander, Kluth, and Lange have completed the calculation of the matching coefficients for EMT up to the two-loop order~\cite{Harlander:2018zpi}. 
Some details of their calculation are given in~\cite{AHLNP}. 
Removing more known small-$t$ behaviors, we may perhaps expect milder $t$-dependence in the $t\rightarrow0$ extrapolation.
The two-loop coefficients were first tested in quenched QCD~\cite{Iritani2019}.
It was found that the results of EoS with one- and two-loop coefficients are well consistent with each other.
It was also noted that the two-loop coefficients lead to a milder $t$-dependence such that systematic errors from the $t\to0$ extrapolation are reduced.

In this paper, we extend the test of two-loop coefficients to QCD with $(2+1)$-flavors of dynamical quarks. 
The lattice setup is the same as in Ref.~\cite{Taniguchi:2016ofw}. 
A point to be noted here is that, unlike the one-loop coefficients of Ref.~\cite{Makino:2014taa}, in the calculation of two-loop coefficients of Ref.~\cite{Harlander:2018zpi}, the equation of motion (EoM) in the continuum,
\begin{equation}
	\bar\psi_f(x)\left(\frac{1}{2}\stackrel{\leftrightarrow}{\not\!\!D} + m_{0,f} \right) \psi_f(x) = 0 ,
\label{eq:EoM}
\end{equation}
is used for quark operators, where
$
   \overleftrightarrow{D}_\mu\equiv D_\mu-\overleftarrow{D}_\mu
$
and $m_{0,f}$ is the bare quark mass for the $f$'th flavor.
With this EoM, we can reduce the number of independent operators and coefficients for EMT.
This should cause no effects after taking the continuum limit when the EMT operators are isolated.
On finite lattices, however, the EoM gets $O(a)$ lattice corrections which may shift the results for the EMT. 

Another point to be addressed in this paper is a technical issue of the choice of the renormalization scale in the matching coefficients of the SF\textit{t}X method.
As shown explicitly in Sec.~\ref{sec:method}, the matching coefficients are written in terms of the flow-time~$t$,  the running coupling $g$ and masses $m_f$ in the $\overline{\textrm{MS}}$ scheme at the renormalization scale~$\mu$, and $\mu$ itself.
Here, $\mu$ is free to choose as far as the perturbative expansion is valid --- the final results for physical observables should be insensitive to the choice of~$\mu$.
In numerical procedures, however, the perturbative expansion is truncated at a finite order and neglected higher-order corrections in the matching coefficients may cause errors in the results.
Because the quality of the perturbative expansion is affected by the choice of~$\mu$, we may control these errors to some extent by an appropriate choice of~$\mu$.
We show that the~$\mu_0$-scale proposed by Harlander \textit{et al.}~\cite{Harlander:2018zpi} helps us to have better signals over the conventional choice~$\mu_d=1/\sqrt{8t}$.

This paper is organized as follows: 
In~Sec.~\ref{sec:method}, we summarize the essence of the SF\textit{t}X method and introduce the one- and two-loop matching coefficients. 
Our simulation parameters are given in~Sec.~\ref{sec:parameters}.
We then discuss the issue of the renormalization scale in~Sec.~\ref{sec:mu0}. 
Our test of two-loop matching coefficients are shown in~Sec.~\ref{sec:test}.
A summary is given in~Sec.~\ref{sec:summary}.
In~Appendix~\ref{sec:notations}, we define the group factors appearing in perturbative expressions of the matching coefficients.
In~Appendix~\ref{sec:confirmation}, we confirm that the one-loop coefficients of~Ref.~\cite{Makino:2014taa} are consistent with the one-loop part of~Ref.~\cite{Harlander:2018zpi}.

%%%%%%%%%%%%%%%%%%%%%%%%%%%%%%%%%%%%%%%%%%%%%%%%%%%%
\section{The $\textrm{SF\textit{t}X}$ method}
\label{sec:method}

%%%%%%%%%%%%%%%%%%%%%%%%%%%%%%%%%%%%%%%%%%%%%%%%%%%%
\subsection{Gradient flow}
\label{sec:GF}
Our flow equations are identical to those given in Refs.~\cite{Luscher:2010iy}
and \cite{Luscher:2013cpa}. That is, for the gauge field, we set%
\footnote{In what follows, summations are always understood over repeated Lorentz indices, $\mu$, $\nu$, \dots, %over $0$, $1$, $2$, and~$3$, 
and adjoint indices, $a$, $b$, \dots. On the other hand, without stated otherwise, summation over repeated flavor indices, $f$, $f'=u$, $d$, $s$ is not assumed.}
\begin{equation}
   \partial_tB_\mu(t,x)=D_\nu G_{\nu\mu}(t,x),\qquad
   B_\mu(t=0,x)=A_\mu(x),
\label{eq:(2.1)}
\end{equation}
where the field strength and the covariant derivative of the flowed gauge field are 
\begin{equation}
   G_{\mu\nu}(t,x)
   =\partial_\mu B_\nu(t,x)-\partial_\nu B_\mu(t,x)
   +[B_\mu(t,x),B_\nu(t,x)],
\label{eq:(2.2)}
\end{equation}
and
\begin{equation}
   D_\nu G_{\nu\mu}(t,x)
   =\partial_\nu G_{\nu\mu}(t,x)+[B_\nu(t,x),G_{\nu\mu}(t,x)],
\label{eq:(2.3)}
\end{equation}
respectively. For the quark fields, we set
\begin{align}
   &\partial_t\chi_f(t,x)=\Delta\chi_f(t,x),\qquad
   \chi_f(t=0,x)=\psi_f(x),
\label{eq:(2.4)}
\\
   &\partial_t\Bar{\chi}_f(t,x)
   =\Bar{\chi}_f(t,x)\overleftarrow{\Delta},
   \qquad\Bar{\chi}_f(t=0,x)=\Bar{\psi}_f(x),
\label{eq:(2.5)}
\end{align}
where $f=u$, $d$, $s$, denotes the flavor index, and
\begin{align}
   &\Delta\chi_f(t,x)\equiv D_\mu D_\mu\chi_f(t,x),\qquad
   D_\mu\chi_f(t,x)\equiv\left[\partial_\mu+B_\mu(t,x)\right]\chi_f(t,x),
\label{eq:(2.6)}
\\
   &\Bar{\chi}_f(t,x)\overleftarrow{\Delta}
   \equiv\Bar{\chi}_f(t,x)\overleftarrow{D}_\mu\overleftarrow{D}_\mu,
   \qquad\Bar{\chi}_f(t,x)\overleftarrow{D}_\mu
   \equiv\Bar{\chi}_f(t,x)\left[\overleftarrow{\partial}_\mu-B_\mu(t,x)\right].
\label{eq:(2.7)}
\end{align}
Note that our flow equations are independent of the flavor.

%%%%%%%%%%%%%%%%%%%%%%%%%%%%%%%%%%%%%%%%%%%%%%%%%%%%
\subsection{Energy-momentum tensor with two-loop matching coefficients}
\label{sec:EMT}

In terms of unflowed operators, the EMT under the dimensional regularization is given by 
\begin{equation}
   T_{\mu\nu}(x)
   =\frac{1}{g_0^2}\left[
   \mathcal{O}_{1,\mu\nu}(x)
   -\frac{1}{4}\mathcal{O}_{2,\mu\nu}(x)
   \right]
   +\frac{1}{4}\mathcal{O}_{3,\mu\nu}(x)
   -\frac{1}{2}\mathcal{O}_{4,\mu\nu}(x)
   -\mathcal{O}_{5,\mu\nu}(x), 
\label{eq:(1.7)}
\end{equation}
where
\begin{align}
   \mathcal{O}_{1,\mu\nu}(x)&\equiv
   F_{\mu\rho}^a(x)F_{\nu\rho}^a(x),
\label{eq:(1.2)}
\\
   \mathcal{O}_{2,\mu\nu}(x)&\equiv
   \delta_{\mu\nu}F_{\rho\sigma}^a(x)F_{\rho\sigma}^a(x),
\label{eq:(1.3)}
\\
   \mathcal{O}_{3,\mu\nu}(x)&\equiv
   \sum_f\Bar{\psi}_f(x)
   \left(\gamma_\mu\overleftrightarrow{D}_\nu
   +\gamma_\nu\overleftrightarrow{D}_\mu\right)
   \psi_f(x),
\label{eq:(1.4)}
\\
   \mathcal{O}_{4,\mu\nu}(x)&\equiv
   \delta_{\mu\nu}
   \sum_f\Bar{\psi}_f(x)
   \overleftrightarrow{\Slash{D}}
   \psi_f(x),
\label{eq:(1.5)}
\\
   \mathcal{O}_{5,\mu\nu}(x)&\equiv
   \delta_{\mu\nu}
   \sum_fm_{f,0}\Bar{\psi}_f(x)
   \psi_f(x),
\label{eq:(1.6)}
\end{align}
with~$F_{\mu\nu}^a(x)$ the field strength of unflowed gauge field~$A^a_\mu(x)$.
Here and in what follows, we assume for notational simplicity that the vacuum expectation value (VEV), i.e., the expectation value at zero-temperature, is subtracted in each operator.

In terms of \textit{flowed} operators, 
\begin{align}
   \Tilde{\mathcal{O}}_{1,\mu\nu}(t,x)&\equiv
   G_{\mu\rho}^a(t,x)G_{\nu\rho}^a(t,x),
\label{eq:(1.8)}
\\
   \Tilde{\mathcal{O}}_{2,\mu\nu}(t,x)&\equiv
   \delta_{\mu\nu}G_{\rho\sigma}^a(t,x)G_{\rho\sigma}^a(t,x),
\label{eq:(1.9)}
\\
   \Tilde{\mathcal{O}}_{3,\mu\nu}(t,x)&\equiv
   \sum_f\mathring{\Bar{\chi}}_f(t,x)
   \left(\gamma_\mu\overleftrightarrow{D}_\nu
   +\gamma_\nu\overleftrightarrow{D}_\mu\right)
   \mathring{\chi}_f(t,x),
\label{eq:(1.10)}
\\
   \Tilde{\mathcal{O}}_{4,\mu\nu}(t,x)&\equiv
   \delta_{\mu\nu}
   \sum_f\mathring{\Bar{\chi}}_f(t,x)
   \overleftrightarrow{\Slash{D}}
   \mathring{\chi}_f(t,x),
\label{eq:(1.11)}
\\
   \Tilde{\mathcal{O}}_{5,\mu\nu}(t,x)&\equiv
   \delta_{\mu\nu}
   \sum_f m_f\mathring{\Bar{\chi}}_f(t,x)
   \mathring{\chi}_f(t,x),
\label{eq:(1.12)}
\end{align}
the EMT is expressed as \cite{Makino:2014taa,Taniguchi:2016ofw}
\begin{align}
   T_{\mu\nu}(x)
   = & \, c_1(t)\left[
   \Tilde{\mathcal{O}}_{1,\mu\nu}(t,x)
   -\frac{1}{4}\Tilde{\mathcal{O}}_{2,\mu\nu}(t,x)
   \right]
   +c_2(t)\Tilde{\mathcal{O}}_{2,\mu\nu}(t,x)
\notag\\
   &
   +c_3(t)\left[
   \Tilde{\mathcal{O}}_{3,\mu\nu}(t,x)
   -2\Tilde{\mathcal{O}}_{4,\mu\nu}(t,x)
   \right]
   +c_4(t)\Tilde{\mathcal{O}}_{4,\mu\nu}(t,x)
   \notag\\
   &
   +c_5(t)\Tilde{\mathcal{O}}_{5,\mu\nu}(t,x)
   +O(t),
\label{eq:(1.15)}
\end{align}
where
\begin{align}
   \mathring{\chi}_f(t,x)
   &\equiv\sqrt{\frac{-2\dim(R)}
   {(4\pi)^2t^2
   \left\langle\Bar{\chi}_f(t,x)\overleftrightarrow{\Slash{D}}\chi_f(t,x)
   \right\rangle}}
   \,\chi_f(t,x),
\label{eq:(1.13)}\\
   \mathring{\Bar{\chi}}_f(t,x)
   &\equiv\sqrt{\frac{-2\dim(R)}
   {(4\pi)^2t^2
   \left\langle\Bar{\chi}_f(t,x)\overleftrightarrow{\Slash{D}}\chi_f(t,x)
   \right\rangle}}
   \,\Bar{\chi}_f(t,x),
\label{eq:(1.14)}
\end{align}
are ``ringed'' quark fields introduced in~Ref.~\cite{Makino:2014taa} to carry out a wave function renormalization of quark fields nonperturbatively, where the operators in the denominator are not VEV-subtracted.
The $O(t)$ term in the right-hand side of~Eq.~\eqref{eq:(1.15)} can be removed by a~$t\to0$ extrapolation.
Explicit forms of the matching coefficients $c_1$, $\cdots$, $c_5$ to the one-loop order (NLO) are given in~Ref.~\cite{Makino:2014taa}.

Recently, two-loop (NNLO) calculation of the matching coefficients has been completed by Harlander, Kluth, and Lange~\cite{Harlander:2018zpi}.
Unlike the calculation of Ref.~\cite{Makino:2014taa}, 
the EoM~\eqref{eq:EoM} was used in Ref.~\cite{Harlander:2018zpi} assuming that the EMT operators are spatially separated from other composite operators. See also Sec.~5 of~Ref.~\cite{Makino:2014taa}.
In terms of unflowed operators, Eq.~\eqref{eq:EoM} reads as
\begin{equation}
   \frac{1}{2}\mathcal{O}_{4,\mu\nu}(x)
   +\mathcal{O}_{5,\mu\nu}(x)=0.
\label{eq:(1.18)}
\end{equation}
Note that Eq.~\eqref{eq:(1.18)} implies that the last two terms of Eq.~\eqref{eq:(1.7)} cancel with each other:
\begin{equation}
   T_{\mu\nu}(x)
   =\frac{1}{g_0^2}\left[
   \mathcal{O}_{1,\mu\nu}(x)
   -\frac{1}{4}\mathcal{O}_{2,\mu\nu}(x)
   \right]
   +\frac{1}{4}\mathcal{O}_{3,\mu\nu}(x).
\label{eq:(1.25)}
\end{equation}
In terms of the flowed operators, the EoM is
\begin{align}
   &\delta_{\mu\nu}
   \sum_f\left[\Bar{\psi}_f(x)
   \left(\overleftrightarrow{\Slash{D}}+2m_{f,0}\right)
   \psi_f(x)\right]
\notag\\
   &=d_2(t)\Tilde{\mathcal{O}}_{2,\mu\nu}(t,x)
   +\mathring{d}_4(t)\Tilde{\mathcal{O}}_{4,\mu\nu}(t,x)
   +\mathring{d}_5(t)\Tilde{\mathcal{O}}_{5,\mu\nu}(t,x)+O(t)=0,
\label{eq:(1.19)}
\end{align}
where, in the one-loop level, the coefficients $d_i(t)$ are given by 
(cf.\ Eqs.~(5.3)--(5.5) of~Ref.~\cite{Makino:2014taa})
\begin{align}
   d_2(t)
   &=\frac{1}{4g^2}\frac{g^2}{(4\pi)^2}\left(-\frac{20}{3}\right)T_F,
\label{eq:(1.72)}\\
   \mathring{d}_4(t)
   &=1+\frac{g^2}{(4\pi)^2}C_F\left(-\frac{1}{2}+\ln432\right),
\label{eq:(1.73)}\\
   \mathring{d}_5(t)
   &=2\left\{
   1
   +\frac{g^2}{(4\pi)^2}C_F\left[3L(\mu,t)+2+\ln432\right]
   \right\}.
\label{eq:(1.74)}
\end{align}
where $g$ is the running coupling in the $\overline{\textrm{MS}}$-scheme at the renormalization scale $\mu$, and 
\begin{equation}
   L(\mu,t)\equiv\ln(2\mu^2t)+\gamma_E
\label{eq:(1.39)}
\end{equation}
with $\gamma_E$ the Euler-Mascheroni constant. 
The definitions of the group factors ($T_F$ etc.) are summarized in~Appendix~\ref{sec:notations}.

Using Eq.~\eqref{eq:(1.19)}, we can eliminate $\Tilde{\mathcal{O}}_{5,\mu\nu}(t,x)$ order by order in perturbation theory (up to $O(t)$ terms), to obtain
\begin{equation}
   T_{\mu\nu}(x)
   = \Check{c}_1(t)\Tilde{\mathcal{O}}_{1,\mu\nu}(t,x)
   +\Check{c}_2(t)\Tilde{\mathcal{O}}_{2,\mu\nu}(t,x)
   +\mathring{\Check{c}}_3(t)\Tilde{\mathcal{O}}_{3,\mu\nu}(t,x)
   +\mathring{\Check{c}}_4(t)\Tilde{\mathcal{O}}_{4,\mu\nu}(t,x)+O(t).
\label{eq:(1.16)}
\end{equation}
The matching coefficients $\Check{c}_1(t)$ and $\Check{c}_2(t)$ to the two-loop order are given in~Ref.~\cite{Harlander:2018zpi} as
\begin{align}
   &\Check{c}_1(t)
\notag\\
   &=\frac{1}{g^2}
   \biggl(1
   +\frac{g^2}{(4\pi)^2}
   \left[-\beta_0L(\mu,t)-\frac{7}{3}C_{\!A}+\frac{3}{2}T_F\right]
\notag\\
   &\quad{}
   +\frac{g^4}{(4\pi)^4}
   \biggl\{
   -\beta_1L(\mu,t)
   +{C_{\!A}}^2
   \left(
   -\frac{14482}{405}
   -\frac{16546}{135}\ln2
   +\frac{1187}{10}\ln3
   \right)
 \notag\\
   &\qquad\qquad\quad{}
   +C_{\!A}T_F
   \left[
   \frac{59}{9}\Li_2\left(\frac{1}{4}\right)
   +\frac{10873}{810}
   +\frac{73}{54}\pi^2
   -\frac{2773}{135}\ln2
   +\frac{302}{45}\ln3
   \right]
\notag\\
   &\qquad\qquad\quad{}
   +C_FT_F
   \left[
   -\frac{256}{9}\Li_2\left(\frac{1}{4}\right)
   +\frac{2587}{108}
   -\frac{7}{9}\pi^2-\frac{106}{9}\ln2-\frac{161}{18}\ln3
   \right]
   \biggr\}
   \biggr),
\label{eq:(1.32)}
\end{align}
\begin{align}
   &\Check{c}_2(t)
\notag\\
   &=\frac{1}{4g^2}
   \biggl(
   -1
   +\frac{g^2}{(4\pi)^2}
   \left[\beta_0L(\mu,t)+\frac{25}{6}C_{\!A}-3T_F\right]
\notag\\
   &\quad{}
   +\frac{g^4}{(4\pi)^4}
   \biggl\{
   \beta_1L(\mu,t)
   +{C_{\!A}}^2
   \left(
   \frac{56713}{1620}
   -\frac{1187}{10}\ln3
   +\frac{16546}{135}\ln2
   \right)
\notag\\
   &\qquad\qquad\quad{}
   +C_{\!A}T_F
   \left[
   -\frac{59}{9}\Li_2\left(\frac{1}{4}\right)
   -\frac{6071}{405}
   -\frac{73}{54}\pi^2
   +\frac{2287}{135}\ln2
   -\frac{361}{90}\ln3
   \right]
\notag\\
    &\qquad\qquad\quad{}
    +C_FT_F
   \left[
   \frac{220}{9}\Li_2\left(\frac{1}{4}\right)
   -\frac{1757}{54}
   +\frac{10}{9}\pi^2
   -\frac{164}{9}\ln2
   +\frac{247}{9}\ln3
   \right]
   \biggr\}
   \biggl),
\label{eq:(1.33)}
\end{align}
where $\Li_2(z)$ is the dilogarithm function with $\Li_2(1/4)=0.26765263908\cdots$,
and $\beta_0$ and $\beta_1$ are the first two coefficients of the beta function,
\begin{align}
   \beta_0&=\frac{11}{3}C_{\!A}-\frac{4}{3}T_F,
\label{eq:(1.30)}
\\
   \beta_1&=\frac{34}{3}{C_{\!A}}^2-\left(4C_F+\frac{20}{3}C_{\!A}\right)T_F.
\label{eq:(1.31)}
\end{align}

The matching coefficients~$\mathring{\Check{c}}_3(t)$ and $\mathring{\Check{c}}_4(t)$ are given by 
\begin{equation}
   \mathring{\Check{c}}_i(t)
   \equiv\Check{c}_i(t)\zeta_\chi(t)^{-1},\qquad\text{for $i=3$, $4$},
\label{eq:(1.46)}
\end{equation}
using 
\begin{align}
   &\Check{c}_3(t)
\notag\\
   &=\frac{1}{4}
   \biggl(1
   +\frac{g^2}{(4\pi)^2}
   \left[\frac{3}{2}C_F+\frac{\gamma_{\chi,0}}{2}L(\mu,t)\right]
\notag\\
   &\quad{}
   +\frac{g^4}{(4\pi)^4}
   \biggl\{
   \frac{\gamma_{\chi,0}}{4}
   \left(\beta_0+\frac{\gamma_{\chi,0}}{2}\right)
   \left[L(\mu,t)^2+L(\mu,t)\right]
   +\frac{\gamma_{\chi,1}}{2}L(\mu,t)
\notag\\
   &\qquad\qquad\quad{}
   +C_F^2
   \left[
   -\frac{137}{9}\Li_2\left(\frac{1}{4}\right)
   -\frac{559}{216}
   +\frac{103}{108}\pi^2
   -\frac{1736}{27}\ln2
   +\frac{122}{3}\ln3
   -4(\ln2)^2
   \right]
\notag\\
   &\qquad\qquad\quad{}
   +C_FT_F
   \left[
   -\frac{136}{9}\Li_2\left(\frac{1}{4}\right)
   -\frac{3377}{810}
   -\frac{7}{9}\pi^2
   +\frac{1232}{135}\ln2
   -\frac{136}{15}\ln3
   \right]
\notag\\
   &\qquad\qquad\quad{}
   +C_{\!A}C_F
   \left[
   -\frac{365}{9}\Li_2\left(\frac{1}{4}\right)
   +\frac{261829}{3240}
   +\frac{77}{108}\pi^2
   +\frac{5788}{45}\ln2
   -\frac{2102}{15}\ln3
   -4(\ln2)^2
   \right]
   \biggr\}
   \biggr),
\label{eq:(1.44)}
\end{align}
and
\begin{align}
   &\Check{c}_4(t)
\notag\\
   &=\frac{C_F}{2}
   \biggl(
   \frac{g^2}{(4\pi)^2}
\notag\\
   &\quad{}
   +\frac{g^4}{(4\pi)^4}
   \biggl\{
   \left[\beta_0+\frac{\gamma_{\chi,0}}{2}\right]L(\mu,t)
\notag\\
   &\qquad\qquad\quad{}
   +C_F
   \left[
   -\frac{161}{18}\Li_2\left(\frac{1}{4}\right)
   -\frac{41}{54}
   -\frac{55}{108}\pi^2
   -\frac{1105}{27}\ln2
   +\frac{101}{6}\ln3
   \right]
\notag\\
   &\qquad\qquad\quad{}
   +T_F
   \left[
   \frac{25}{9}\Li_2\left(\frac{1}{4}\right)
   -\frac{20573}{1620}
   +\frac{5}{18}\pi^2
   +\frac{6559}{135}\ln2
   -\frac{679}{30}\ln3
   \right]
\notag\\
   &\qquad\qquad\quad{}
   +C_{\!A}
   \left[
   \frac{257}{36}\Li_2\left(\frac{1}{4}\right)
   -\frac{137}{405}
   +\frac{11}{216}\pi^2
   -\frac{419}{90}\ln2
   +\frac{1157}{60}\ln3
   \right]
   \biggr\}
   \biggr),
\label{eq:(1.45)}
\end{align}
given in Ref.~\cite{Harlander:2018zpi},
where
\begin{align}
  \gamma_{\chi,0}&=6C_F,
\label{eq:(1.35)}
\\
  \gamma_{\chi,1}
  &=C_{\!A}C_F\left(\frac{223}{3}-16\ln2\right)
  -C_F^2(3+16\ln2)-\frac{44}{3}C_FT_F.
\label{eq:(1.36)}
\end{align}
Here, $\zeta_\chi(t)$ adjusts the normalization of quark fields to that of the ringed variables in~Eqs.~\eqref{eq:(1.13)} and~\eqref{eq:(1.14)},
and is given by 
\begin{align}
   &\zeta_\chi(t)
\notag\\
   &=
   1
   +\frac{g^2}{(4\pi)^2}
   \left[\frac{\gamma_{\chi,0}}{2}L(\mu,t)-3C_F\ln3-4C_F\ln2\right]
\notag\\
   &\qquad{}
   +\frac{g^4}{(4\pi)^4}
   \Bigl\{
   \frac{\gamma_{\chi,0}}{4}
   \left(\beta_0+\frac{\gamma_{\chi,0}}{2}\right)L(\mu,t)^2
\notag\\
   &\qquad\qquad\qquad{}
   +\left[
   \frac{\gamma_{\chi,1}}{2}
   -\frac{\gamma_{\chi,0}}{2}
   \left(\beta_0+\frac{\gamma_{\chi,0}}{2}\right)\ln3
   -\frac{2}{3}\gamma_{\chi,0}
   \left(\beta_0+\frac{\gamma_{\chi,0}}{2}\right)\ln2
   \right]L(\mu,t)
\notag\\
   &\qquad\qquad\qquad{}
   +C_2
   \Bigr\},
\label{eq:(1.47)}
\end{align}
with
\begin{equation}
   C_2\equiv-23.8C_{\!A}C_F+30.4C_F^2-3.92C_FT_F.
\label{eq:(1.37)}
\end{equation}
Its inverse reads as
\begin{align}
   &\zeta_\chi(t)^{-1}
\notag\\
   &=1
   +\frac{g^2}{(4\pi)^2}
   \left[-\frac{\gamma_{\chi,0}}{2}L(\mu ,t)+C_F\ln432\right]
\notag\\
   &\qquad{}
   +\frac{g^4}{(4\pi)^4}
   \Bigl\{
   \frac{\gamma_{\chi,0}}{4}
   \left(-\beta_0+\frac{\gamma_{\chi,0}}{2}\right)L(\mu,t)^2
\notag\\
   &\qquad\qquad\qquad{}
   +\left[
   -\frac{\gamma_{\chi,1}}{2}
   +\frac{\gamma_{\chi, 0}}{6}
   \left(\beta_0+\frac{\gamma_{\chi,0}}{2}-6C_F\right)\ln432
   \right]
   L(\mu,t)
\notag\\
   &\qquad\qquad\qquad{}
   -C_2+C_F^2(\ln432)^2
   \Bigr\}.
\label{eq:(1.48)}
\end{align}
We then obtain 
\begin{align}
   &\mathring{\Check{c}}_3(t)
\notag\\
   &=\frac{1}{4}
   \biggl(
   1
   +\frac{g^2}{(4\pi)^2}C_F\left(\frac{3}{2}+\ln432\right)
\notag\\
   &\quad{}
   +\frac{g^4}{(4\pi)^4}
   \biggl\{
   \frac{\gamma_{\chi,0}}{6}
   \left(\beta_0+\frac{\gamma_{\chi,0}}{2}-3C_F\right)
   \left(\frac{3}{2}+\ln432\right)L(\mu,t)
\notag\\
   &\qquad\qquad\quad{}
   +C_F^2\left(\frac{3}{2}+\ln432\right)\ln432-C_2
\notag\\
   &\qquad\qquad\quad{}
   +C_F^2
   \left[
   -\frac{137}{9}\Li_2\left(\frac{1}{4}\right)
   -\frac{559}{216}
   +\frac{103}{108}\pi^2
   -\frac{1736}{27}\ln2
   +\frac{122}{3}\ln3
   -4(\ln2)^2
   \right]
\notag\\
   &\qquad\qquad\quad{}
   +C_FT_F
   \left[
   -\frac{136}{9}\Li_2\left(\frac{1}{4}\right)
   -\frac{3377}{810}
   -\frac{7}{9}\pi^2
   +\frac{1232}{135}\ln2
   -\frac{136}{15}\ln3
   \right]
\notag\\
   &\qquad\qquad\quad{}
   +C_{\!A}C_F
   \left[
   -\frac{365}{9}\Li_2\left(\frac{1}{4}\right)
   +\frac{261829}{3240}
   +\frac{77}{108}\pi^2
   +\frac{5788}{45}\ln2
   -\frac{2102}{15}\ln3
   -4(\ln2)^2
   \right]
   \biggr\}
   \biggr),
\label{eq:(1.34)}
\end{align}
where
\begin{align}
   &\mathring{\Check{c}}_4(t)
\notag\\
   &=\frac{C_F}{2}
   \biggl(
   \frac{g^2}{(4\pi)^2}
\notag\\
   &\quad{}
   +\frac{g^4}{(4\pi)^4}
   \biggl\{
   \beta_0L(\mu,t)
\notag\\
   &\qquad\qquad\quad{}
   +C_F\ln432
\notag\\
   &\qquad\qquad\quad{}
   +C_F
   \left[
   -\frac{161}{18}\Li_2\left(\frac{1}{4}\right)
   -\frac{41}{54}
   -\frac{55}{108}\pi^2
   -\frac{1105}{27}\ln 2
   +\frac{101}{6}\ln3
   \right]
\notag\\
   &\qquad\qquad\quad{}
   +T_F
   \left[
   \frac{25}{9}\Li_2\left(\frac{1}{4}\right)
   -\frac{20573}{1620}
   +\frac{5}{18}\pi^2
   +\frac{6559}{135}\ln2
   -\frac{679}{30}\ln3
   \right]
\notag\\
    &\qquad\qquad\quad{}
    +C_{\!A}
   \left[
   \frac{257}{36}\Li_2\left(\frac{1}{4}\right)
   -\frac{137}{405}
   +\frac{11}{216}\pi^2
   -\frac{419}{90}\ln2
   +\frac{1157}{60}\ln3
   \right]
   \biggr\}
   \biggr).
\label{eq:(1.38)}
\end{align}

In Appendix~\ref{sec:confirmation}, we confirm that the results of~Ref.~\cite{Makino:2014taa} are consistent with the one-loop parts of~Eqs.~\eqref{eq:(1.32)}--\eqref{eq:(1.38)}. 

From the diagonal components of the EMT, we compute the pressure and the energy density as
\begin{equation}
p/T^4  = \sum_{i=1}^3 \langle T_{ii}\rangle /(3T^4), \hspace{5mm}
\epsilon/T^4 = - \langle T_{00}\rangle/T^4 .
\end{equation}
The entropy density and trace anomaly are then computed as $(\epsilon+p)/T$ and $\epsilon-3p$, respectively.

%%%%%%%%%%%%%%%%%%%%%%%%%%%%%%%%%%%%%%%%%%%%%%%%%%%%
\subsection{Extrapolation to $t\to0$}
\label{sec:t0extrapolation}

To extract physical results of EMT in~Eqs.~\eqref{eq:(1.15)}, \eqref{eq:(1.16)}, etc., we remove contamination of $O(t)$ terms in the right-hand side of these equations by extrapolating them to~$t\to0$. 
On finite lattices, lattice artifacts contaminate additionally.
With $O(a)$-improved Wilson quarks we adopt, the lattice artifacts start with $O(a^2)$ and we expect the EMT, for example, to be 
\begin{eqnarray}
   T_{\mu\nu}(t,x,a)
   &=&T_{\mu\nu}(x) + t S_{\mu\nu}(x) 
   +A_{\mu\nu}\frac{a^2}{t}+\sum_f B^f_{\mu\nu}(am_f)^2+C_{\mu\nu}(aT)^2 \nonumber\\  &&
   +D_{\mu\nu}(a\Lambda_{\mathrm{QCD}})^2
   +a^2S'_{\mu\nu}(x)+O(a^4,t^2),
\label{eq:a2overt}
\end{eqnarray}
where $T_{\mu\nu}(x)$ is the physical EMT, $S_{\mu\nu}$ and $S'_{\mu\nu}$ are contaminations of dimension-six operators with the same quantum number, and $A_{\mu\nu}$, $B^f_{\mu\nu}$, $C_{\mu\nu}$, and $D_{\mu\nu}$ are those from dimension-four operators. 
Though both $a\to0$ and $t\to0$ extrapolations are needed to extract the physical EMT, it is often attractive to reserve the $a\to0$ extrapolation for a late stage of numerical analyses.
To perform a $t\to0$ extrapolation on finite lattices~$a\ne0$, the influence of singular terms such as $a^2/t$ must be suppressed. 
This will be possible when we have a window in~$t$ in which the linear terms dominate (``linear window'')~\cite{Taniguchi:2016ofw}.

We found in the study of~Ref.~\cite{Taniguchi:2016ofw} that, depending on the observable and simulation parameters, we do have ranges of $t$ in which the data show well linear behavior.
We performed linear~$t\to0$ extrapolation of observables when a linear window is available and obtained reasonable results, as introduced in~Sec.~\ref{sec:intro}.
We think that the success of the SF\textit{t}X method in~Ref.~\cite{Taniguchi:2016ofw} is largely due to the fineness of the lattices studied.
In this paper, we adopt the same strategy. 
We also discuss a method which may be used to improve linear behaviors in~Sec.~\ref{sec:mu0}.

\begin{figure}[htb]
\centering
\includegraphics[width=7cm]{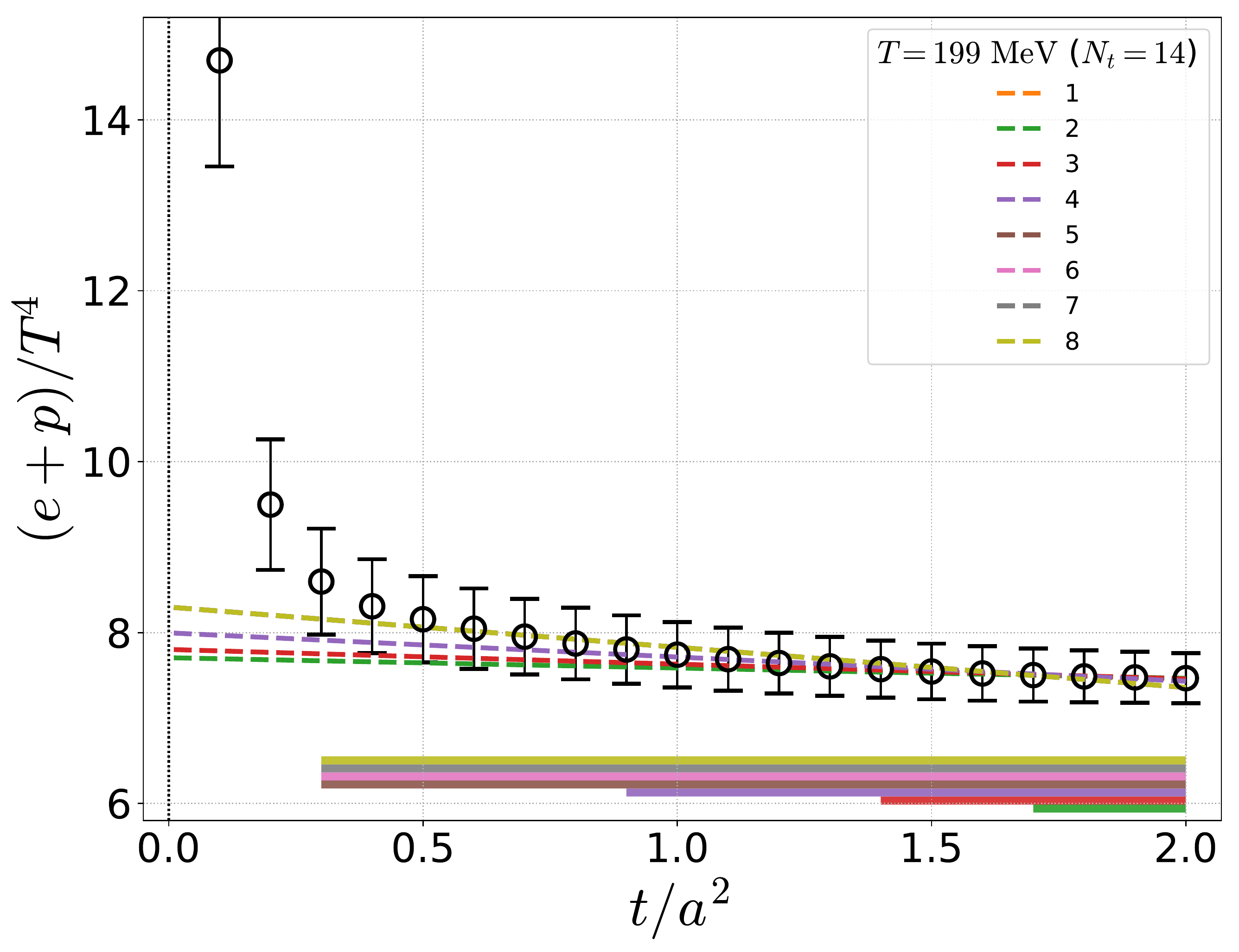}
\includegraphics[width=7cm]{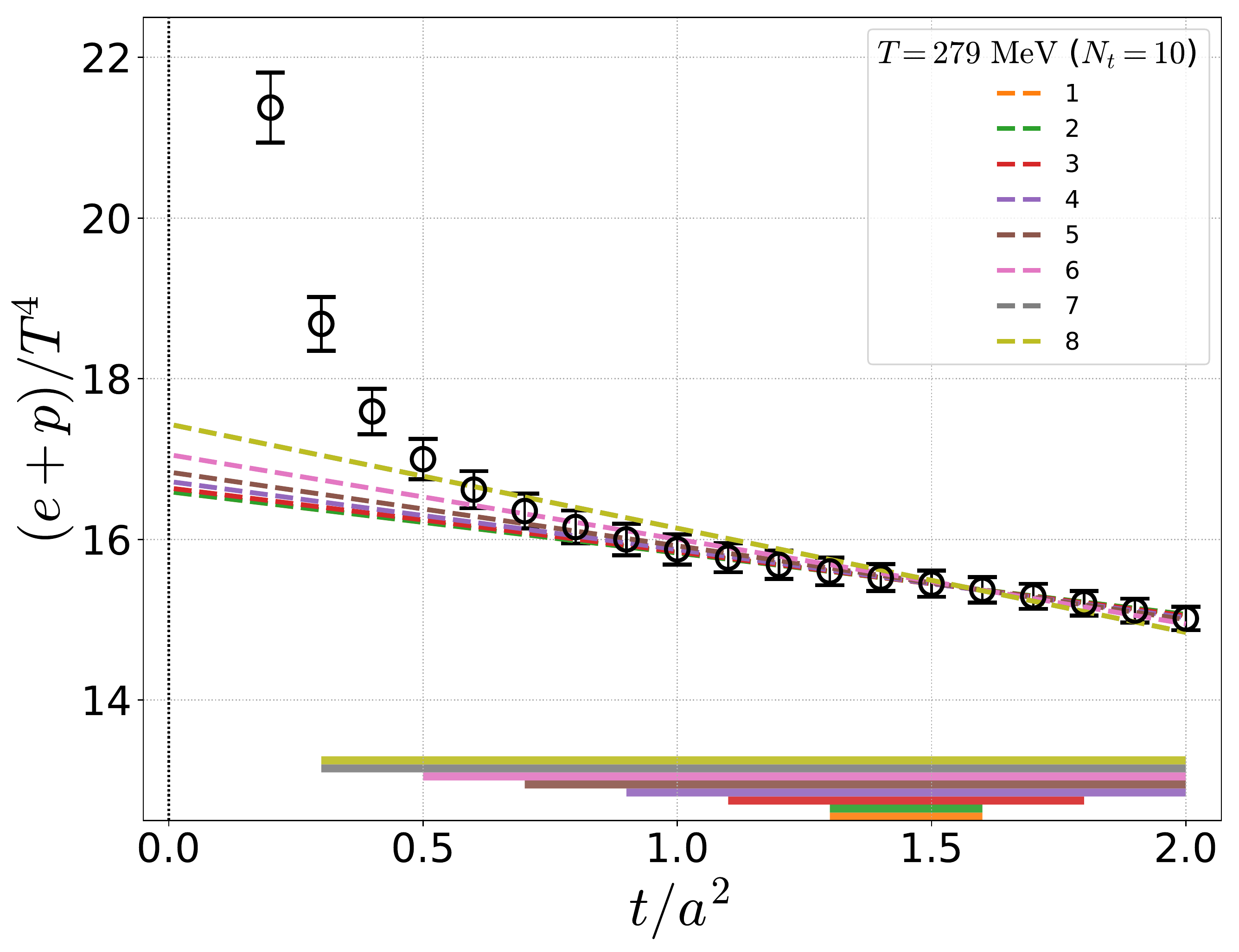}
\includegraphics[width=7cm]{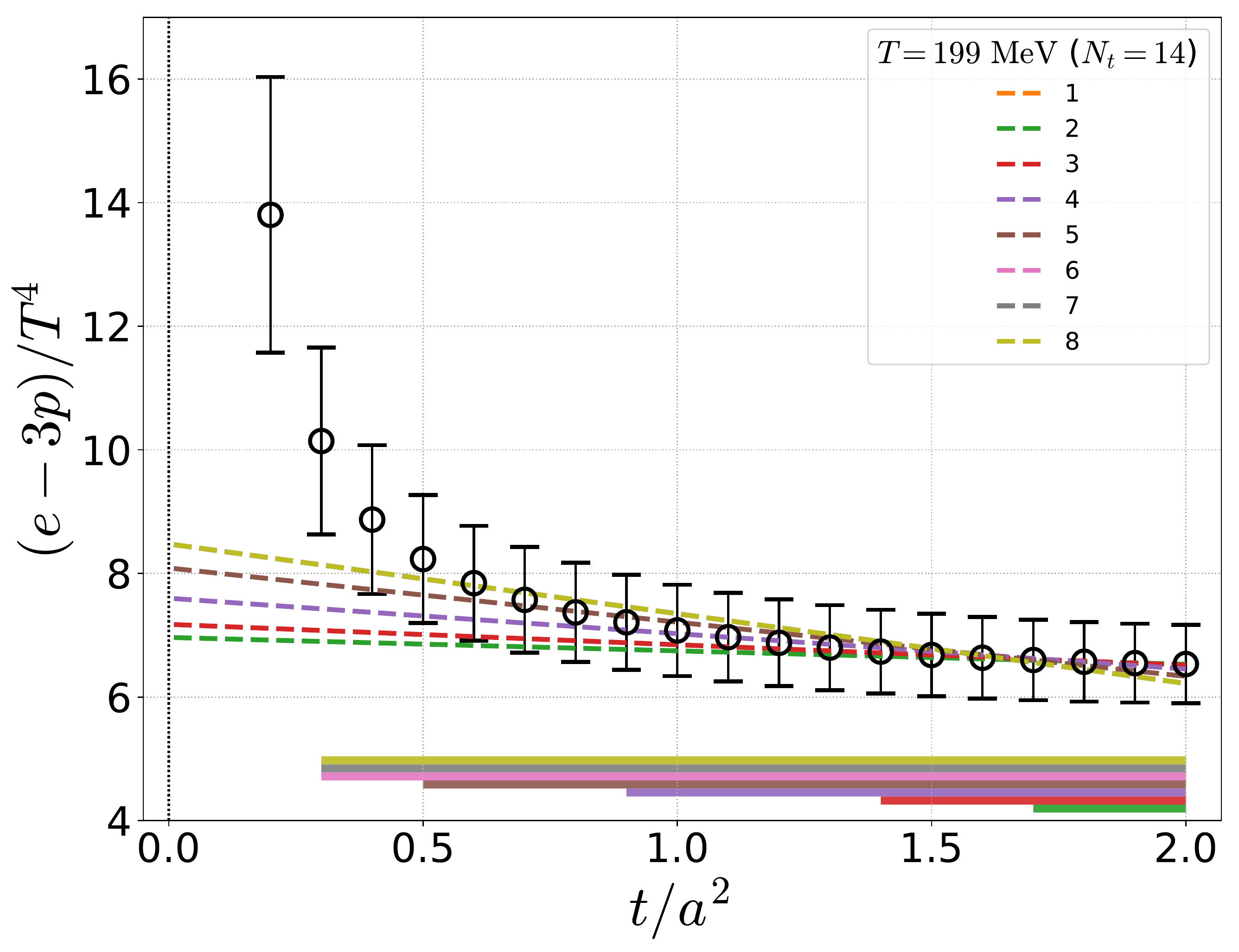}
\includegraphics[width=7cm]{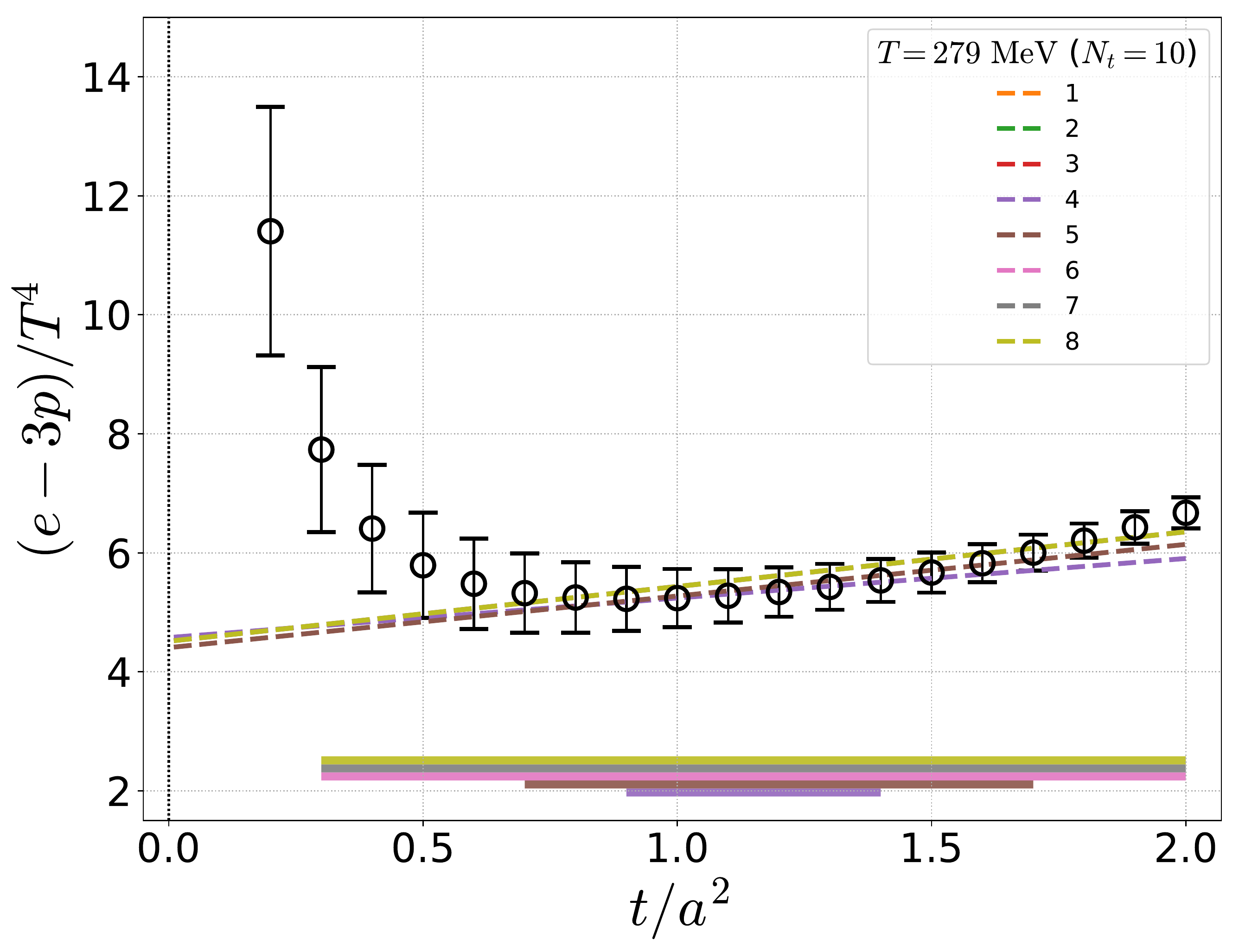}
\includegraphics[width=7cm]{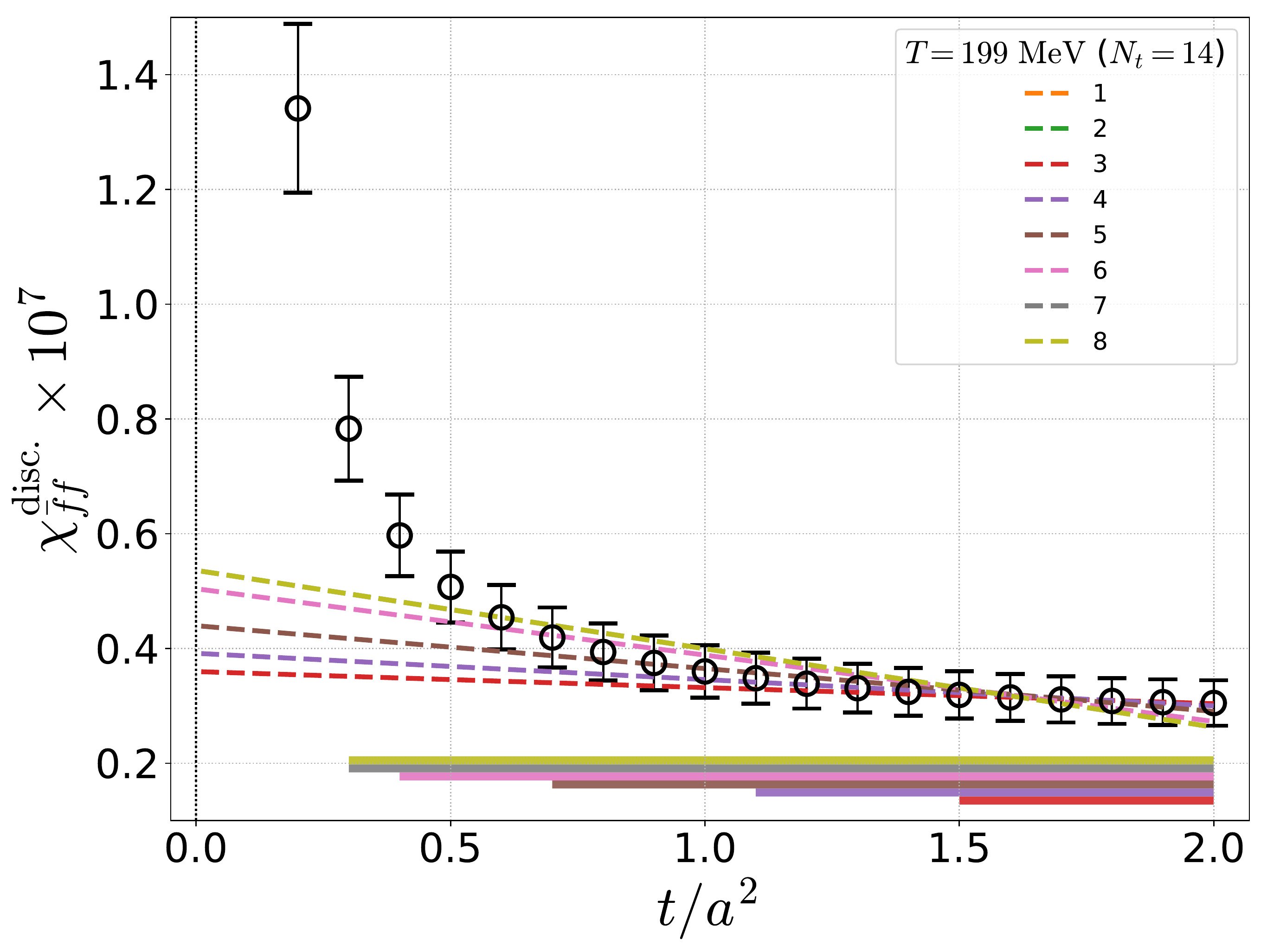}
\includegraphics[width=7cm]{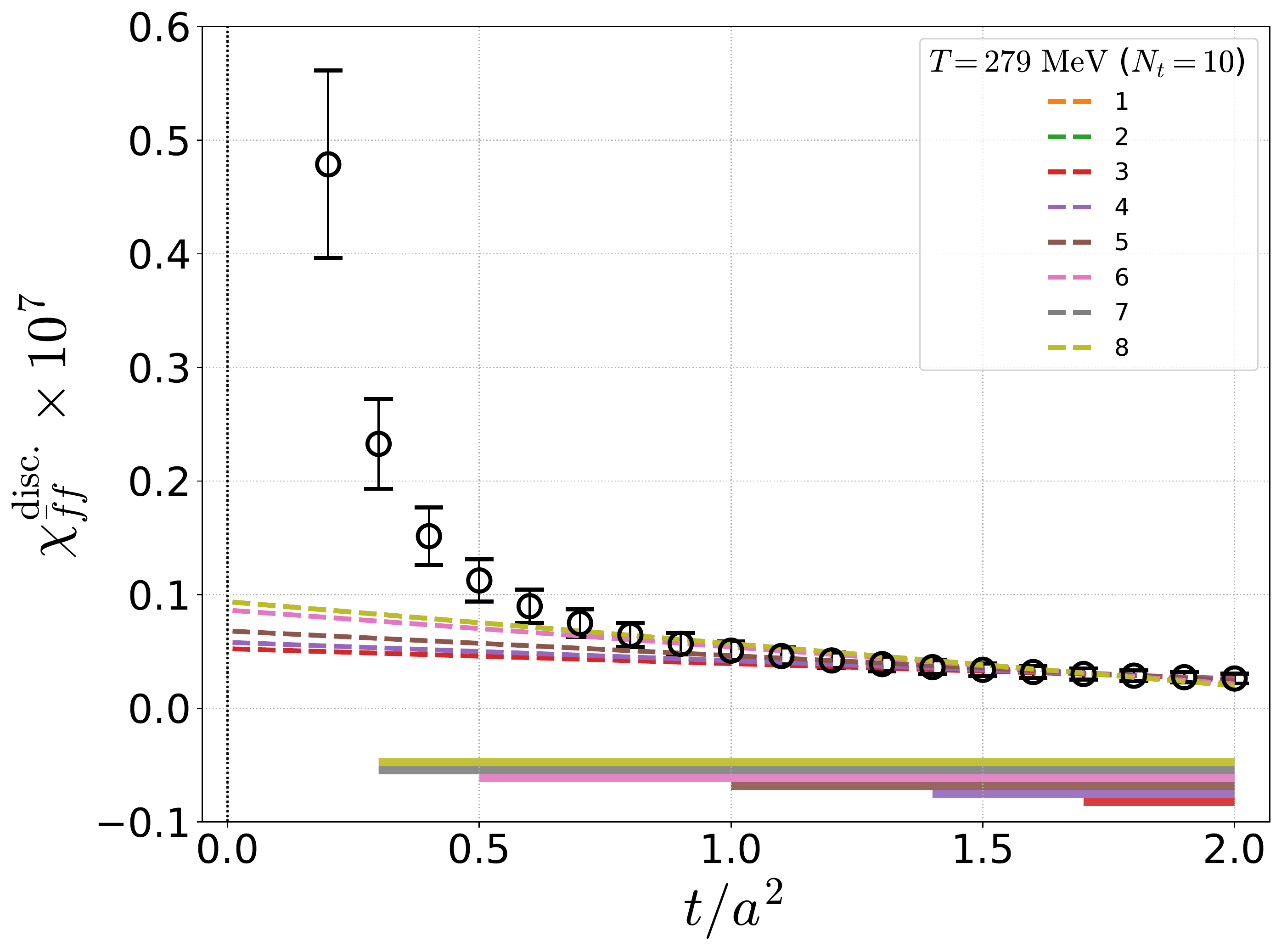}
\vspace{-5mm}
\caption{Typical results of the linear fits with the fitting ranges shown by the bands at the bottom of each plot.
These fitting ranges are selected by the procedure discussed in~Sec.~\ref{sec:t0extrapolation}.
The cutoff values of $\chi^2/N_\textrm{dof}$ for the linear fits 1, 2, 3, $\cdots$, and 8 are $10^{-5}$, $10^{-4}$, $10^{-3}$, $\cdots$, and $10^{+2}$, respectively.
Results of corresponding linear fits are shown by dashed lines with the same color.
The dashed lines with the same fitting range are overlapped with each other, and only the color with the largest fit number is visible.
One-loop matching coefficients of~Ref.~\cite{Makino:2014taa} and the~$\mu_0$-scale are adopted.
Errors are statistical only.
}
\label{fig0}
\end{figure}

We identify linear windows as follows:
First of all, we require the flow-time to satisfy 
$
     \sqrt{2} a \le \sqrt{8t} \le \min(N_t a/2,N_s a/2),
$
\emph{i.e.}, the smearing range $\sqrt{8t}$ by the gradient flow covers the minimal lattice separations to make the smearing effective,
and, simultaneously, is smaller than the half of the smallest lattice extent to avoid finite-size effects due to overlapped smearing.
In terms of the dimensionless flow-time $t/a^2$, these conditions read%
\footnote{In practice, $t/a^2$ is bounded also by the maximum value of $t/a^2$ in the calculation of flowed fields. In this study, we calculate them up to $t/a^2=2.0$.}
\begin{equation}
 \frac{1}{4} \; \le \; \frac{t}{a^2} \; \le \;  t_{1/2} \equiv \frac{1}{8}\left[
   \min\left(\frac{N_t}{2},\frac{N_s}{2}\right)\right]^2 .
\label{eq:t-half}
\end{equation} 
For each observable $\mathcal{O}$, we then look for a range of $t$ in which terms linear in~$t$ look dominating, and try linear extrapolations of the form 
\begin{equation}
 \langle \mathcal{O}(t,a) \rangle = \langle \mathcal{O} \rangle + t\,S_{\mathcal{O}}  %+ O(a^2,t^2) 
\label{eqn:5.3}
\end{equation}
with various choices of the fitting range. 
We then select a temporally best linear fit whose fitting range is the widest within the range~(\ref{eq:t-half}), under the condition that $\chi^2/N_\textrm{dof}$ is smaller than a cutoff value.
In this study, due to limitation of the statistics, we could not obtain a statistically reliable correlation matrix among data at different flow-times $t$.
We thus disregard the correlation among different $t$'s in the calculation of $\chi^2/N_\textrm{dof}$.
This means that the absolute value of $\chi^2/N_\textrm{dof}$ does not have a strong sense --- we can reduce it below any desired value by adding correlated data at intermediate $t$'s.
To obtain reasonable fits, we thus repeat the test varying the cutoff value widely.
In~Fig.~\ref{fig0}, we show some typical results of this test. 
One-loop matching coefficients of~Ref.~\cite{Makino:2014taa} and the~$\mu_0$-scale discussed in~Sec.~\ref{sec:mu0} are adopted.
The cutoffs for the linear fits 1, 2, 3, $\cdots$, and 8 are $10^{-5}$, $10^{-4}$, $10^{-3}$, $\cdots$, and $10^{+2}$, respectively.
The selected fitting range for each fit is shown by a line at the bottom of each plot with the same color.
From~Fig.~\ref{fig0}, we note that the linear fits are stable when the cutoff value is large.
On the other hand, when we require $\chi^2/N_\textrm{dof} < 10^{-3}$ or smaller, our selection procedure for the temporal linear window becomes sometimes unstable and fails to give a window.
Consulting these plots and also requiring that the resulting linear windows are common among similar observables, we decide to choose the fits 5 which require $\chi^2/N_\textrm{dof} < 0.1$ to select optimum linear windows, for all observables we study in this paper. 

To confirm the validity of the linear window and to estimate a systematic error due to the fit ansatz, we also make 
additional fits using the data within the same window:
One is a ``nonlinear fit'' inspired from~Eq.~\eqref{eq:a2overt},
\begin{equation}
 \langle \mathcal{O}(t,a) \rangle
  =\langle \mathcal{O} \rangle +A_{\mathcal{O}} \frac{a^2}{t}+t\,S_{\mathcal{O}}
  + t^2 R_{\mathcal{O}}  .
\label{eqn:5.4}
\end{equation}
Another is a ``linear+log fit'' to estimate the effects of neglected higher-order loop corrections in the matching coefficients.
For the case of one-loop matching coefficients, possible corrections are $O(g^4) \sim O(1/\log^2\mu)$. 
Because $\mu \sim 1/\sqrt{t}$ (see~Sec.~\ref{sec:mu0}), the leading $t$-dependence of $O(g^4)$ terms may be taken by fits of the form 
\begin{equation}
 \langle \mathcal{O}(t,a) \rangle
  =\langle \mathcal{O} \rangle + t\,S_{\mathcal{O}}
  +  \frac{Q_{\mathcal{O}}}{\log^2(\sqrt{8t} \Lambda_{\overline{\mathrm{MS}}})} ,
\label{eqn:5.4l}
\end{equation}
where $\Lambda_{\overline{\mathrm{MS}}} = 332(17)$ MeV is the QCD lambda parameter for three-flavor QCD quoted in the Particle Data Group~\cite{PDG}.
For the case of two-loop matching coefficients, we instead try  
\begin{equation}
 \langle \mathcal{O}(t,a) \rangle
  =\langle \mathcal{O} \rangle + t\,S_{\mathcal{O}}
  +  \frac{Q'_{\mathcal{O}}}{\log^3(\sqrt{8t} \Lambda_{\overline{\mathrm{MS}}})} 
\label{eqn:5.4ll}
\end{equation}
to estimate the~$O(g^6)$ contaminations.
We take the difference between the linear fit and the nonlinear or linear+log fits as an estimate of the systematic error due to the choice of the fit ansatz.

%%%%%%%%%%%%%%%%%%%%%%%%%%%%%%%%%%%%%%%%%%%%%%%%%%%%
\section{Simulation parameters and our numerical methods}
\label{sec:parameters}

The numerical setup for this study is the same as that of Ref.~\cite{Taniguchi:2016ofw}. 
We study $2+1$ flavor QCD with slightly heavy $u$ and $d$ quarks ($m_\pi/m_\rho\simeq0.63$) and approximately physical $s$ quark ($m_{\eta_{ss}}/m_\phi\simeq0.74$) on a relatively fine lattice with the lattice spacing $a\simeq0.07$~fm ($a^{-1}\simeq2.79$~GeV)~\cite{Umeda:2012er,Ishikawa:2007nn}. 
To reduce the finite lattice spacing effects, we adopt the renormalization-group improved Iwasaki gauge action~\cite{Iwasaki:2011np,Iwasaki:1985we} and the $O(a)$-improved Wilson quark action~\cite{Sheikholeslami:1985ij}.
For the clover coefficients of the improved Wilson quark action, we adopt nonperturbatively evaluated values \cite{SAoki2006}.
The bare gauge coupling parameter and the hopping parameters are set to $\beta=2.05$, $\kappa_{ud}=0.1356$, and $\kappa_s=0.1351$. 

Finite temperature configurations in the range of $T\simeq174$--348 MeV ($N_t=16$--8) have been generated adopting the fixed-scale approach \cite{Umeda:2012er,Umeda:2008bd}.
The values of temperature at each~$N_t$ are given in Table~\ref{table:parameters}.
The spatial box size is $32^3$ for $T>0$ and $28^3$ for $T=0$. 
Although we have configurations also at $T\simeq464$ MeV ($N_t=6$) and $T\simeq697$ MeV ($N_t=4$)~\cite{Taniguchi:2016ofw}, limitations by~$t_{1/2}=1.125$ and 0.5, respectively, are too severe to obtain a stable linear window. 
It was also noted that the EoS on $N_t \simle 8$ lattices has large $O\!\left((aT)^2=1/N_t^2\right)$ lattice artifacts~\cite{Taniguchi:2016ofw}.
We thus just omit these configurations in this study.

\begin{table}[tbh]
\centering
\caption{Simulation parameters: Temperature in MeV,
$T/T_{\mathrm{pc}}$ assuming $T_{\mathrm{pc}}=190$ MeV, the temporal lattice size $N_t$, $t_{1/2}$ defined by Eq.~(\ref{eq:t-half}), and the number of configurations used in gauge and fermion measurements. 
Spatial box size is $32^3$ for~$T>0$ and $28^3$ for~$T=0$.
}
\label{table:parameters}
\begin{tabular}{cccccc}
 $T$[MeV] & $T/T_{\mathrm{pc}}$ & $N_t$ & $t_{1/2}$ & gauge configurations & fermion configurations \\
\hline
 $0$   & $0$    & $56$ &$24.5$ & $650$ & $65$ \\
 $174$ & $0.92$ & $16$ &$8$ & $1440$ & $144$ \\
 $199$ & $1.05$ & $14$ &$6.125$ & $1270$ & $127$ \\
 $232$ & $1.22$ & $12$ &$4.5$ & $1290$ & $129$ \\
 $279$ & $1.47$ & $10$ &$3.125$ & $780$ & $78$ \\
 $348$ & $1.83$ &  $8$ &$2$ & $510$ & $51$ \\
\hline
\end{tabular}
\end{table}

Our numerical algorithm for gradient flow is given in~Ref.~\cite{Taniguchi:2016ofw}.
We adopt the third order Runge-Kutta method with the step size of~$\epsilon=0.02$. 
For the quadratic terms of the field strength tensor $G_{\mu\nu}(x)$, we adopt clover operator with four plaquette Wilson loops 
and that with eight $1\times2$ rectangle Wilson loops such that 
the tree-level improved field strength squared is obtained~\cite{AliKhan:2001ym}. 
To calculate the~$\overline{\textrm{MS}}$ running coupling and masses in the matching coefficients, we adopt five-loop beta and gamma functions~\cite{FiveLoop}, instead of the four-loop functions adopted in our previous study~\cite{Taniguchi:2016ofw}.

We evaluate fermionic observables by the noisy estimator method. 
The number of noise vectors we adopt is 20 for each color and spinor component.
To measure fermionic bilinear observables at~$t>0$, instead of the adjoint Runge-Kutta integration adopted in Ref.~\cite{Taniguchi:2016ofw} (see Appendix~B of Ref.~\cite{Taniguchi:2016ofw}), we apply an alternative method using usual forward Runge-Kutta integrations only by locating the noise vectors at $t=0$.
See Appendix~\ref{sec:normalflow} for details.
We confirm that the both methods give consistent results within statistical errors of the noise method.
This alternative method is applicable to fermionic bilinear observables and reduces the computational cost.
On the other hand, because the data at all~$t$'s are estimated with the same noise vector at~$t=0$, the correlation among different~$t$'s is stronger than the study of~Ref.~\cite{Taniguchi:2016ofw} in which independent noise vectors were generated at~each~$t$. 
The gauge observables are measured every 5~trajectories at~$T>0$ and every 10~trajectories at~$T=0$, while the fermionic observables are measured every 50~trajectories at~$T>0$ and every 100~trajectories at~$T=0$. 
The statistical errors are estimated by the standard jackknife analysis
%After a study of the bin size dependence, we choose 
with the bin size of 100~trajectories for the energy-momentum tensor and 300~trajectories for the chiral condensates and susceptibilities.

%%%%%%%%%%%%%%%%%%%%%%%%%%%%%%%%%%%%%%%%%%%%%%%%%%%%
\section{Renormalization scale}
\label{sec:mu0}

Our matching coefficients, Eqs.~\eqref{eq:(1.32)}--\eqref{eq:(1.38)}, are functions of the renormalization scale $\mu$ and the~ $\overline{\textrm{MS}}$ running coupling constant~$g$ at the renormalization scale~$\mu$.
Here, the renormalization scale $\mu$ is free to choose as long as the perturbative expansion of the matching coefficients is valid because the final physical observables are independent of~$\mu$.

A conventional choice of $\mu$ is 
\begin{equation}
   \mu=\mu_d(t) \equiv\frac{1}{\sqrt{8t}},
\label{eq:mu_d}
\end{equation}
which is a natural scale of flowed operators because $\sqrt{8t}$ is the physical smearing extent of flowed fields.
On the other hand, the authors of~Ref.~\cite{Harlander:2018zpi} argued that the choice
\begin{equation}
   \mu=\mu_0(t)\equiv\frac{e^{-\gamma_E/2}}{\sqrt{2t}},
\label{eq:(1.40)}
\end{equation}
which sets $L(\mu,t)=0$ in Eqs.~\eqref{eq:(1.32)}--\eqref{eq:(1.38)}, keeps the relative contribution of two-loop terms small in a similar level as the $\mu_d$-scale.

We note that, since
\begin{equation}
   \mu_0(t) \simeq1.4986 \times \mu_d(t),
\label{eq:(1.41)}
\end{equation}
the $\mu_0$-scale is more perturbative than the $\mu_d$-scale in asymptotically free theories.
Thus, the $\mu_0$-scale may improve the quality of perturbative expressions, in particular at~large~$t$.%
\footnote{One may try even larger $\mu$, such as $2\mu_d$ or $3\mu_d$, as the renormalization scale. On the other hand, adopting a too big value for $\mu$ will make $L(\mu,t)$ large and thus may invalidate the perturbative expansion, \emph{i.e.}, $\mu$ should be $O(\mu_d)$.}

\subsection{Test of renormalization scales using one-loop matching coefficients}
\label{sec:1loop_mu}

\begin{figure}[tbh]
\centering
\includegraphics[width=7cm]{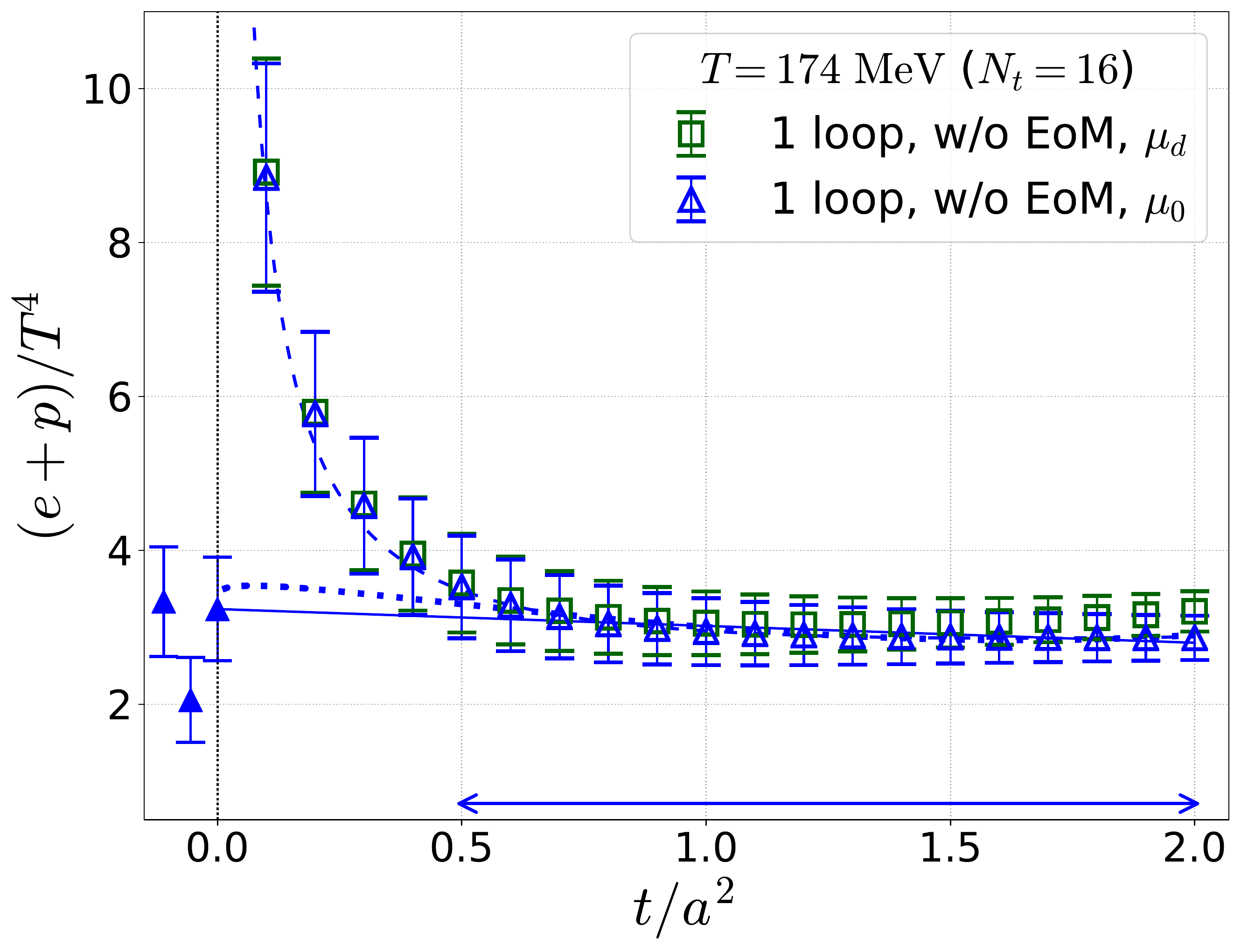}
\includegraphics[width=7cm]{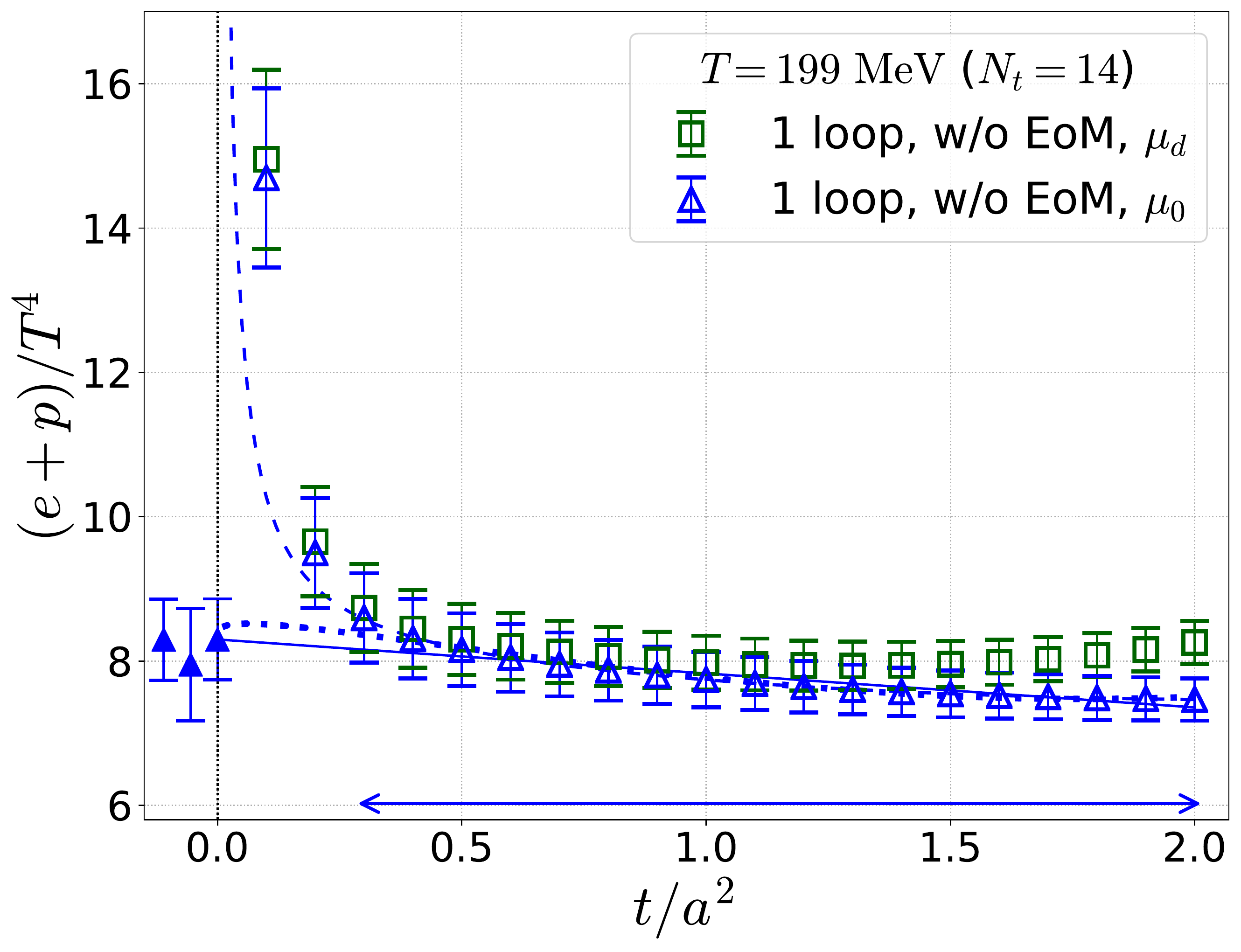}
\includegraphics[width=7cm]{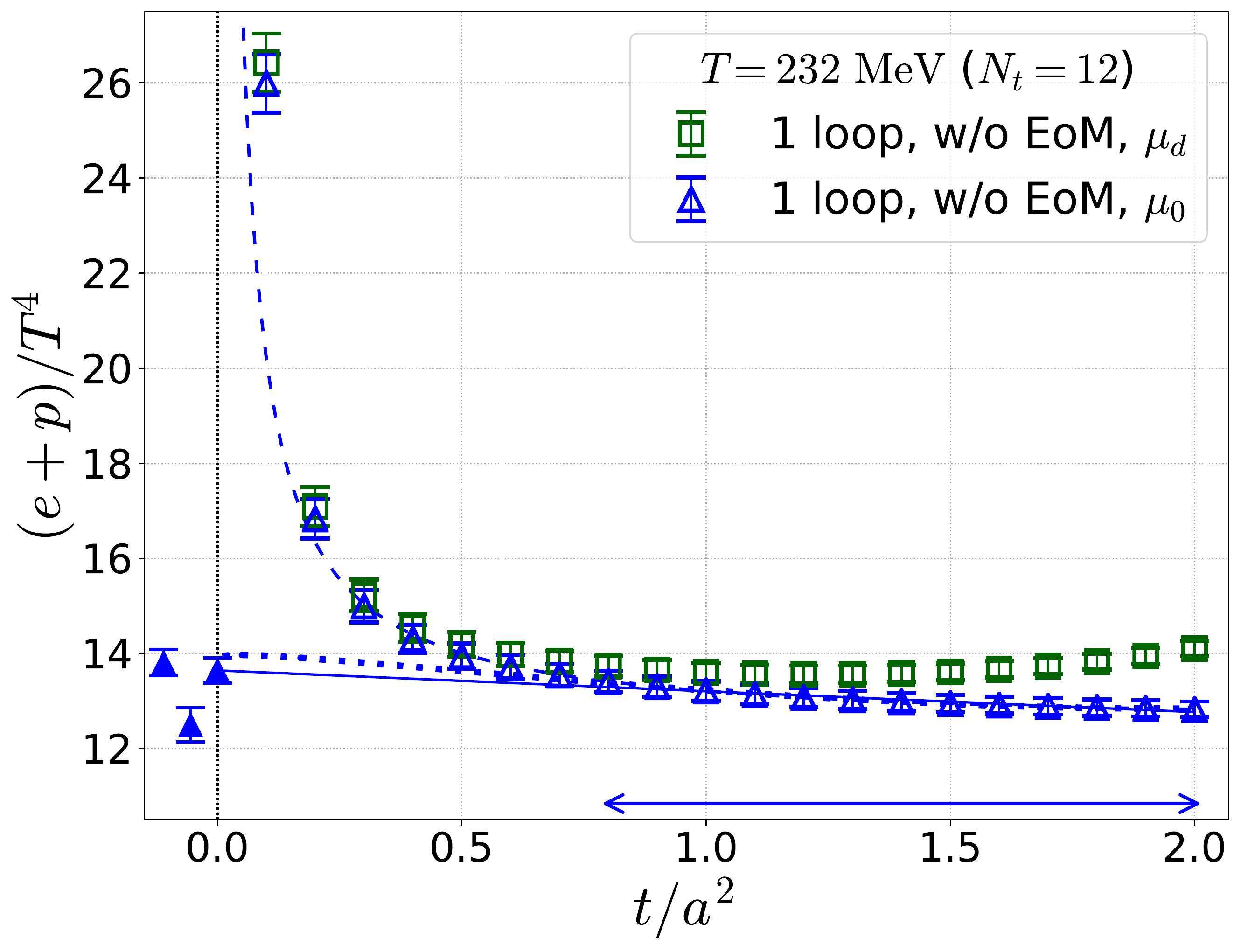}
\includegraphics[width=7cm]{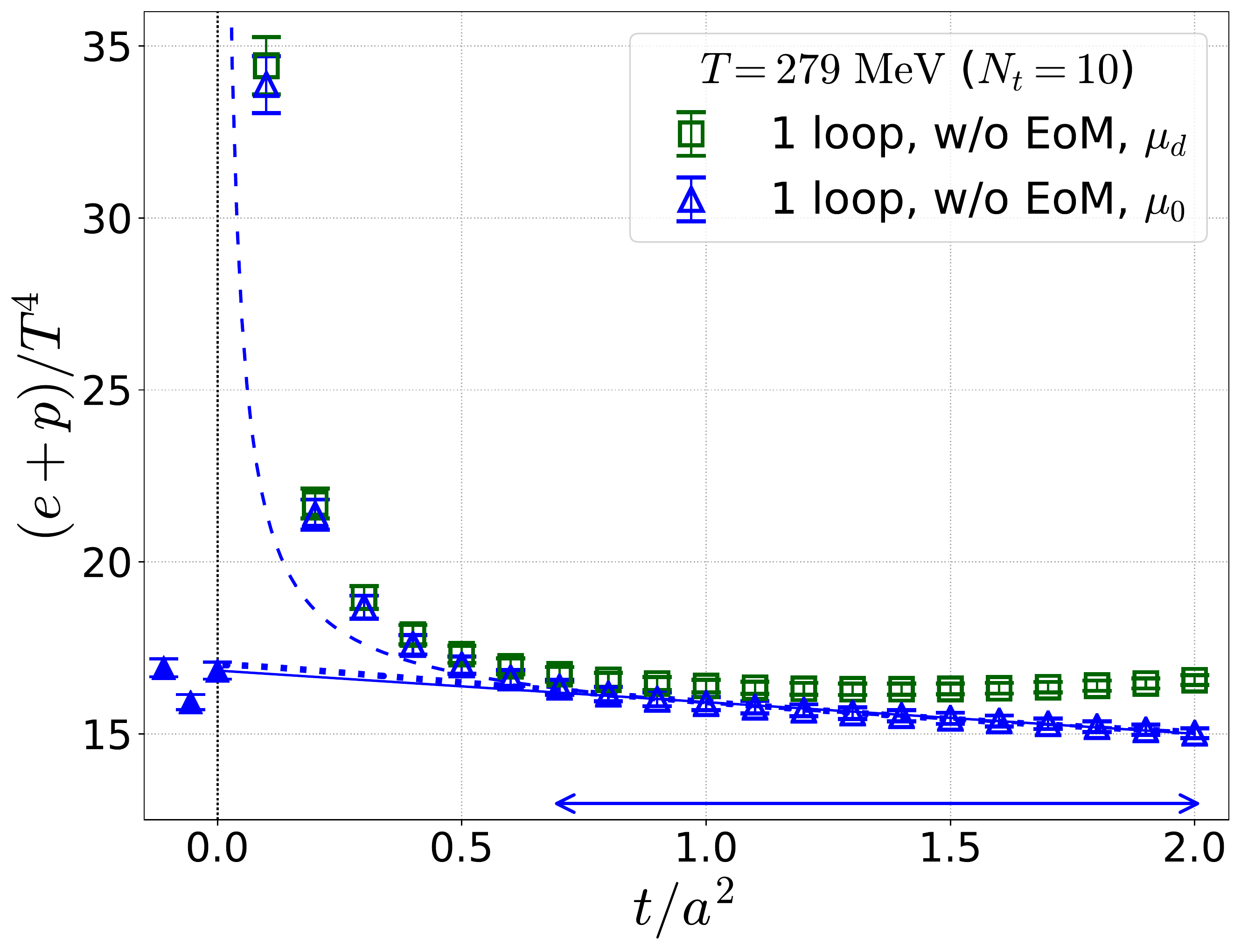}
\includegraphics[width=7cm]{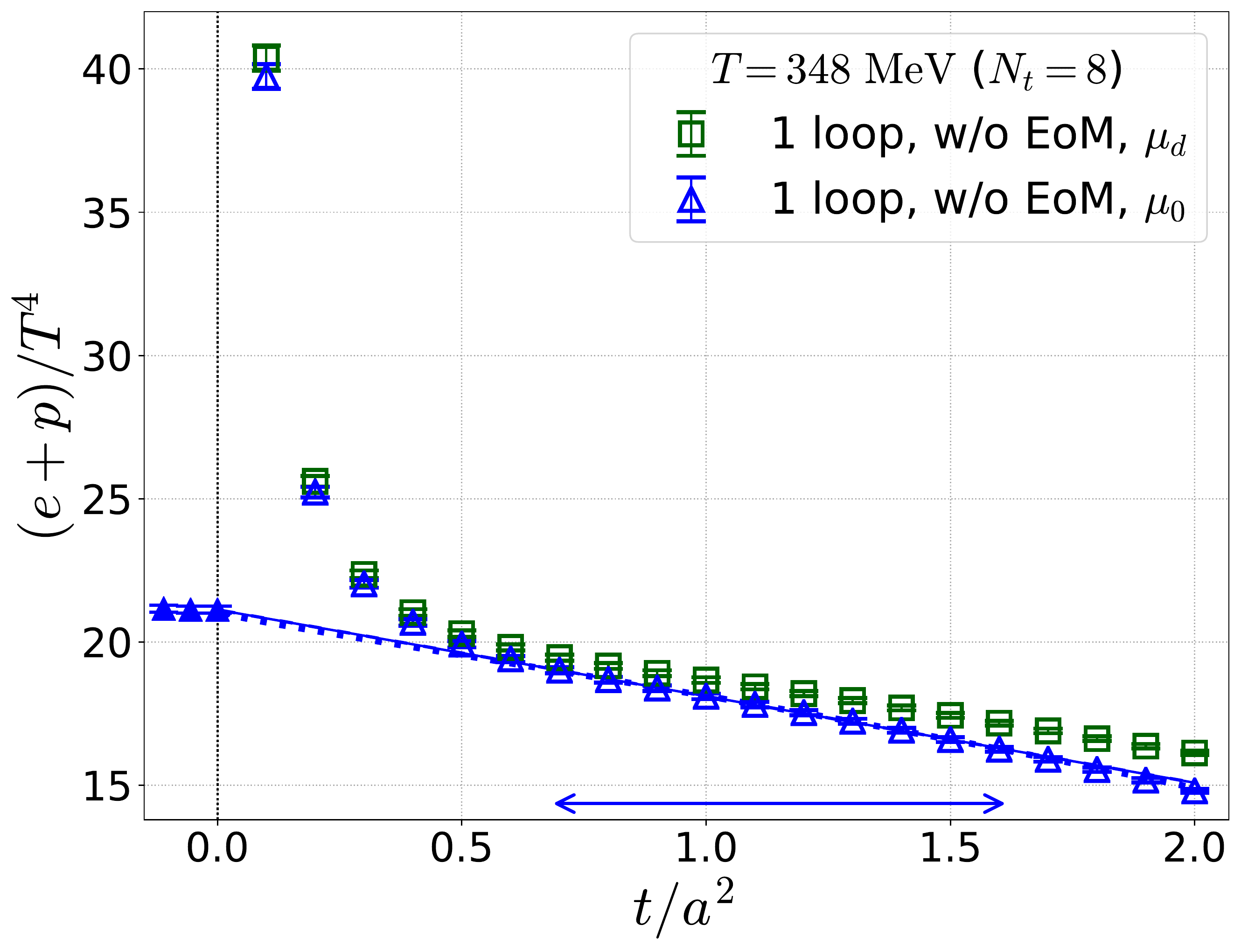}
\vspace{-5mm}
\caption{Entropy density $(\epsilon+p)/T^4$ with $\mu_0$ (blue) and $\mu_d$ (green) scales as function of the flow-time. 
One-loop matching coefficients of Ref.~\cite{Makino:2014taa} are used.
Also shown are the results of the $t\to0$ extrapolations using the data with the $\mu_0$-scale:
Solid line is the linear fit using the linear window indicated by the arrow at the bottom of each plot, 
and the symbol at $t/a^2 = 0$ is the result of the linear fit for the physical entropy density in the $t\to0$ limit.
Fit results with the nonlinear ansatz~\eqref{eqn:5.4} and linear+log ansatz~\eqref{eqn:5.4l} are shown by dashed and dotted curves together with the symbols at $t/a^2 < 0$ to the right and to the left, respectively.
Errors are statistical only.
}
\label{figmu:eplusp}
\end{figure}

\begin{figure}[tbh]
\centering
\includegraphics[width=7cm]{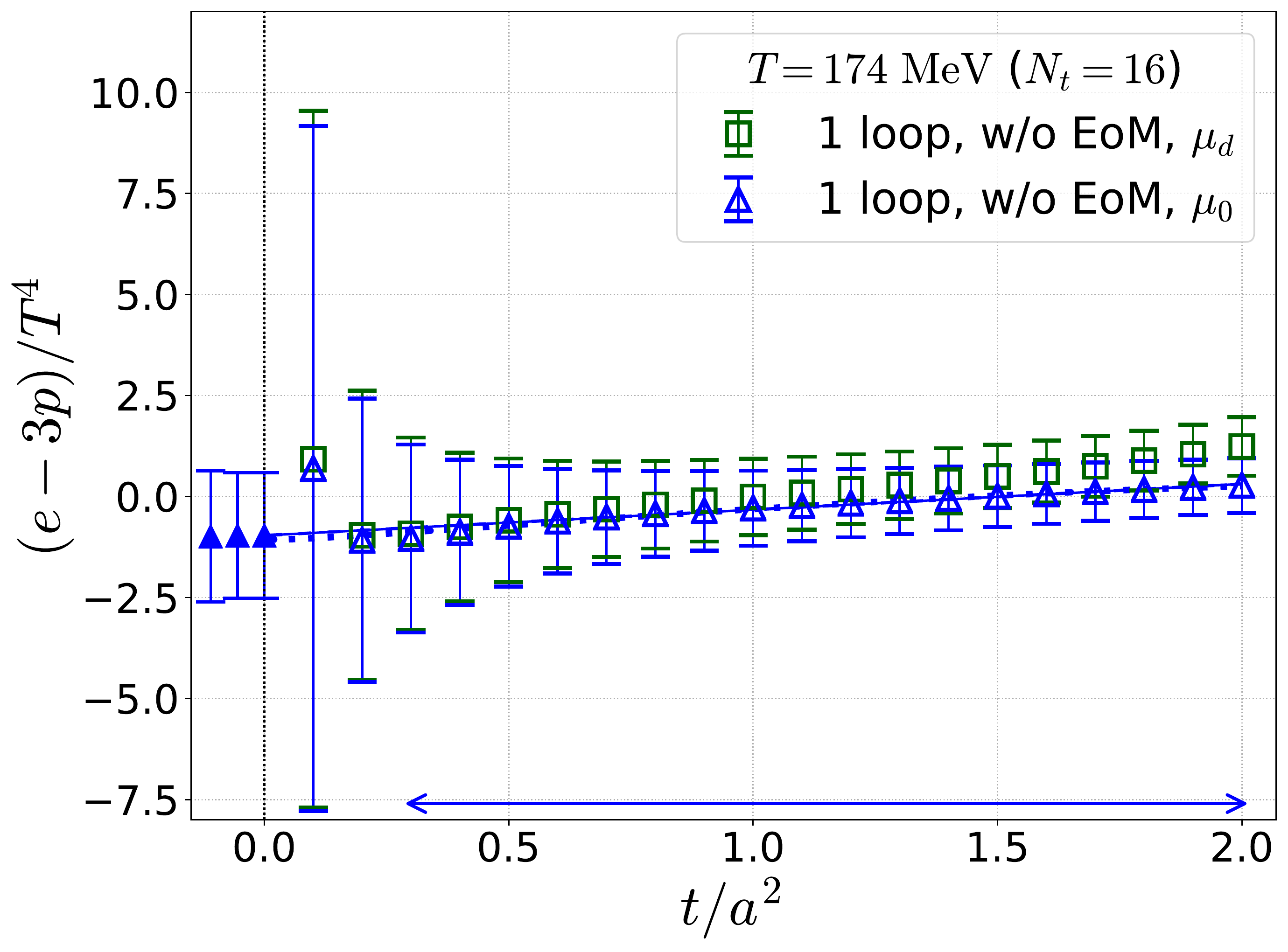}
\includegraphics[width=7cm]{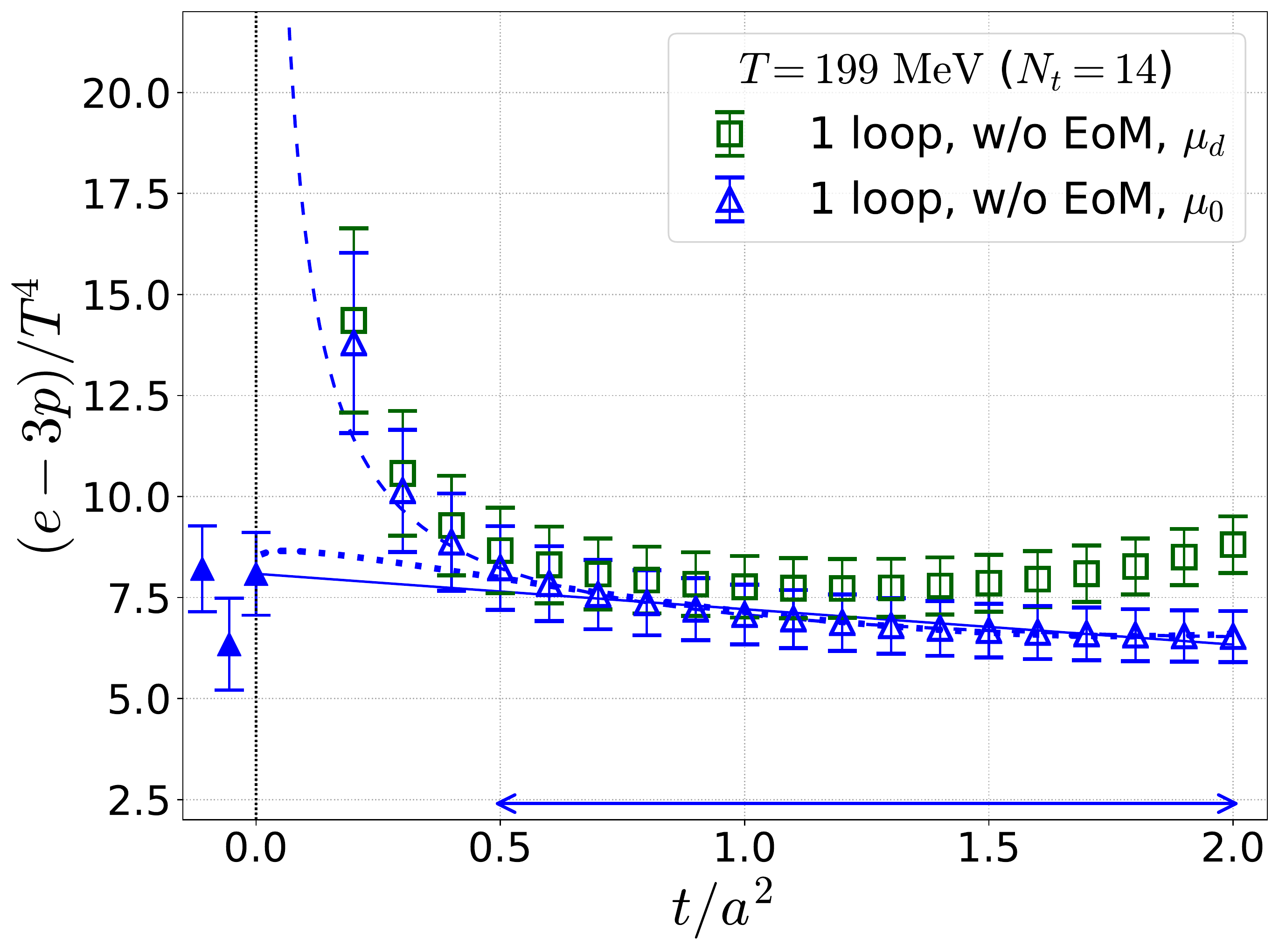}
\includegraphics[width=7cm]{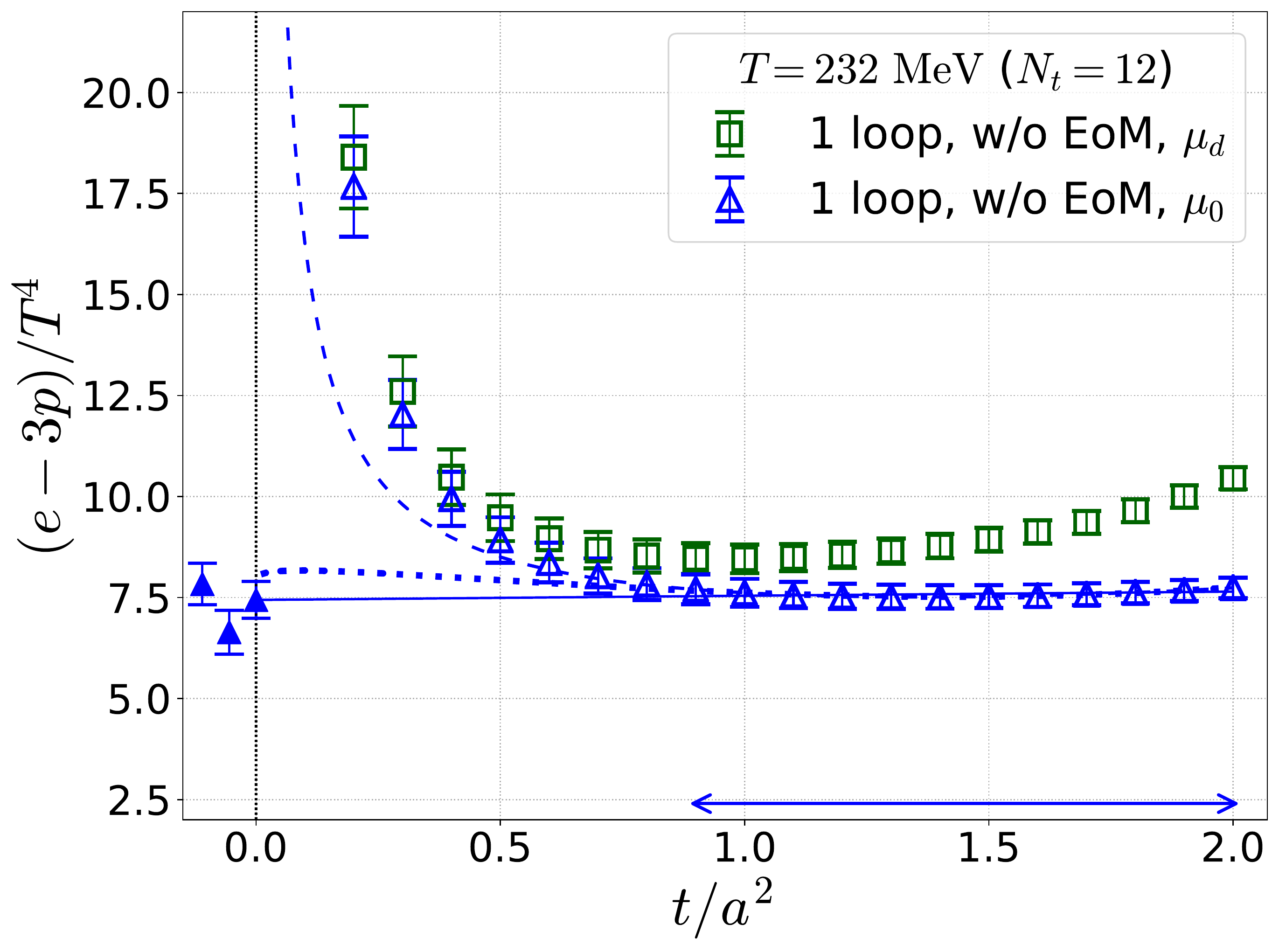}
\includegraphics[width=7cm]{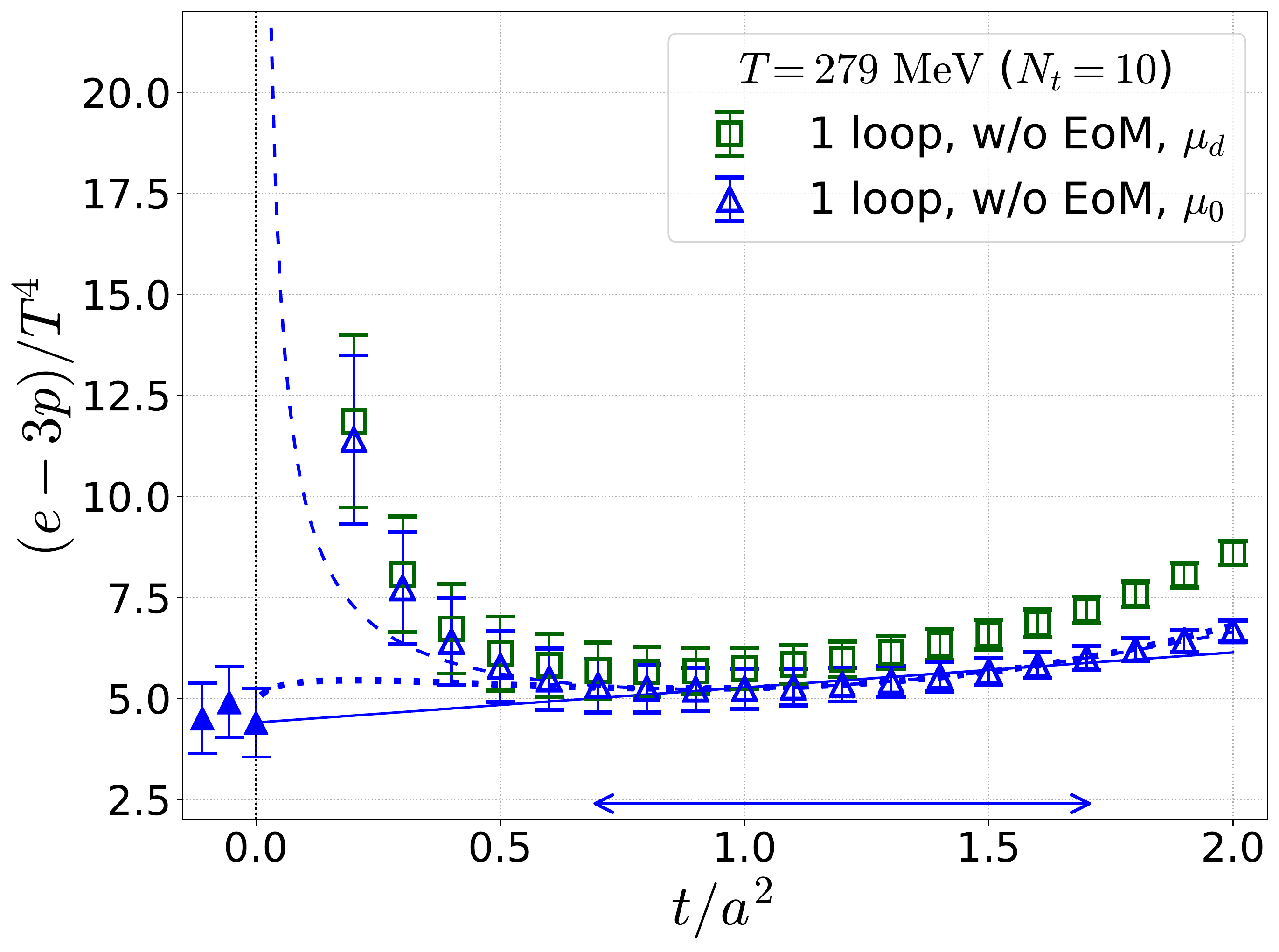}
\includegraphics[width=7cm]{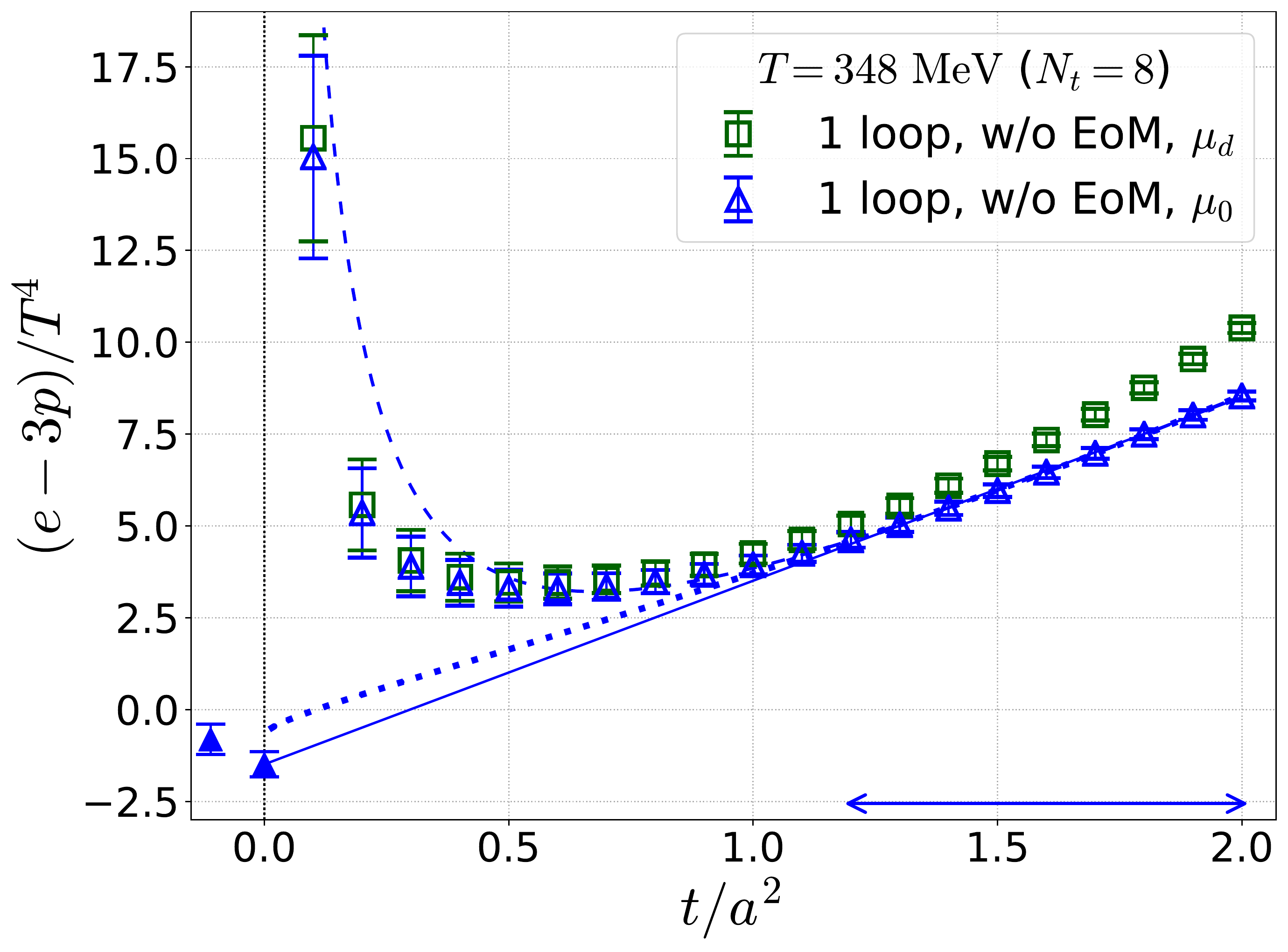}
\vspace{-5mm}
\caption{The same as Fig.~\ref{figmu:eplusp} but for the trace anomaly~$(\epsilon-3p)/T^4$.
}
\label{figmu:eminus3p}
\end{figure}

In Fig.~\ref{figmu:eplusp}, we examine effects of the renormalization scale by comparing the results for the entropy density,
\begin{equation}
   \frac{\epsilon+p}{T^4}
   =-\frac{4}{3 T^4}\left\langle T_{00}-\frac{1}{4}T_{\mu\mu}\right\rangle,
\label{eq:(4.3)}
\end{equation}
at finite $t$ with $\mu_0$- and $\mu_d$-scales. 
Corresponding results for the trace anomaly,
\begin{equation}
   \frac{\epsilon-3p}{T^4}
   =-\frac{1}{T^4}\langle T_{\mu\mu}\rangle,
\label{eq:(4.4)}
\end{equation}
are shown in~Fig.~\ref{figmu:eminus3p}. 
We adopt one-loop matching coefficients of~Ref.~\cite{Makino:2014taa} in this test.

Consulting Figs.~\ref{figmu:eplusp} and~\ref{figmu:eminus3p}, we find that, though some dependence on the choice of the renormalization scale is visible, the difference become smaller as we decrease $t/a^2$. 
We also note that the slight curvatures visible sometimes with the $\mu_d$-scale data become smaller with the $\mu_0$-scale, and the linear behavior is much improved by adopting the $\mu_0$-scale in particular at large~$t/a^2$. 
This is consistent with our expectation that the $\mu_0$-scale extends the perturbative region over the $\mu_d$-scale toward larger~$t/a^2$. 
The $\mu_0$-scale enables us to carry out $t\to0$ extrapolations based on the linear window more confidently.

In Figs.~\ref{figmu:eplusp} and~\ref{figmu:eminus3p}, we also show our $t\to0$ extrapolations using the data with the $\mu_0$-scale.
The arrow at the bottom of each plot is the linear window we adopt determined by the procedure discussed in Sec.~\ref{sec:t0extrapolation}.
Solid lines, dashed curves, and dotted curves are the results of linear, nonlinear, and linear+log fits, rspectively.
The symbols at $t\sim0$ shows the results of $t\to0$ extrapolations, from the right to the left, using linear, nonlinear, and linear+log fits, respectively. 

In Table~\ref{table:eos1}, we summarize our results for  the physical values of $(\epsilon+p)/T^4$ and $(\epsilon-3p)/T^4$, obtained in the $t\to0$ limit using the one-loop matching coefficients of Ref.~\cite{Makino:2014taa} with the $\mu_0$-scale.
We take the result of the linear fit as our central value and take the difference with other fits as an estimate of the systematic error due to the $t\to0$ extrapolation. 
In the same table, we also list the results for~$\epsilon/T^4$ and $p/T^4$ obtained by independent~$t\to0$ extrapolations using data of $ -\langle T_{00} \rangle / T^4 $ and $\sum_i \langle T_{ii} \rangle / (3T^4) $, respectively.
Corresponding results of the equation of state with the $\mu_d$-scale are summarized in~Table~\ref{table:eos1d}. 
We find that the results with the $\mu_d$-scale are consistent within the errors with both those given in~Table~\ref{table:eos1} and also in~Ref.~\cite{Taniguchi:2016ofw}.

\begin{table}[tbh]
\centering
\caption{Results for EoS by the SF\textit{t}X method with the $\mu_0$-scale
using the one-loop matching coefficients of Ref.~\cite{Makino:2014taa}.
The first parenthesis is for the statistical error, and the second for the systematic error due to the fit ansatz.
}
\label{table:eos1}
\begin{tabular}{ccccc}
 $T$[MeV] & $(\epsilon+p)/T^4$ & $(\epsilon-3p)/T^4$ & $\epsilon/T^4$ & $p/T^4$ \\
\hline
174 & 3.24(68)($^{+10}_{-1.19}$) & $-$0.96(1.56)($^{+00}_{-03}$) & 2.27(65)($^{+06}_{-95}$) & 1.10(43)($^{+03}_{-00}$) \\
199 & 8.30(57)($^{+00}_{-35}$) & 8.09(1.03)($^{+13}_{-1.75}$) & 8.25(56)($^{+03}_{-67}$) & $-$0.00(27)($^{+43}_{-02}$) \\
232 & 13.64(27)($^{+17}_{-1.15}$) & 7.44(46)($^{+40}_{-80}$) & 12.05(23)($^{+23}_{-1.14}$) & 1.48(15)($^{+39}_{-05}$) \\
279 & 16.84(25)($^{+08}_{-92}$) & 4.41(86)($^{+51}_{-00}$) & 13.46(28)($^{+20}_{-1.29}$) & 3.07(25)($^{+02}_{-15}$) \\
348 & 21.13(13)($^{+04}_{-00}$) & $-$1.49(35)($^{+68}_{-3.52}$) & 15.91(12)($^{+00}_{-03}$) & 4.87(20)($^{+21}_{-1.44}$) \\
\hline
\end{tabular}
\end{table}

\begin{table}[tbh]
\centering
\caption{The same as the Table~\ref{table:eos1} but with the $\mu_d$-scale.
}
\label{table:eos1d}
\begin{tabular}{ccccc}
 $T$[MeV] & $(\epsilon+p)/T^4$ & $(\epsilon-3p)/T^4$ & $\epsilon/T^4$ & $p/T^4$ \\
\hline
174 & 3.14(66)($^{+14}_{-1.41}$) & -1.15(1.70)($^{+10}_{-01}$) & 2.14(65)($^{+09}_{-82}$) & 1.14(43)($^{+02}_{-00}$) \\
199 & 8.22(56)($^{+34}_{-12}$) & 7.80(1.04)($^{+88}_{-03}$) & 8.02(54)($^{+77}_{-02}$) & 0.10(26)($^{+00}_{-17}$) \\
232 & 13.45(27)($^{+1.13}_{-00}$) & 7.43(52)($^{+1.86}_{-00}$) & 12.03(24)($^{+96}_{-00}$) & 1.53(15)($^{+00}_{-19}$) \\
279 & 16.19(24)($^{+1.31}_{-00}$) & 4.76(99)($^{+05}_{-59}$) & 13.38(31)($^{+12}_{-03}$) & 3.06(26)($^{+06}_{-13}$) \\
348 & 21.25(13)($^{+00}_{-04}$) & 2.36(75)($^{+91}_{-4.63}$) & 15.57(12)($^{+14}_{-34}$) & 4.92(20)($^{+28}_{-1.54}$) \\
\hline
\end{tabular}
\end{table}

\begin{figure}[htb]
\centering
\includegraphics[width=7cm]{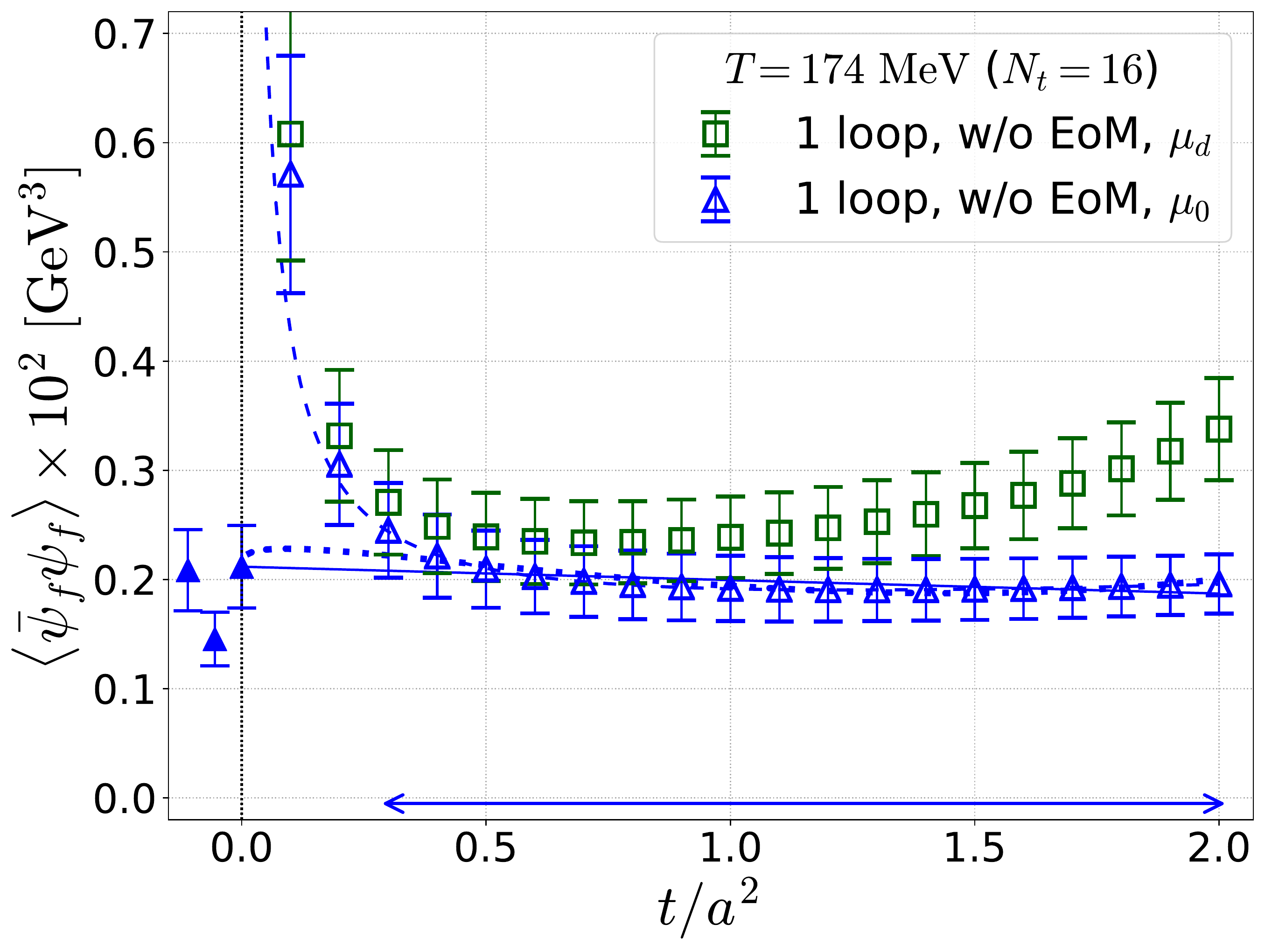}
\includegraphics[width=7cm]{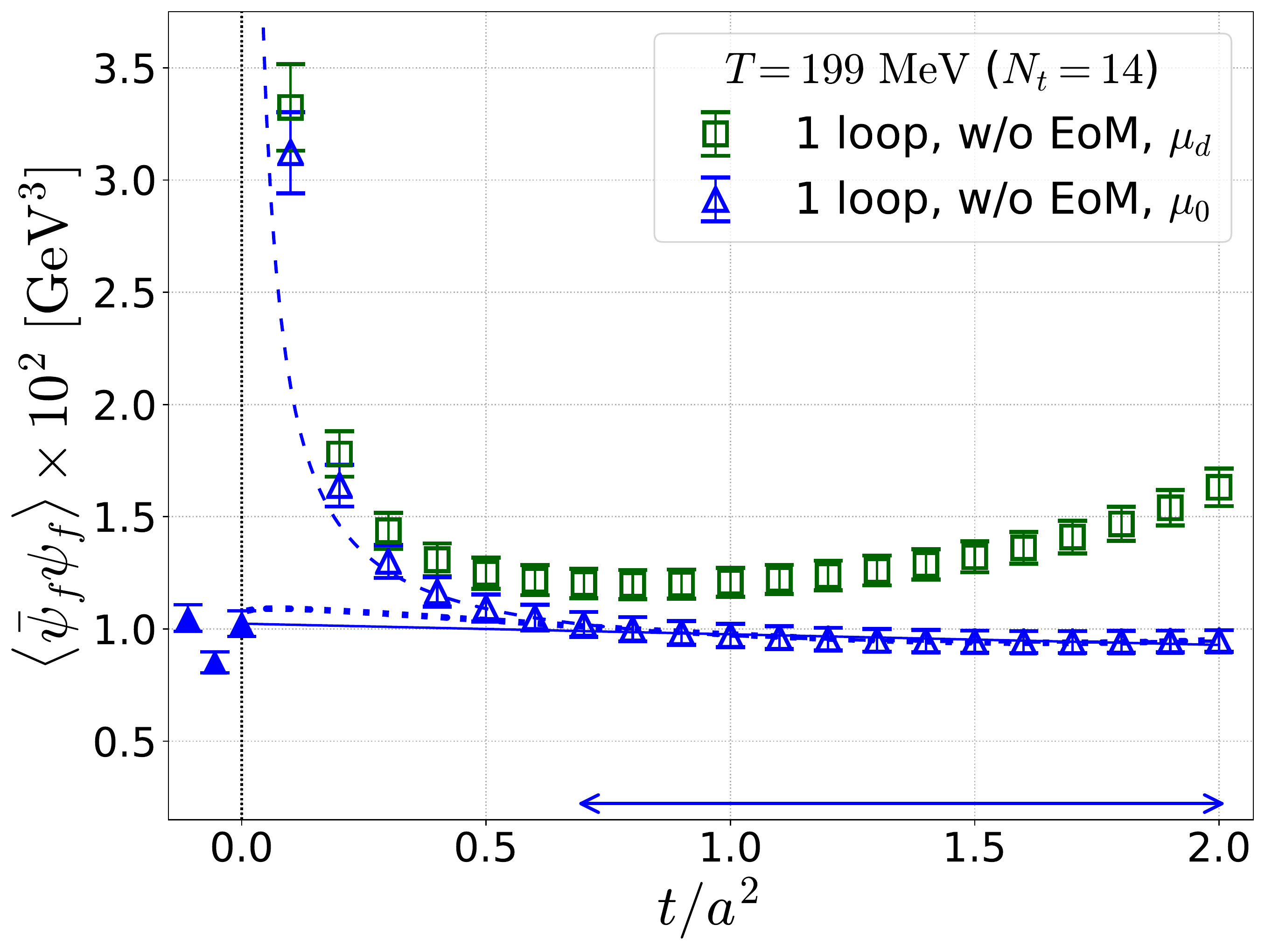}
\includegraphics[width=7cm]{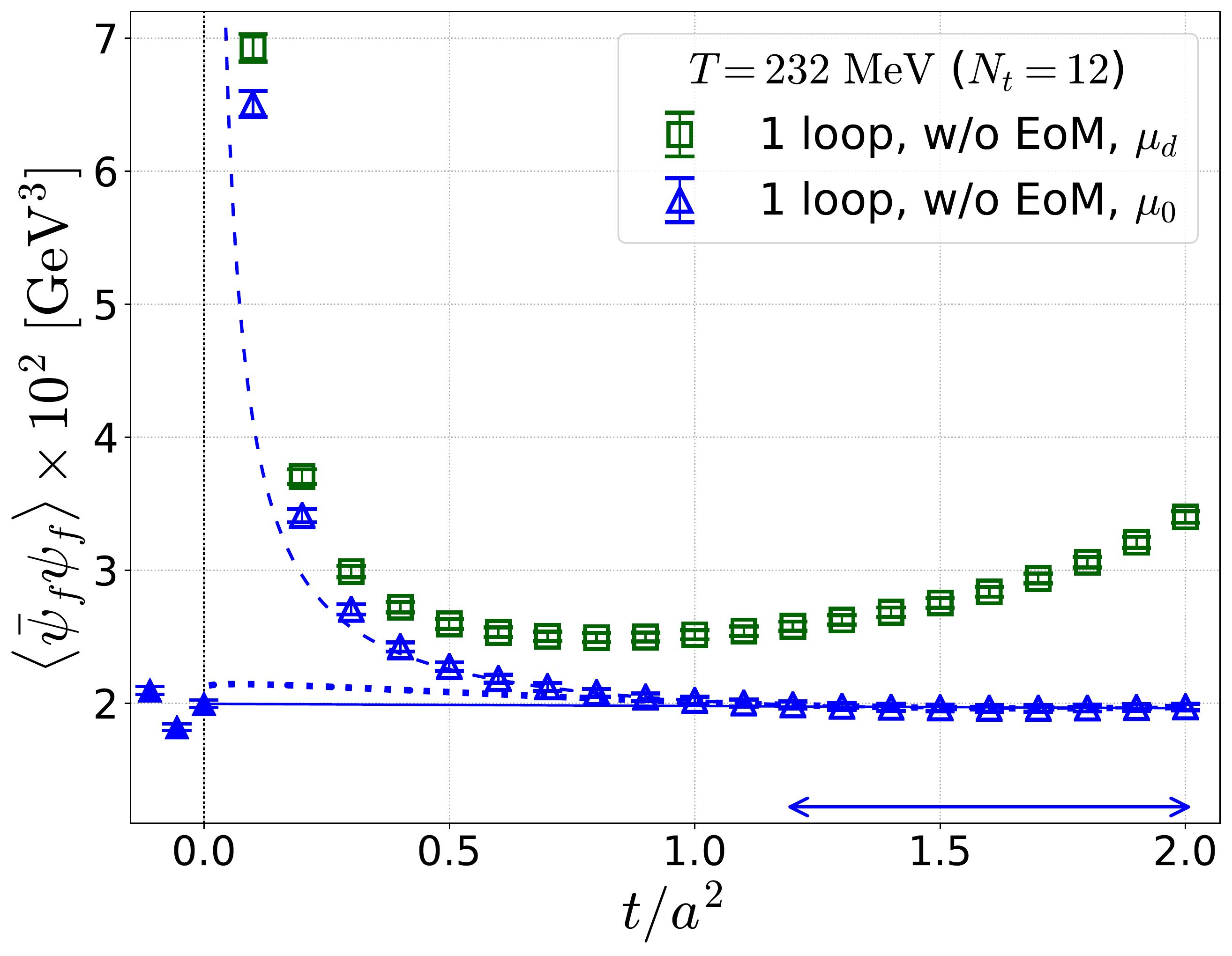}
\includegraphics[width=7cm]{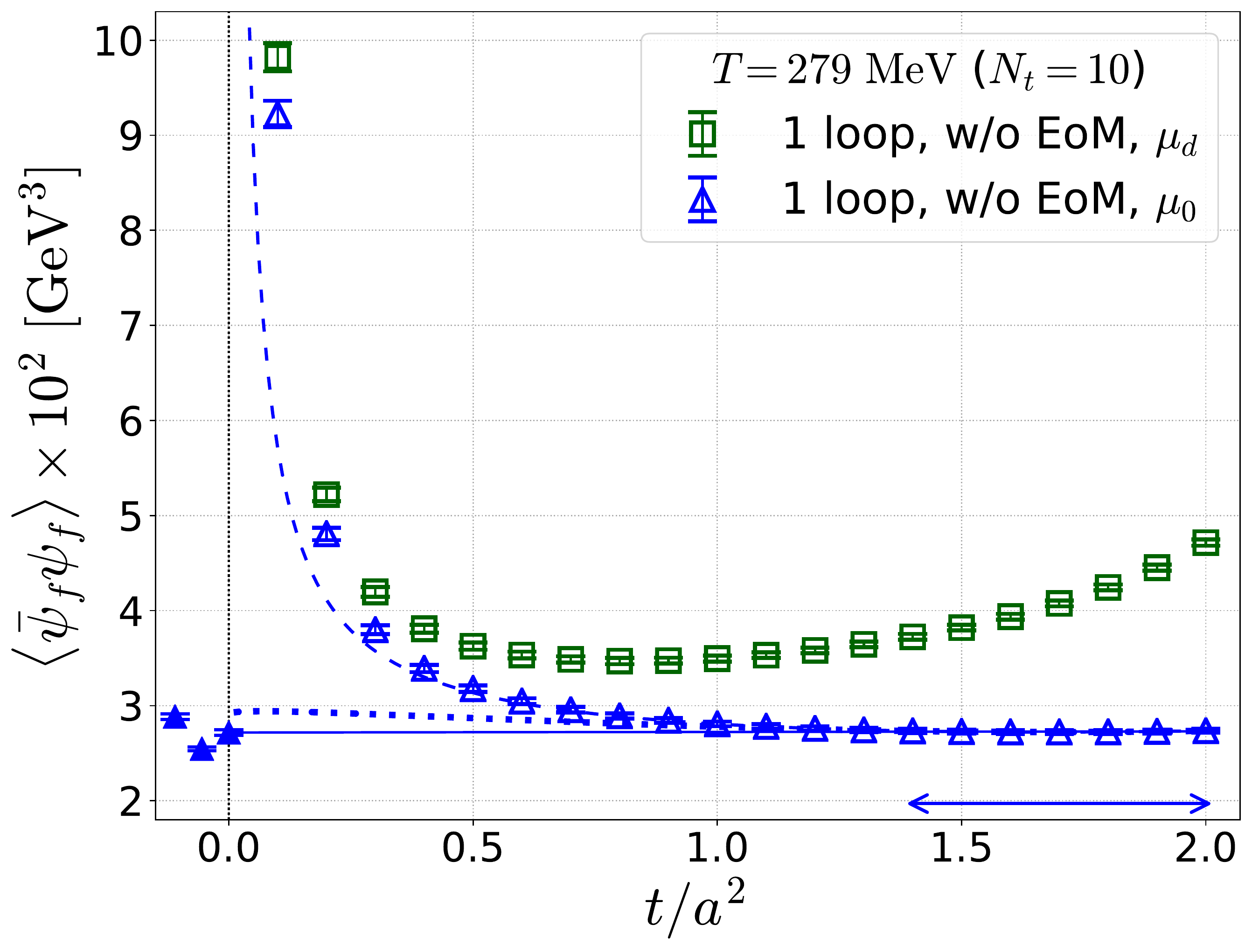}
\includegraphics[width=7cm]{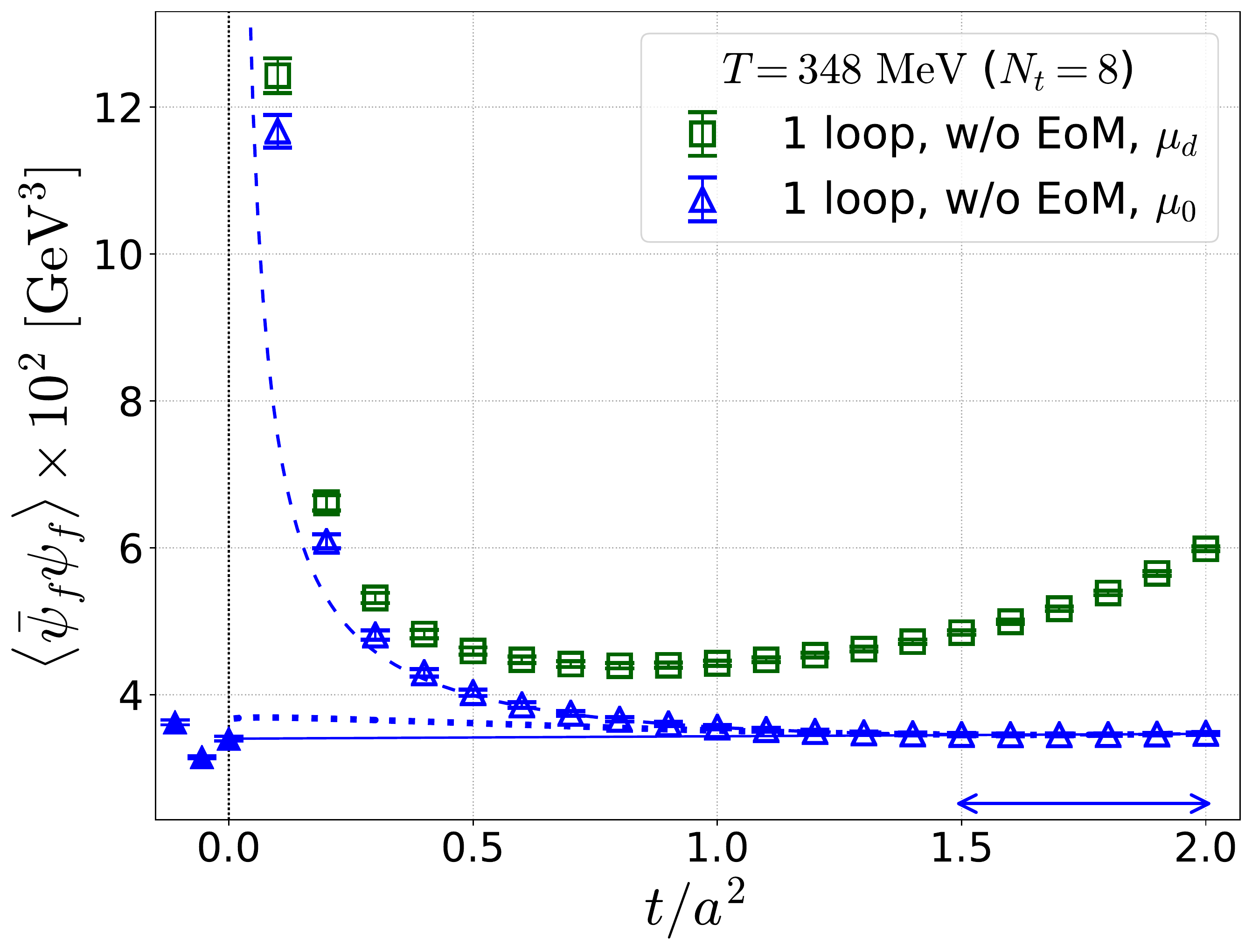}
\vspace{-5mm}
\caption{The same as Fig.~\ref{figmu:eplusp} but for the chiral condensate~$\left\langle\{\Bar{\psi}_f\psi_f\}\right\rangle$ for $f=u$ or $d$ with VEV subtraction. 
 The vertical axis is in units of GeV$^{3}$.  
}
\label{figmu:bpsipsi_subvev_ud}
\end{figure}

\begin{figure}[htb]
\centering
\includegraphics[width=7cm]{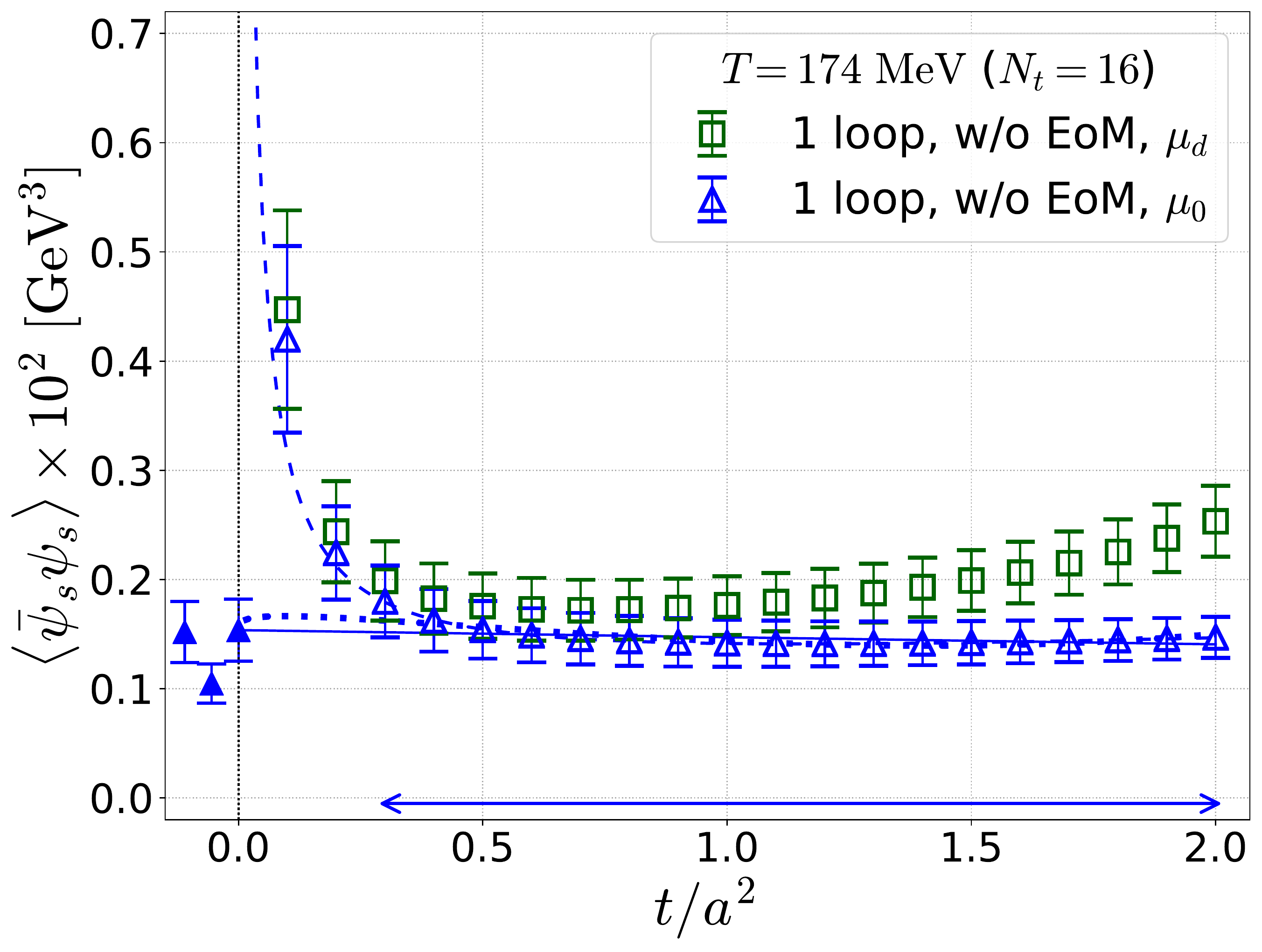}
\includegraphics[width=7cm]{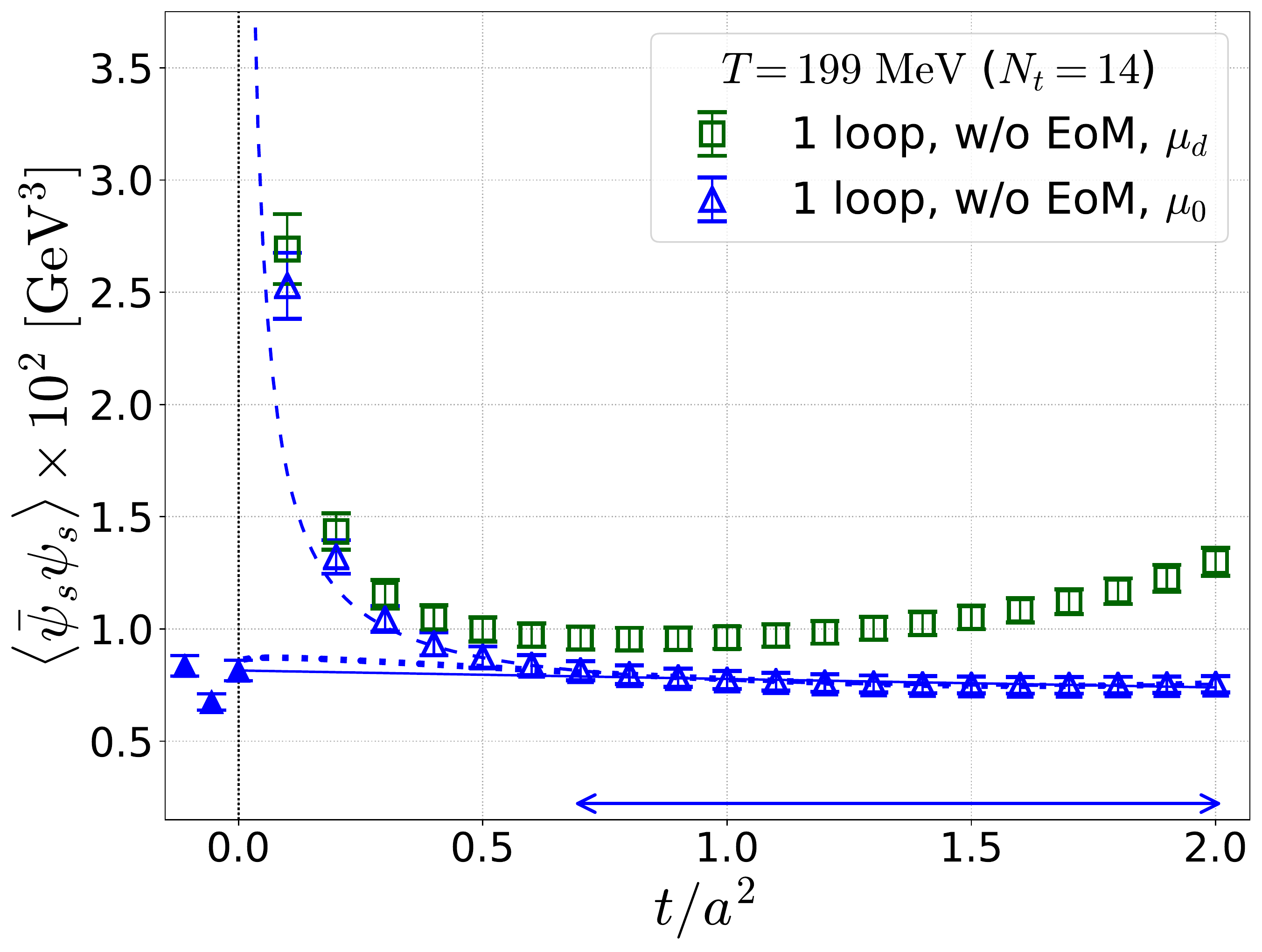}
\includegraphics[width=7cm]{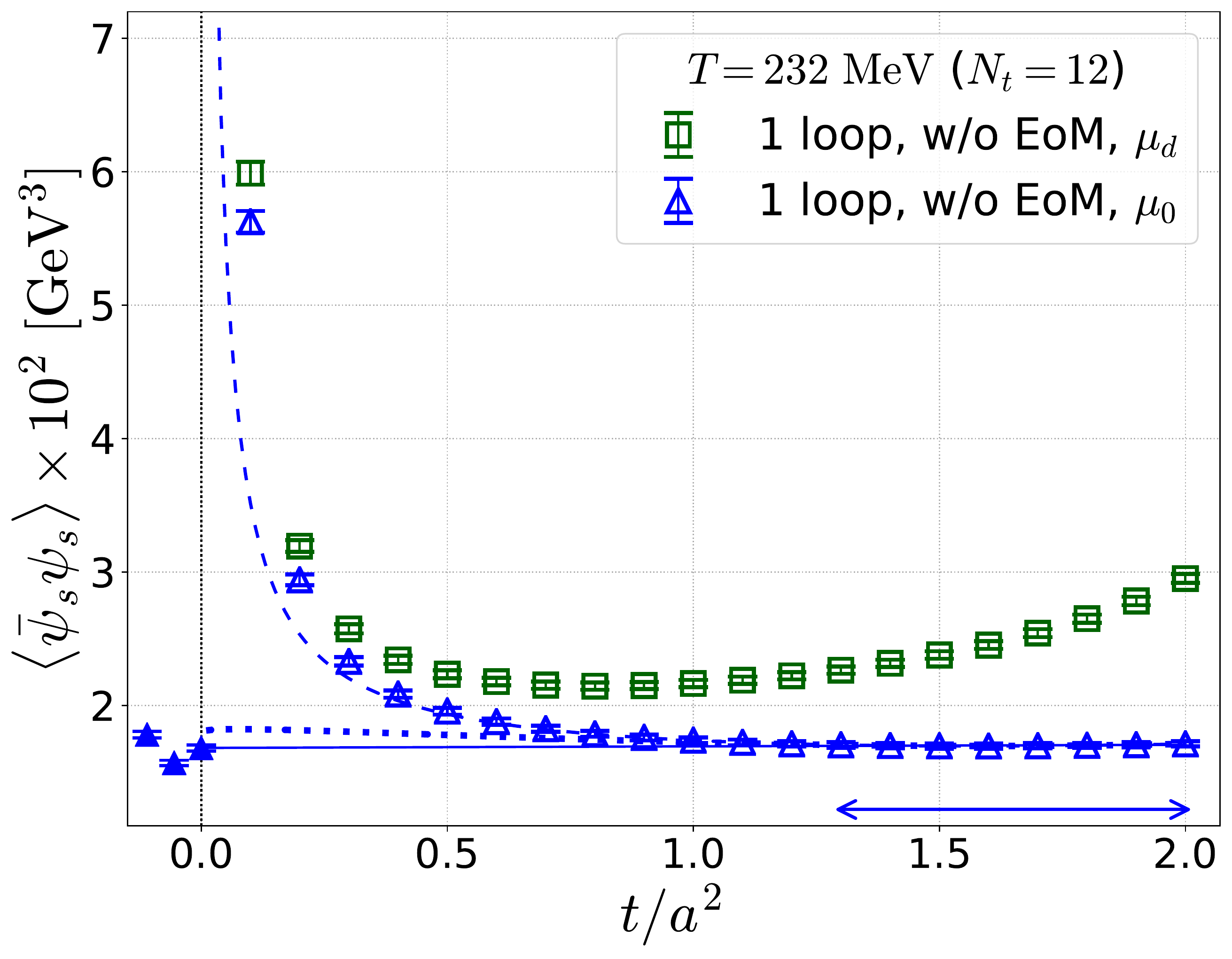}
\includegraphics[width=7cm]{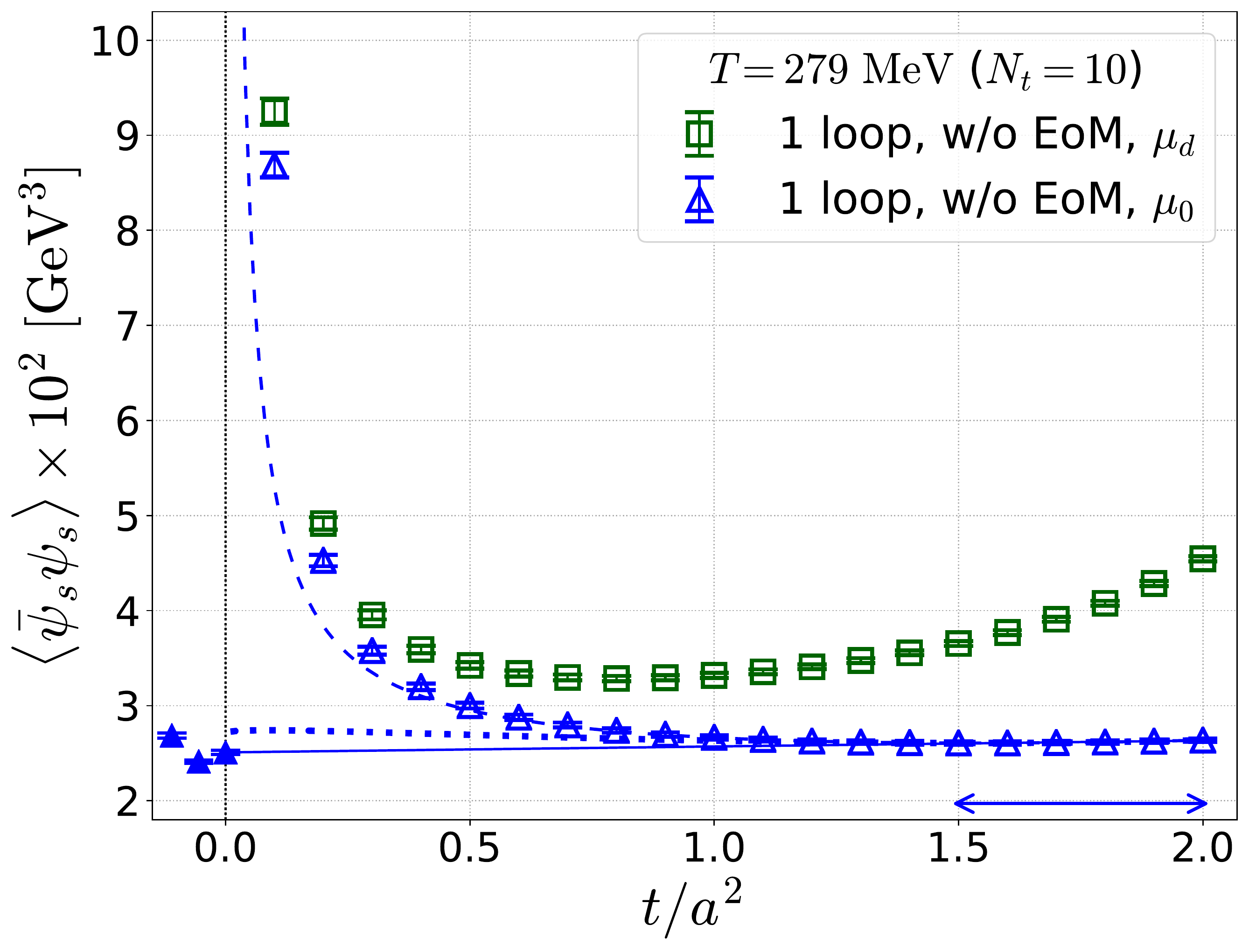}
\includegraphics[width=7cm]{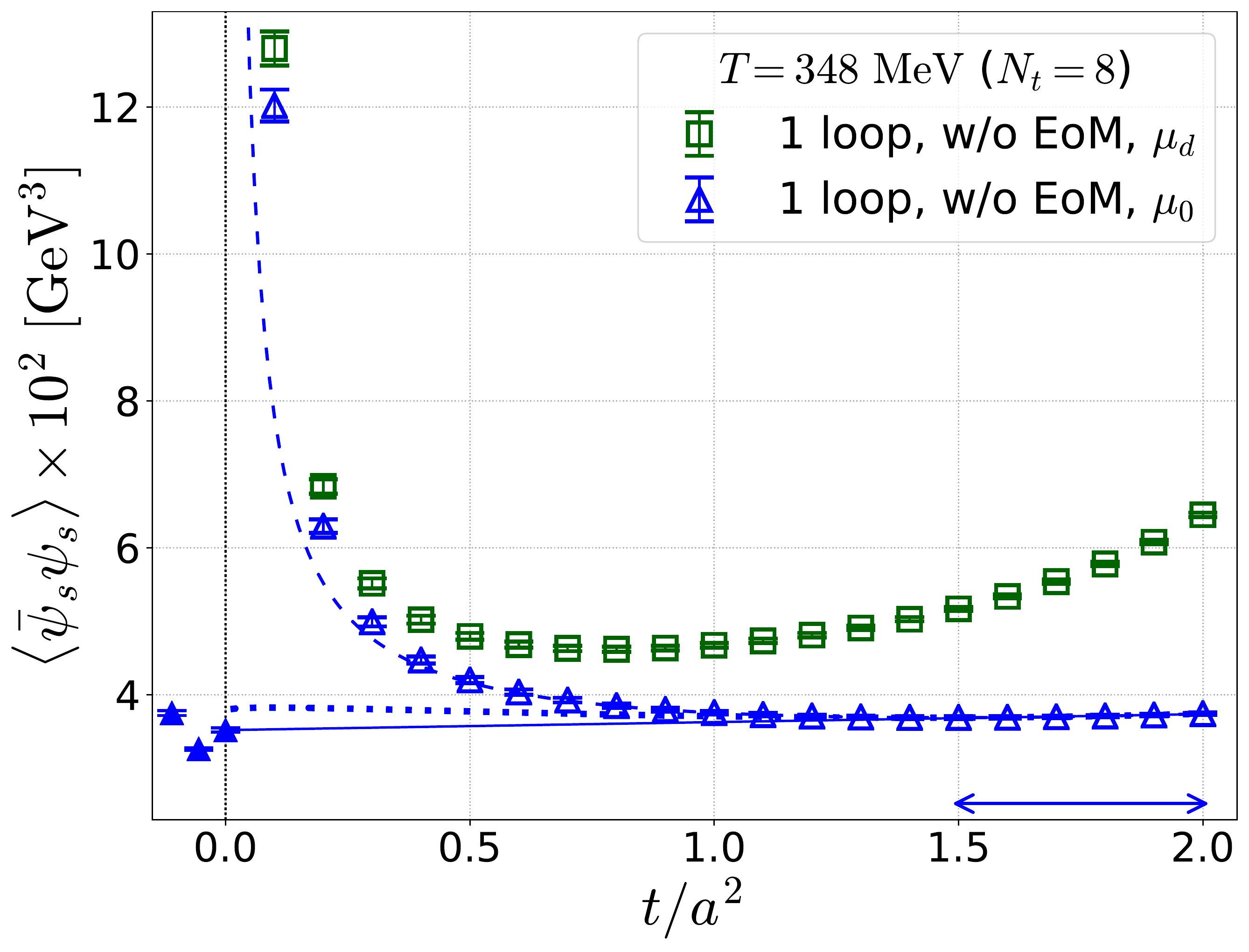}
\vspace{-5mm}
\caption{The same as Fig.~\ref{figmu:bpsipsi_subvev_ud} but for the $s$ quark ($f=s$). 
}
\label{figmu:bpsipsi_subvev_s}
\end{figure}

\begin{figure}[htb]
\centering
\includegraphics[width=7cm]{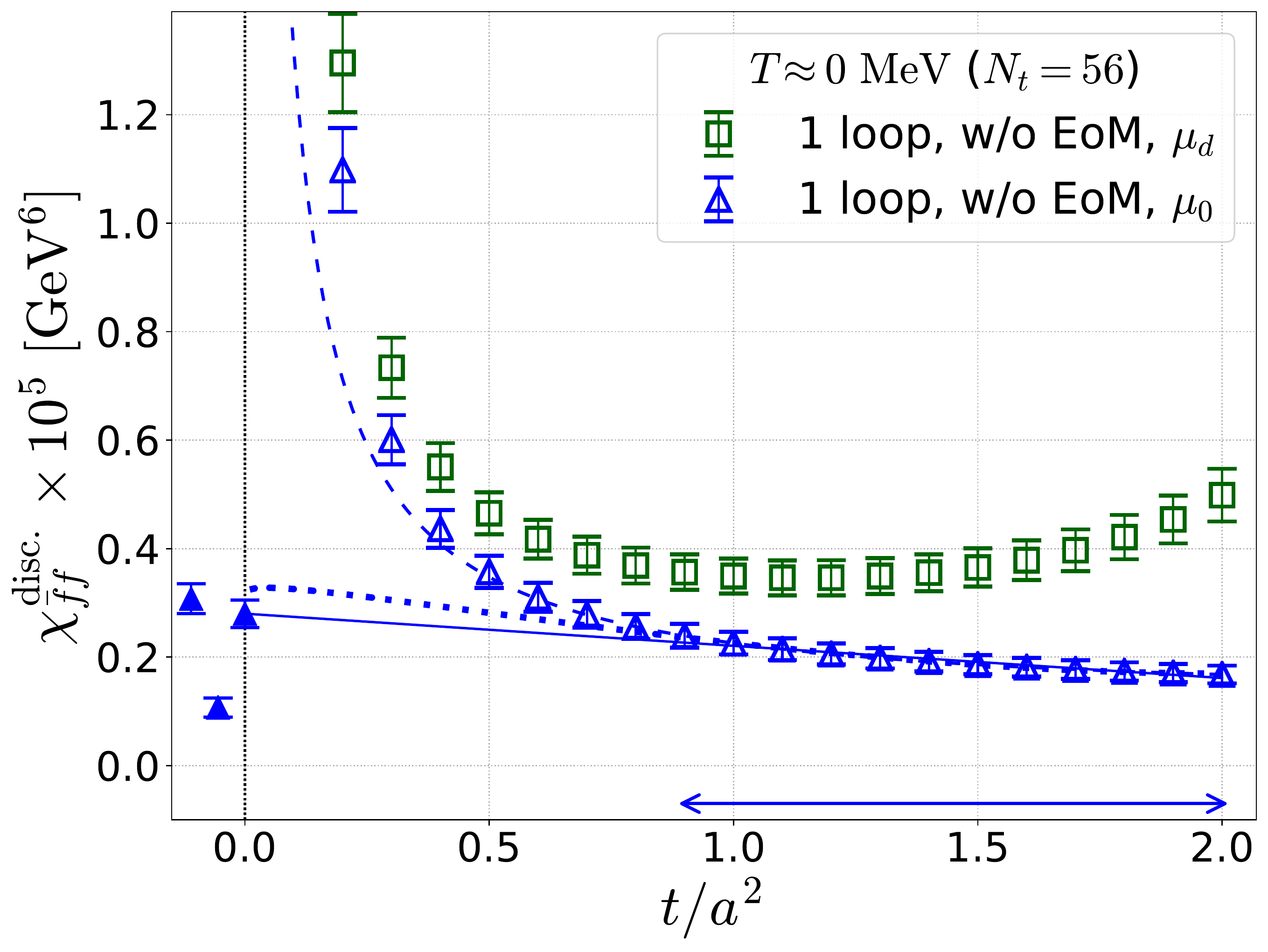}
\includegraphics[width=7cm]{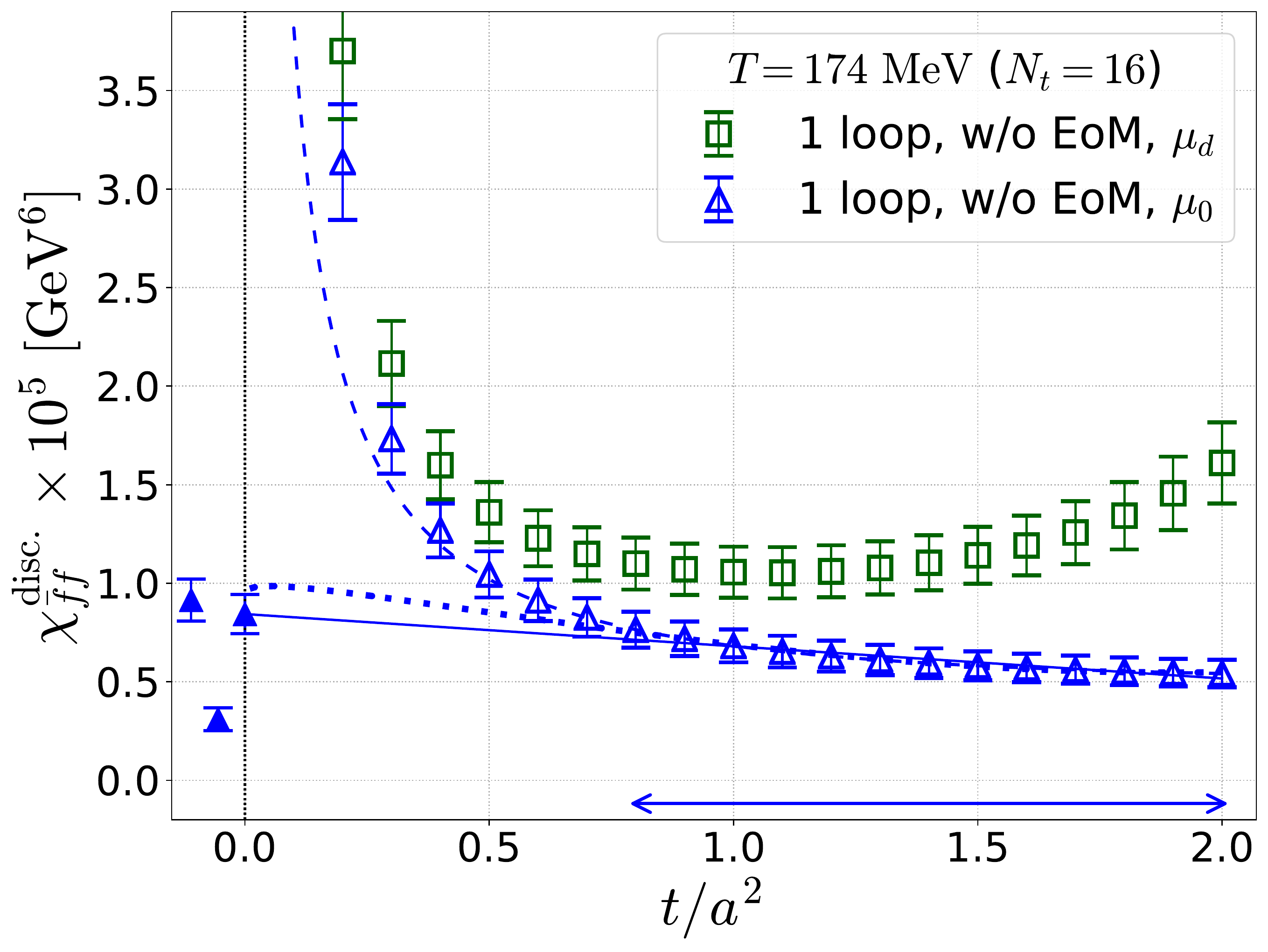}
\includegraphics[width=7cm]{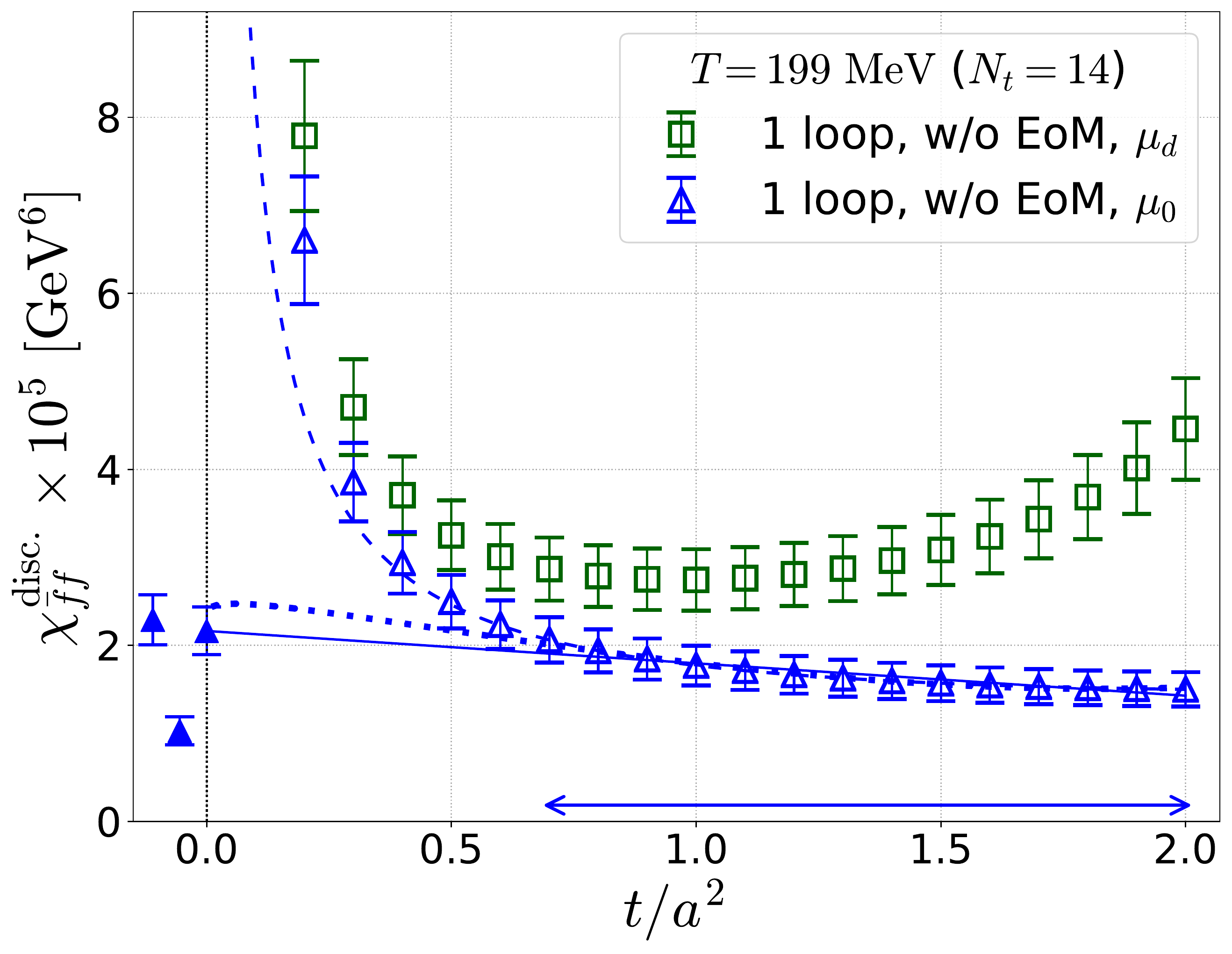}
\includegraphics[width=7cm]{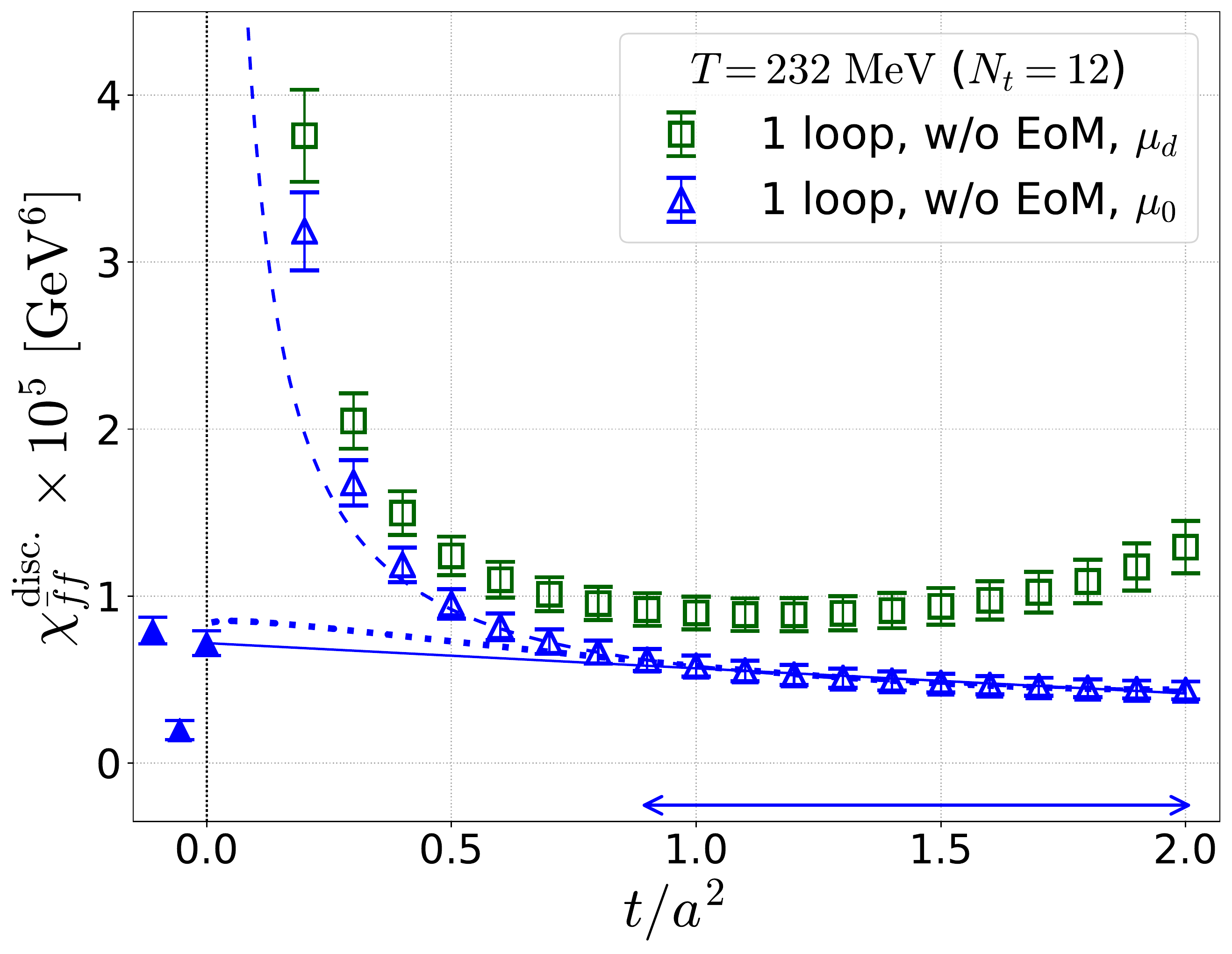}
\includegraphics[width=7cm]{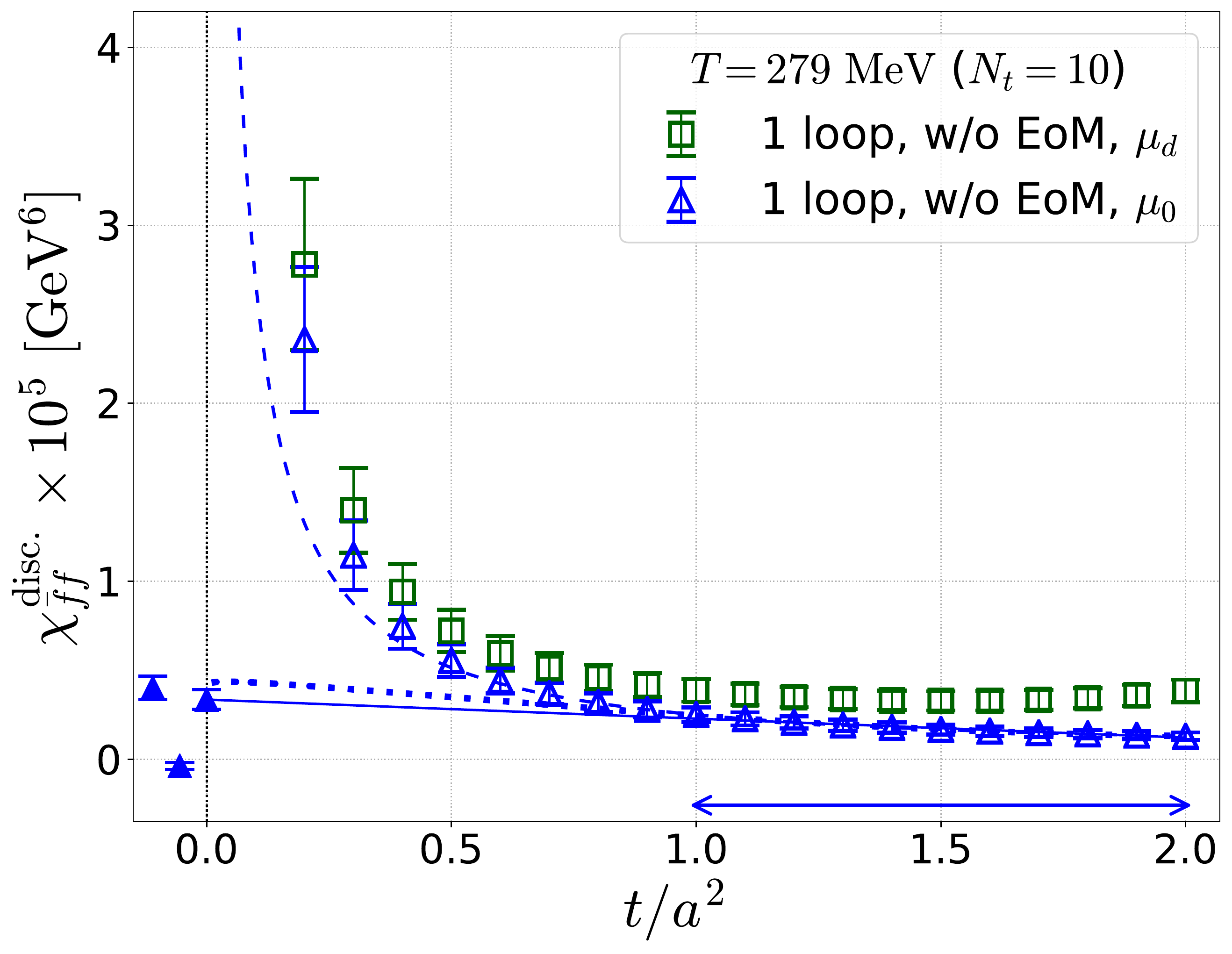}
\includegraphics[width=7cm]{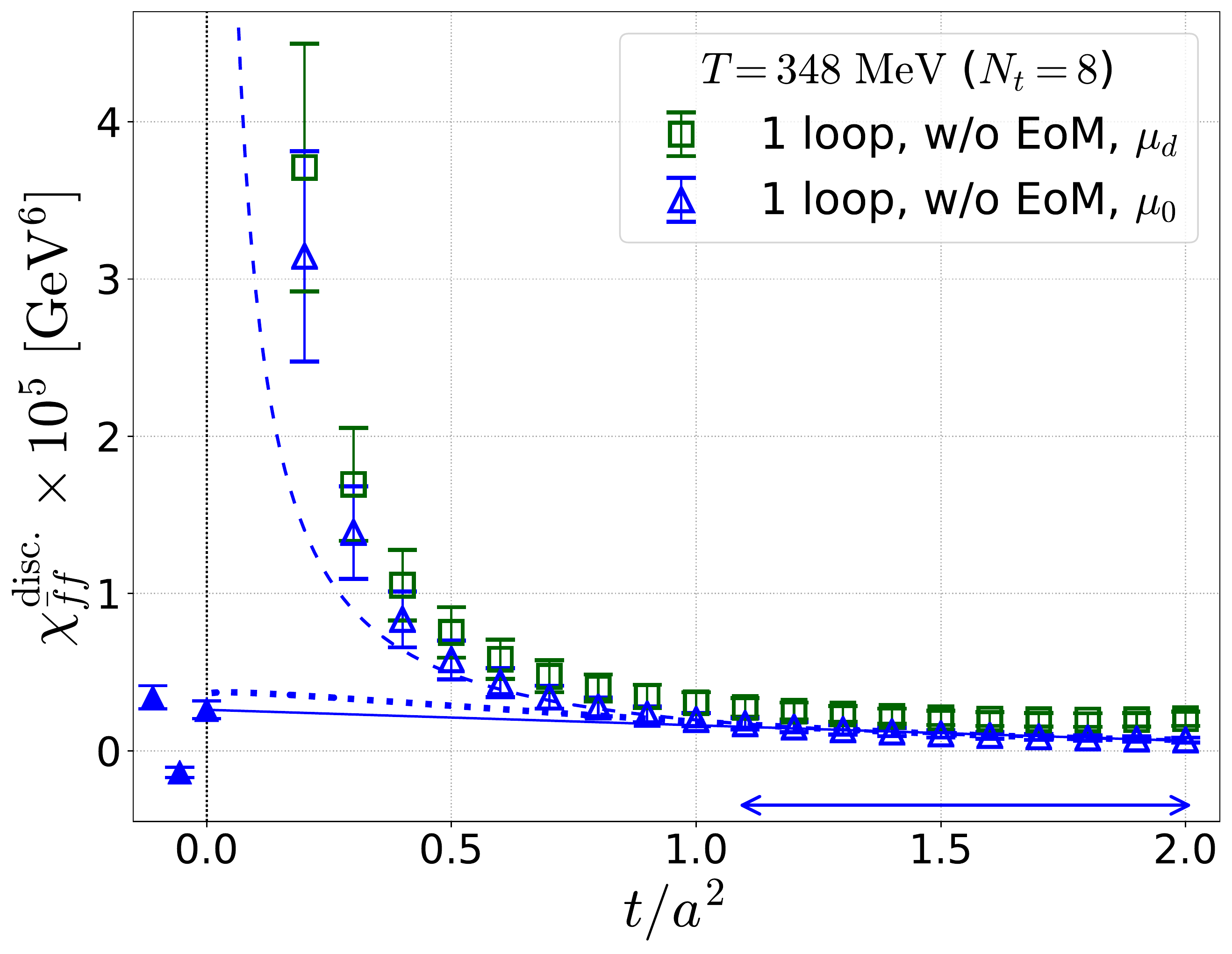}
\vspace{-5mm}
\caption{The same as Fig.~\ref{figmu:eplusp} but for the disconnected chiral susceptibility $\chi_{\Bar{f}f}^{\mathrm{disc.}}$ in the $\overline{\textrm{MS}}$-scheme at 2 GeV for $f=u$ or $d$ quark.
The vertical axis is in units of GeV$^{6}$.  
}
\label{figmu:chiral_susceptibility_ud}
\end{figure}

\begin{figure}[htb]
\centering
\includegraphics[width=7cm]{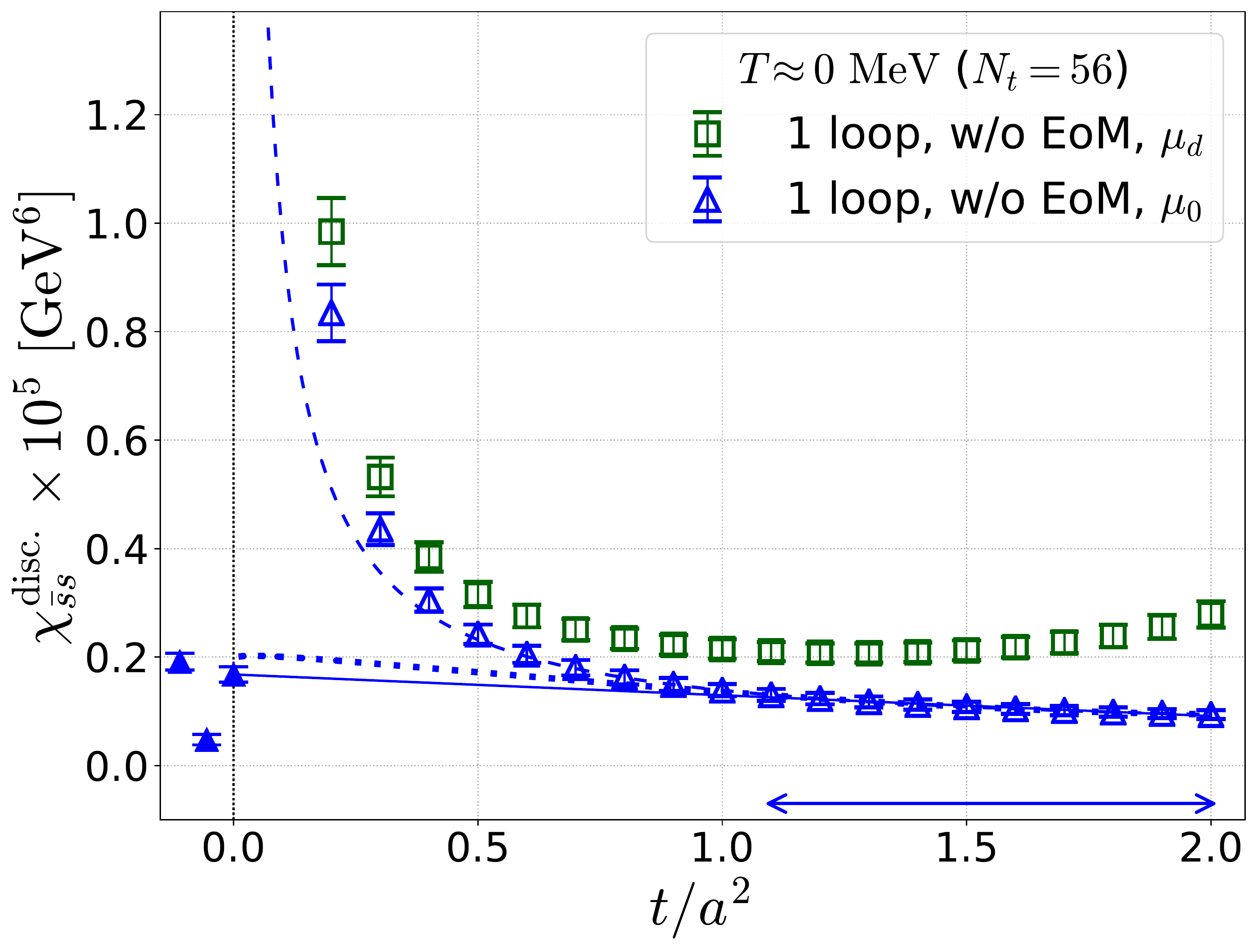}
\includegraphics[width=7cm]{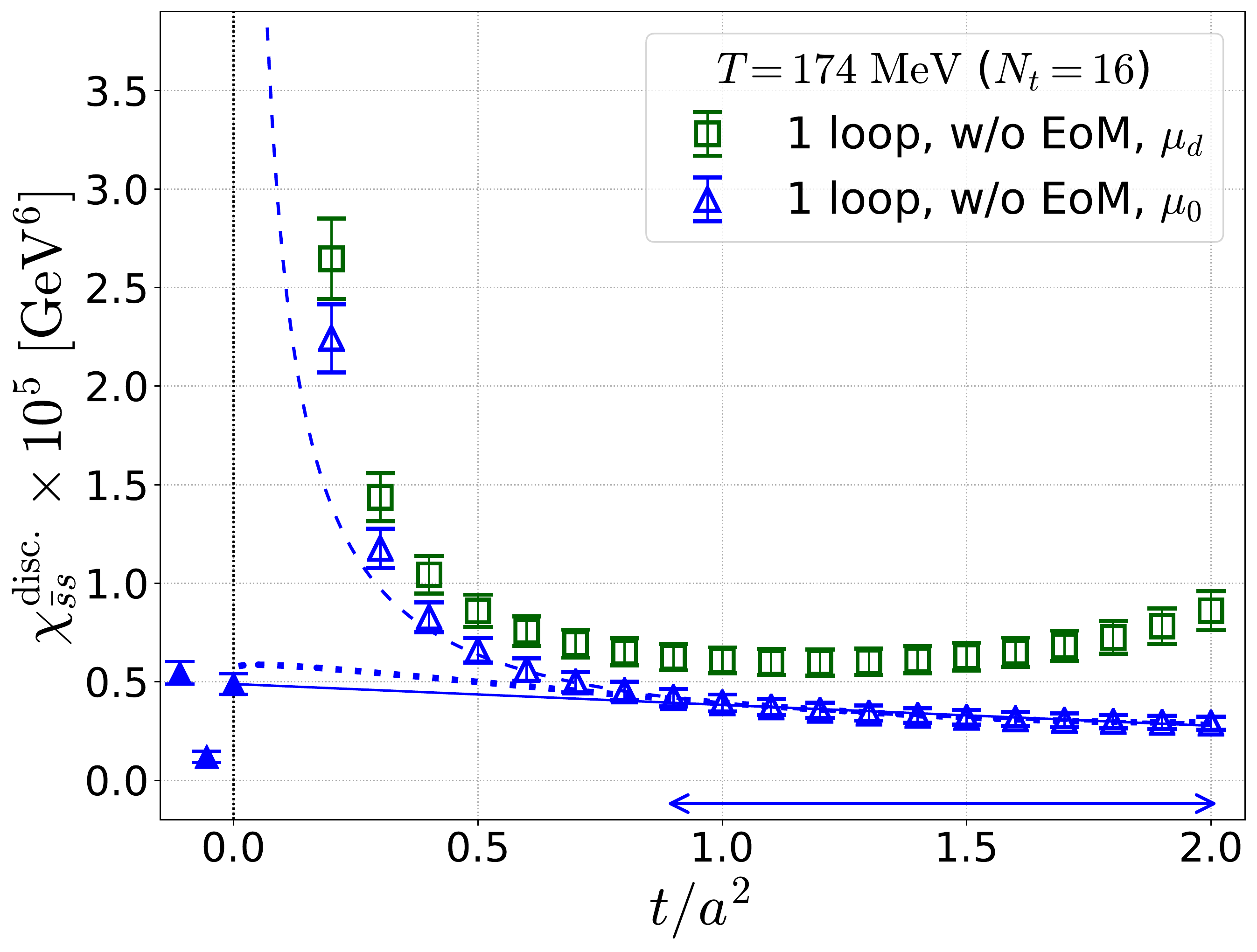}
\includegraphics[width=7cm]{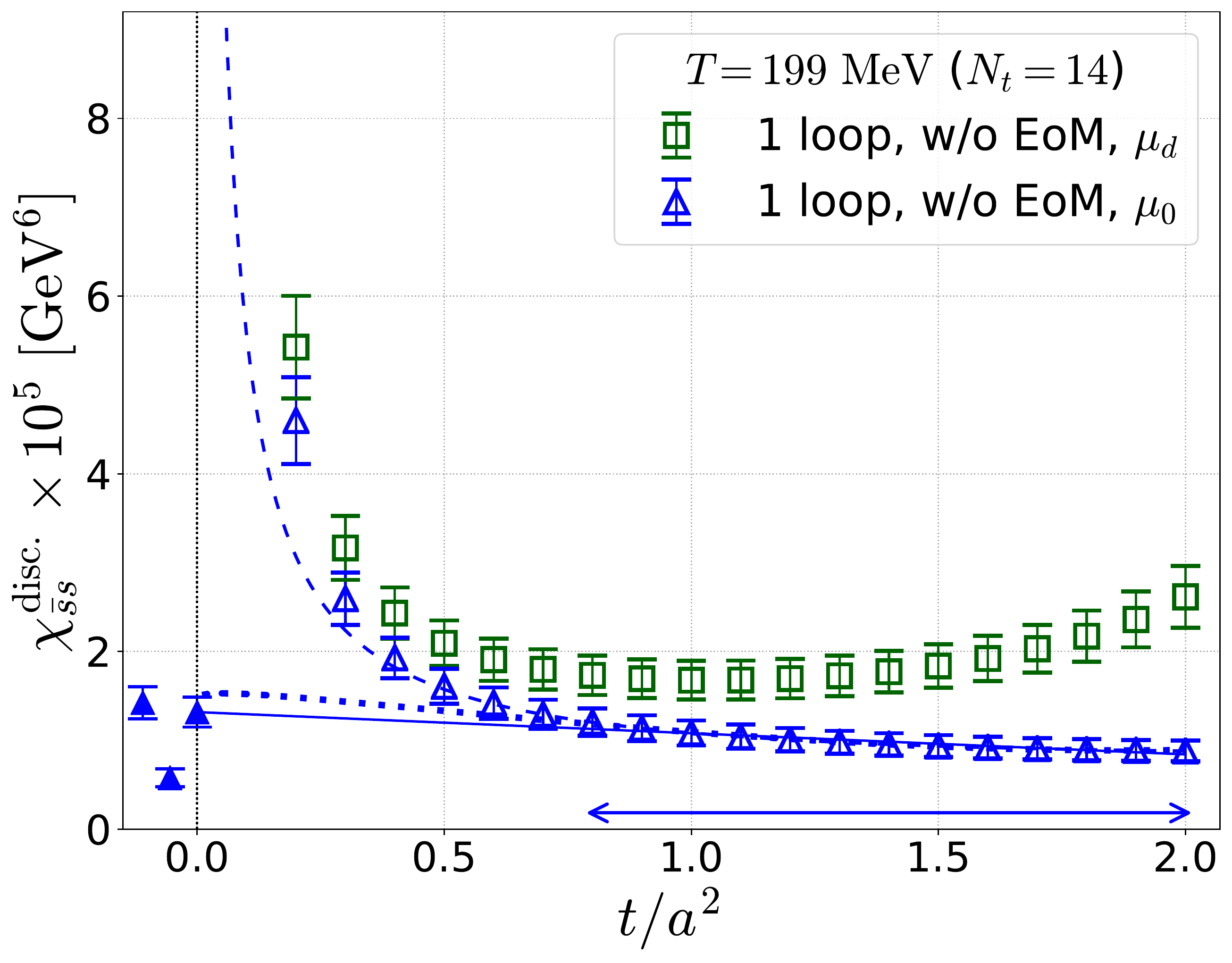}
\includegraphics[width=7cm]{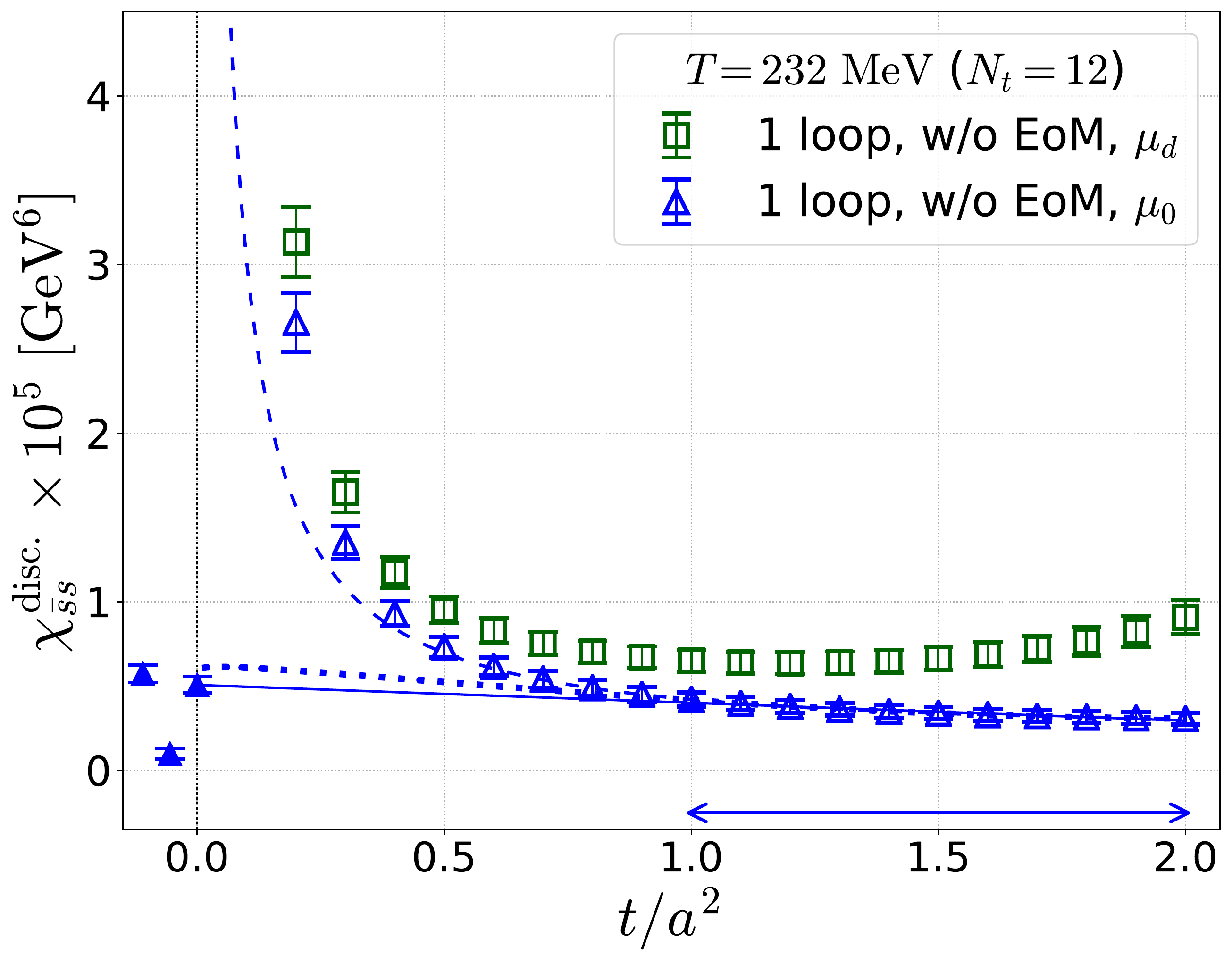}
\includegraphics[width=7cm]{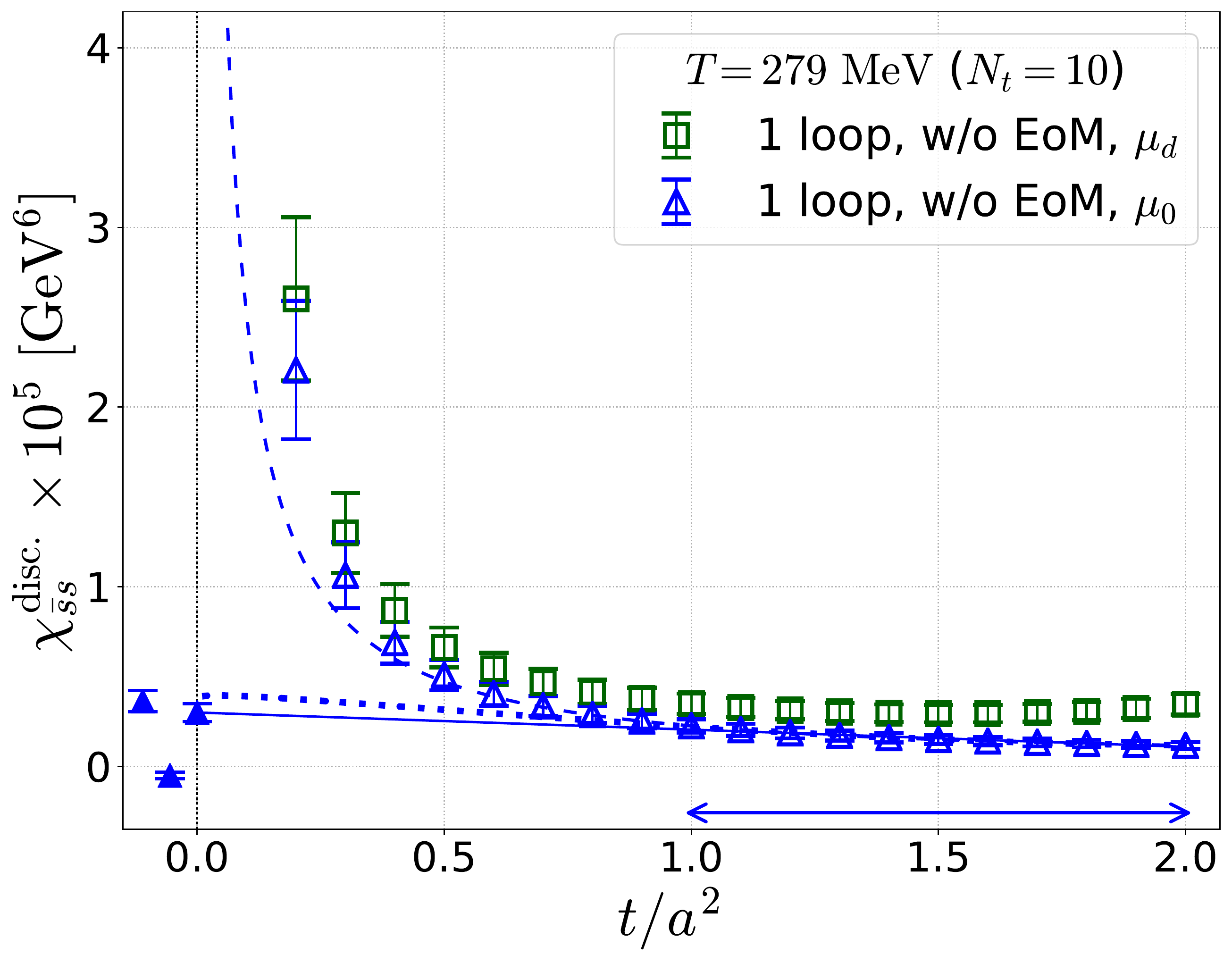}
\includegraphics[width=7cm]{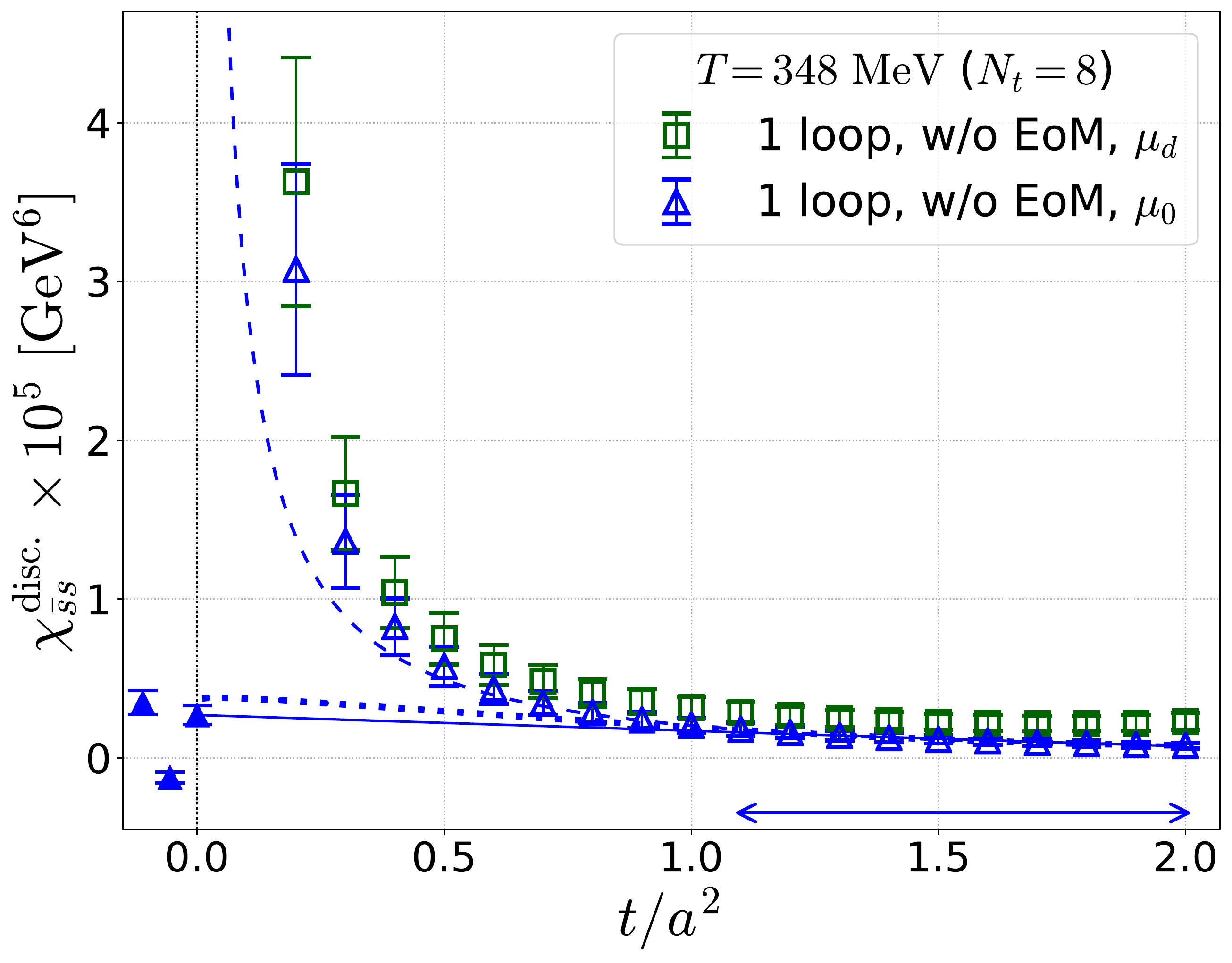}
\vspace{-5mm}
\caption{The same as Fig.~\ref{figmu:chiral_susceptibility_ud} but for the $s$ quark.
}
\label{figmu:chiral_susceptibility_s}
\end{figure}

\begin{table}[htb]
\centering
\caption{Results for chiral condensates and disconnected chiral susceptibilities by the SF\textit{t}X method with the $\mu_0$-scale
using the one-loop matching coefficients of Ref.~\cite{Makino:2014taa}.
The chiral condensates are in units of GeV$^3$, and the disconnected chiral susceptibilities are in units of GeV$^6$.
The first parenthesis is for the statistical error, and the second for the systematic error due to the fit ansatz.
}
\label{table:chiral1}
\begin{tabular}{ccccc}
 $T$[MeV] 
 & $\left\langle\{\Bar{\psi}_u\psi_u\}(x)\right\rangle\times10^2$ 
 & $\left\langle\{\Bar{\psi}_s\psi_s\}(x)\right\rangle\times10^2$ 
 & $\chi_{\Bar{u}u}^{\mathrm{disc.}}\times10^5$ 
 & $\chi_{\Bar{s}s}^{\mathrm{disc.}}\times10^5$ \\
\hline
174 & 0.21(4)($^{+0}_{-7}$) & 0.15(3)($^{+0}_{-5}$) & 0.84(11)($^{+08}_{-54}$) & 0.49(6)($^{+06}_{-37}$) \\
199 & 1.02(6)($^{+03}_{-18}$) & 0.81(5)($^{+03}_{-14}$) & 2.16(28)($^{+13}_{-1.14}$) & 1.32(17)($^{+11}_{-75}$) \\
232 & 2.00(3)($^{+11}_{-18}$) & 1.68(3)($^{+10}_{-12}$) & 0.72(8)($^{+08}_{-53}$) & 0.51(5)($^{+07}_{-41}$) \\
279 & 2.72(3)($^{+17}_{-18}$) & 2.51(3)($^{+18}_{-10}$) & 0.33(6)($^{+07}_{-38}$) & 0.30(5)($^{+07}_{-36}$) \\
348 & 3.40(4)($^{+23}_{-26}$) & 3.52(3)($^{+24}_{-26}$) & 0.26(6)($^{+08}_{-40}$) & 0.27(6)($^{+08}_{-40}$) \\
\hline
\end{tabular}
\end{table}

Results for the chiral condensate $\left\langle\{\Bar{\psi}_f\psi_f\}\right\rangle$ for $f=u$ or $d$ quark and for $s$ quark are shown in Figs.~\ref{figmu:bpsipsi_subvev_ud} and \ref{figmu:bpsipsi_subvev_s}, respectively,
where the VEV's are subtracted from the chiral condensates to remove singularities like $m^2/t$~\cite{Hieda:2016lly,Taniguchi:2016ofw}.  
Figures~\ref{figmu:chiral_susceptibility_ud} and \ref{figmu:chiral_susceptibility_s} show the results for the disconnected chiral susceptibility, 
\begin{equation}
   \chi_{\Bar{f}f}^{\mathrm{disc.}}
   = \left\langle \left[ \frac{1}{N_\Gamma} \sum_x \{\Bar{\psi}_f\psi_f\}(x) \right]^2 \right\rangle_{\!\mathrm{disc.}}
   - \left[ \left\langle\frac{1}{N_\Gamma} \sum_x \{\Bar{\psi}_f\psi_f\}(x)\right\rangle \right]^2,
\end{equation}
where connected quark loop contribution is dropped from the scalar density two-point function and $N_\Gamma$ is the lattice volume. 
Here, because the VEV-subtraction is not required for~$\chi_{\Bar{f}f}^{\mathrm{disc.}}$, we also study the case of $T=0$.

We find that, though a linear window is sometimes not clear with the conventional $\mu_d$-scale, the linear behavior is much improved by adopting the $\mu_0$-scale in particular at large~$t/a^2$. 
The $\mu_0$-scale extends the applicability of the $t\to0$ extrapolation method based on the linear window. 
In the followings, we perform $t\to0$ extrapolations with the $\mu_0$-scale only.
\footnote{
Similar but more drastic improvement with the $\mu_0$-scale was observed in our preliminary study of $2+1$ flavor QCD at the physical point on a less fine lattice~\cite{Lat2019-kanaya}.
On the other hand, no apparent improvement with the $\mu_0$-scale was reported in the study of quenched QCD~\cite{Iritani2019}. 
This may be understood as follows: Because $T_\textrm{c}$ in quenched QCD is higher than~$T_\textrm{pc}$ in full QCD, the effective coupling in quenched QCD is smaller at similar~$T/T_\textrm{c/pc}$, and thus the relevant range of $t$ is well perturbative already with the~$\mu_d$-scale.
We also note that~$T_\textrm{pc}$ in full QCD decreases as we decrease the quark mass toward the physical point.
}

In Figs.~\ref{figmu:bpsipsi_subvev_ud}, \ref{figmu:bpsipsi_subvev_s}, \ref{figmu:chiral_susceptibility_ud}, and~\ref{figmu:chiral_susceptibility_s},
the arrow at the bottom of each plot is the linear window we adopt, and the symbols at $t\sim0$ shows the results of $t\to0$ extrapolations, from the right to the left, using linear, nonlinear, and linear+log fits, respectively. 
In~Table~\ref{table:chiral1}, we summarize the final results for the chiral condensates and disconnected chiral susceptibilities, obtained in the $t\to0$ limit using the one-loop matching coefficients of Ref.~\cite{Makino:2014taa} with the $\mu_0$-scale.

\clearpage

\begin{figure}[tbh]
\centering
\includegraphics[width=8cm]{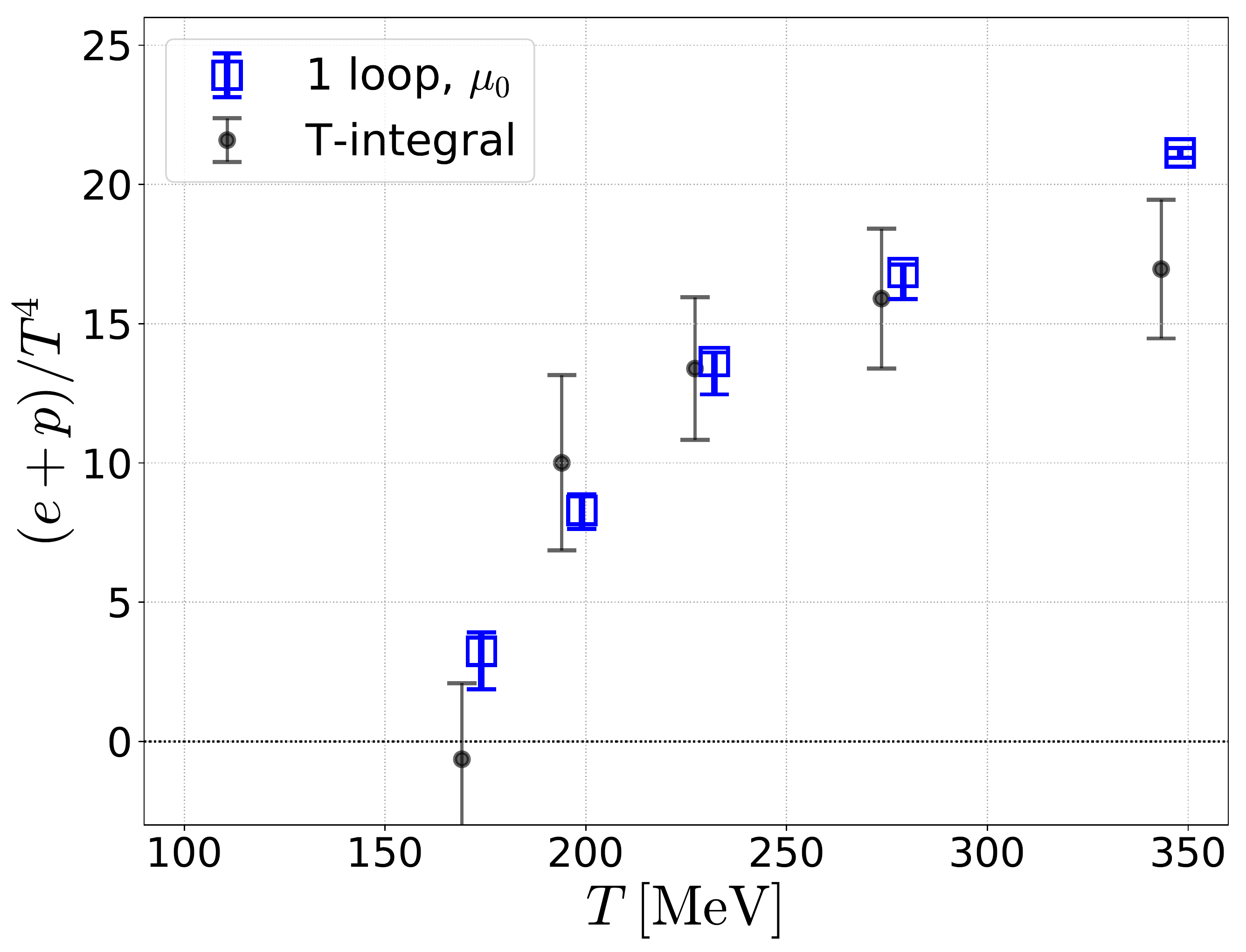}
\includegraphics[width=8cm]{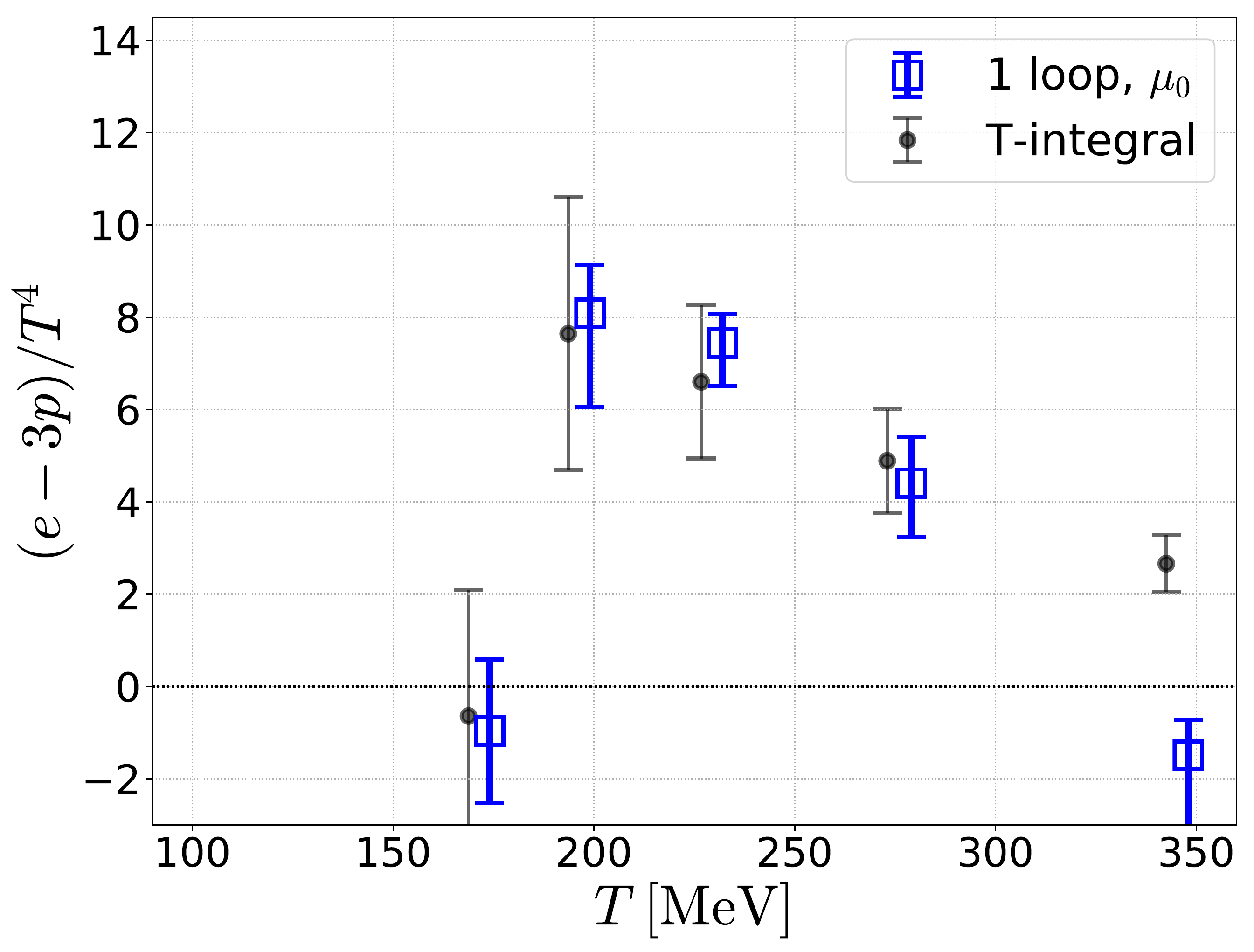}
\caption{Results of the SF\textit{t}X method for EoS with the $\mu_0$-scale as function of temperature. 
One-loop matching coefficients of~Ref.~\cite{Makino:2014taa} are used.
Black circles are the results of the $T$-integration method \cite{Umeda:2012er}.
Left panel: entropy density. Right panel: trace anomaly.
Errors include systematic error due to the fit ansatz for the $t\to0$ extrapolation.
The symbols are slightly shifted horizontally to avoid overlapping.
}
\label{figmu:eos}
\end{figure}

\subsection{Results with the $\mu_0$-scale using one-loop matching coefficients}
\label{sec:1loop_mu_t0}

In Fig.~\ref{figmu:eos}, we summarize the physical results for EoS as function of temperature, obtained by $t\to0$ extrapolations of the data with the $\mu_0$-scale shown in Sec.~\ref{sec:1loop_mu}.
The central values are taken from the linear fits and difference with the results of nonlinear and linear+log fits are taken as estimates of the systematic error due to the fit ansatz for the $t\to0$ extrapolation. 

Black dots in Fig.~\ref{figmu:eos} show the results of EoS obtained previously by the conventional $T$-integration method using the same configurations~\cite{Umeda:2012er}.
Our conclusions are the same as~Ref.~\cite{Taniguchi:2016ofw}:
At~$T\simle279$ MeV ($N_t \ge 10$), the SF\textit{t}X method leads to EoS which is well consistent with the result of the conventional method,
while $a$-independent lattice artifacts of $O\!\left((aT)^2=1/N_t^2\right)$ are suggested for $N_t\simle8$. 
Because the continuum extrapolation is not taken yet, the good agreement of different estimations at $N_t \ge 10$ suggests that the remaining $O(a^{2} T^{2},\; a^{2} m^{2},\; a^{2} \Lambda_{\mathrm{QCD}}^{2})$ lattice artifacts are small with our lattice action at $a\simeq0.07$ fm.

\begin{figure}[htb]
\centering
\includegraphics[width=8cm]{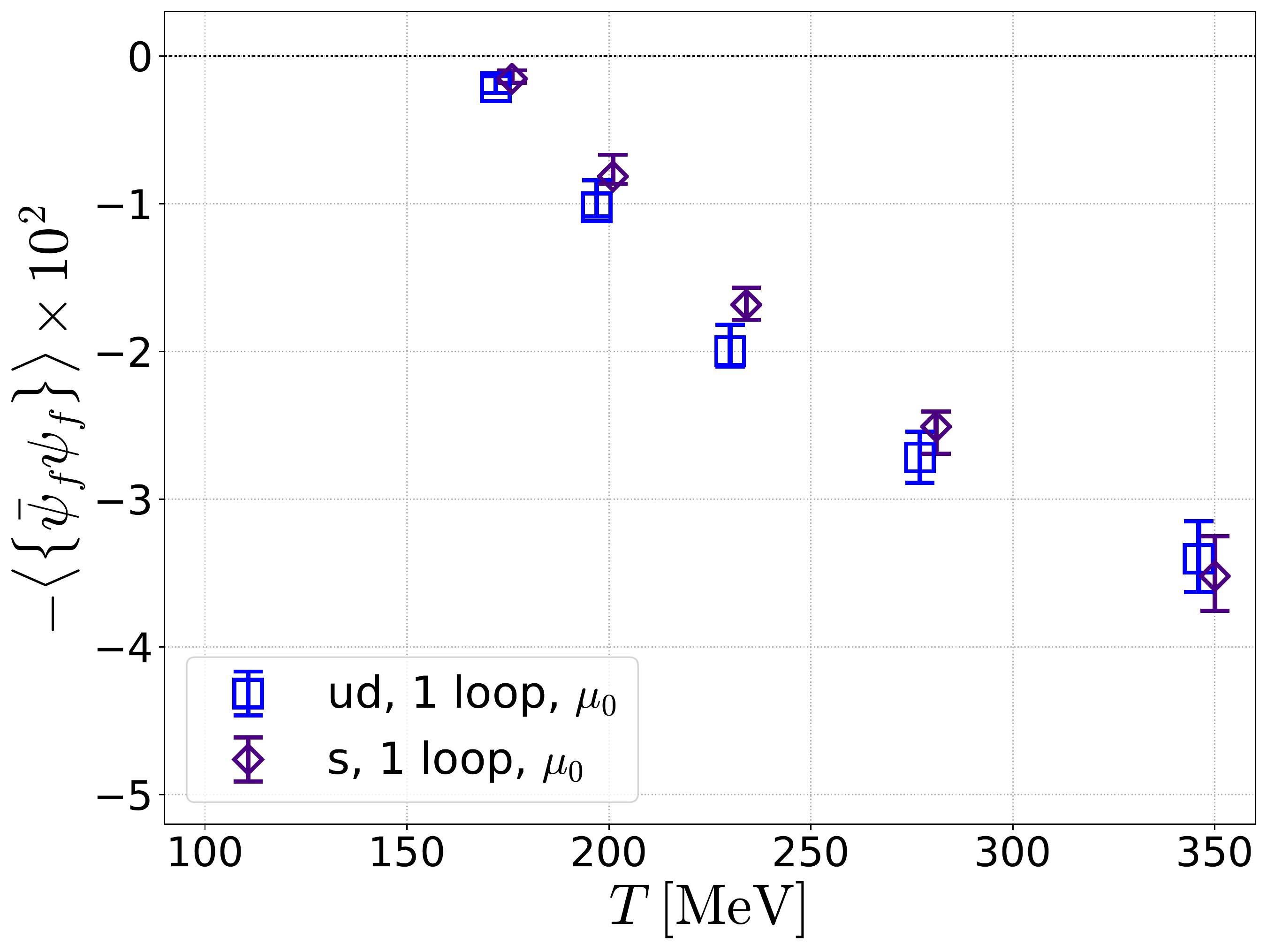}
\includegraphics[width=8cm]{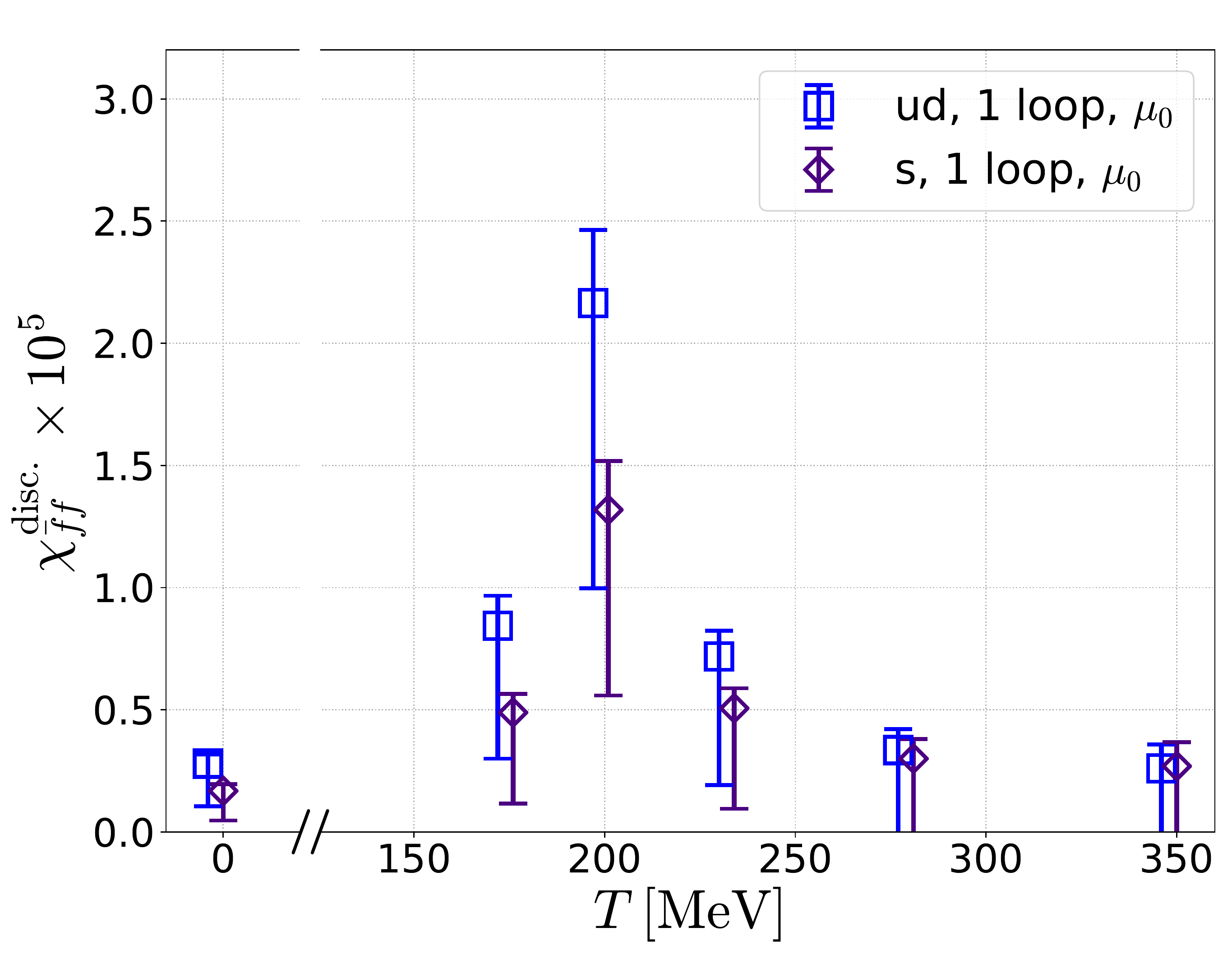}
\caption{
Results of the SF\textit{t}X method with the $\mu_0$-scale for the VEV-subtracted chiral condensates~$\left\langle\{\Bar{\psi}_f\psi_f\}\right\rangle$ and disconnected chiral susceptibilities $\chi_{\Bar{f}f}^{\mathrm{disc.}}$ in the $\overline{\textrm{MS}}$-scheme at 2~GeV, as function of temperature. 
One-loop matching coefficients of Ref.~\cite{Makino:2014taa} are used.
Left panel: $-\left\langle\{\Bar{\psi}_f\psi_f\}\right\rangle$ with the vertical axis in units of GeV$^3$. Following a convention, we plot $-\left\langle\{\Bar{\psi}_f\psi_f\}\right\rangle$. 
Right panel: $\chi_{\Bar{f}f}^{\mathrm{disc.}}$ with the vertical axis in units of GeV$^6$.
Errors include systematic error due to the fit ansatz for the $t\to0$ extrapolation.
The symbols are slightly shifted horizontally to avoid overlapping.
}
\label{figmu:ccond}
\end{figure}

Results for the VEV-subtracted chiral condensates are shown in the left panel of~Fig.~\ref{figmu:ccond}.
In the right panel of Fig.~\ref{figmu:ccond}, we show the results for disconnected chiral susceptibilities as function of temperature.
Because the VEV-subtraction has no effects in this quantity, we also show the results at $T=0$.
We find a clear peak at $T\simeq199$ MeV, which may be indicating the pseudocritical point around 
$T_{\mathrm{pc}}\sim190$ MeV previously suggested using the Polyakov loop etc.\ \cite{Umeda:2012er}. 
We also note that, although the errors are large yet, the height of the peak looks increasing as we decrease the
valence quark mass from $s$ quark to $u$ (or $d$) quark.

\clearpage 
%%%%%%%%%%%%%%%%%%%%%%%%%%%%%%%%%%%%%%%%%%%%%%%%%%%%
\section{Test of two-loop matching coefficients}
\label{sec:test}

We now test the effects of two-loop matching coefficients for EMT by Harlander \textit{et al.}~\cite{Harlander:2018zpi} in QCD with $2+1$ flavors of dynamical quarks.
Following the discussion in~Sec.~\ref{sec:mu0}, we adopt the $\mu_0$-scale in this test.
As mentioned in Sec.~\ref{sec:intro}, unlike the one-loop coefficients of~Ref.~\cite{Makino:2014taa}, the EoM is used in the two-loop coefficients of~Refs.~\cite{Harlander:2018zpi}.

\begin{figure}[htb]
\centering
\includegraphics[width=7cm]{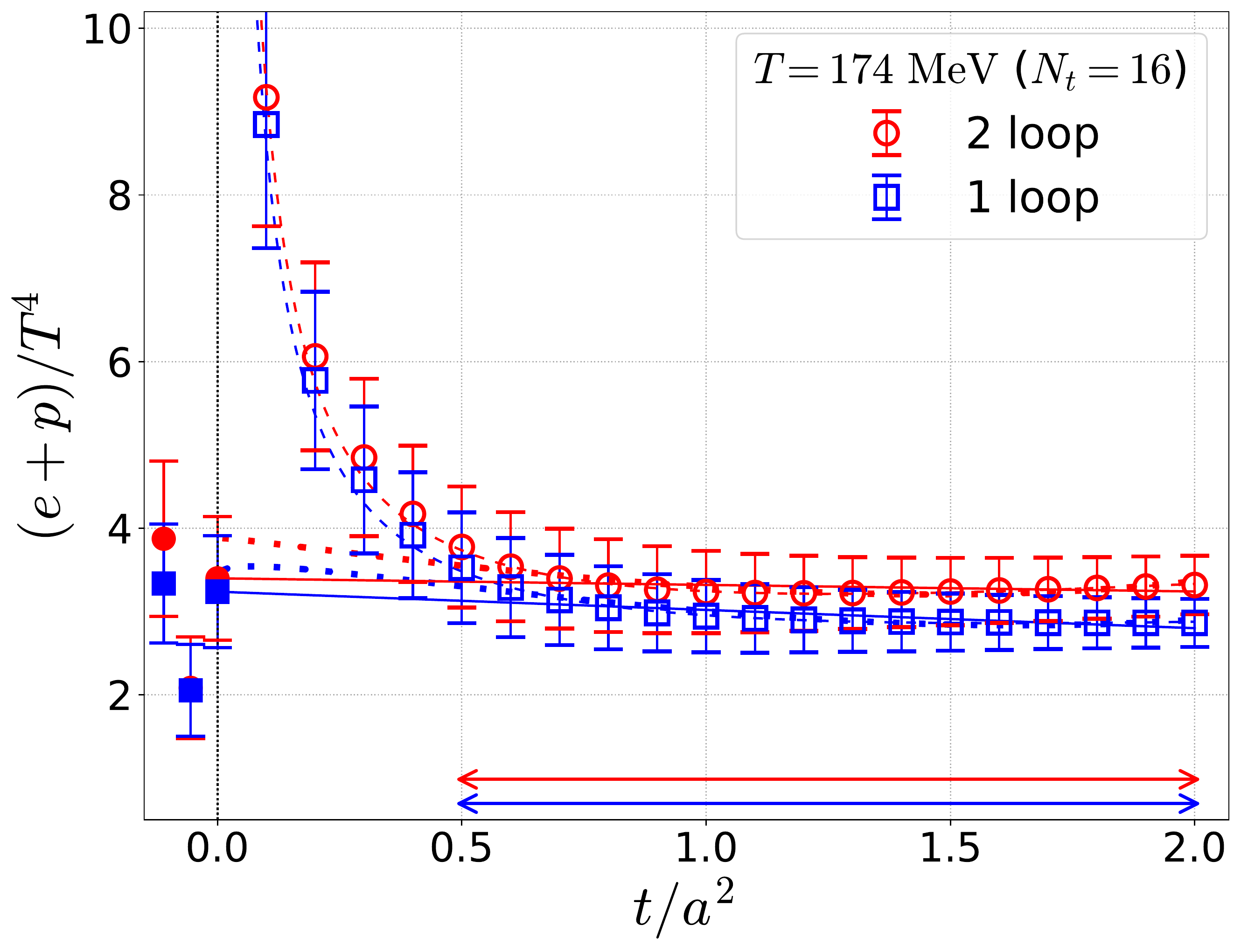}
\includegraphics[width=7cm]{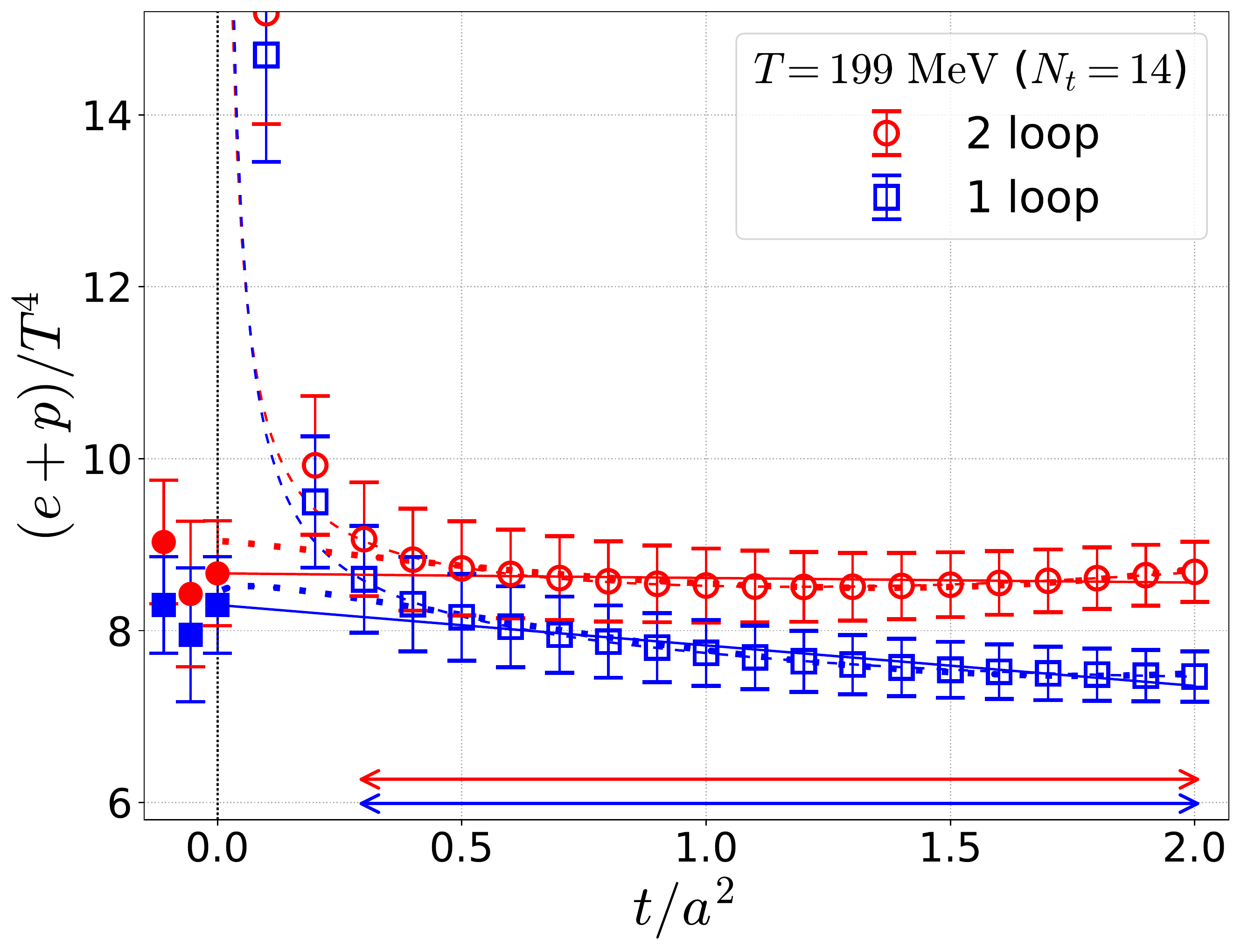}
\includegraphics[width=7cm]{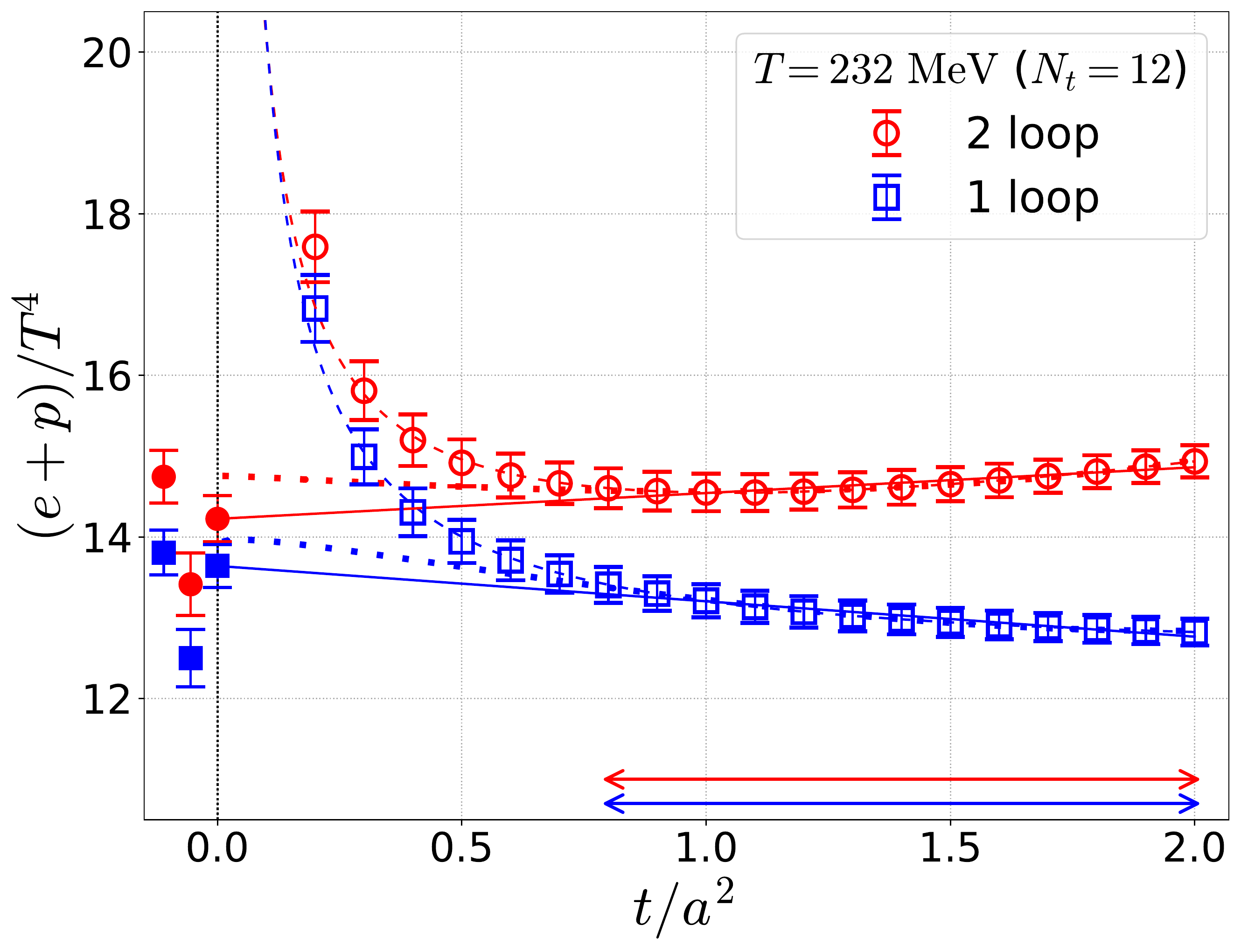}
\includegraphics[width=7cm]{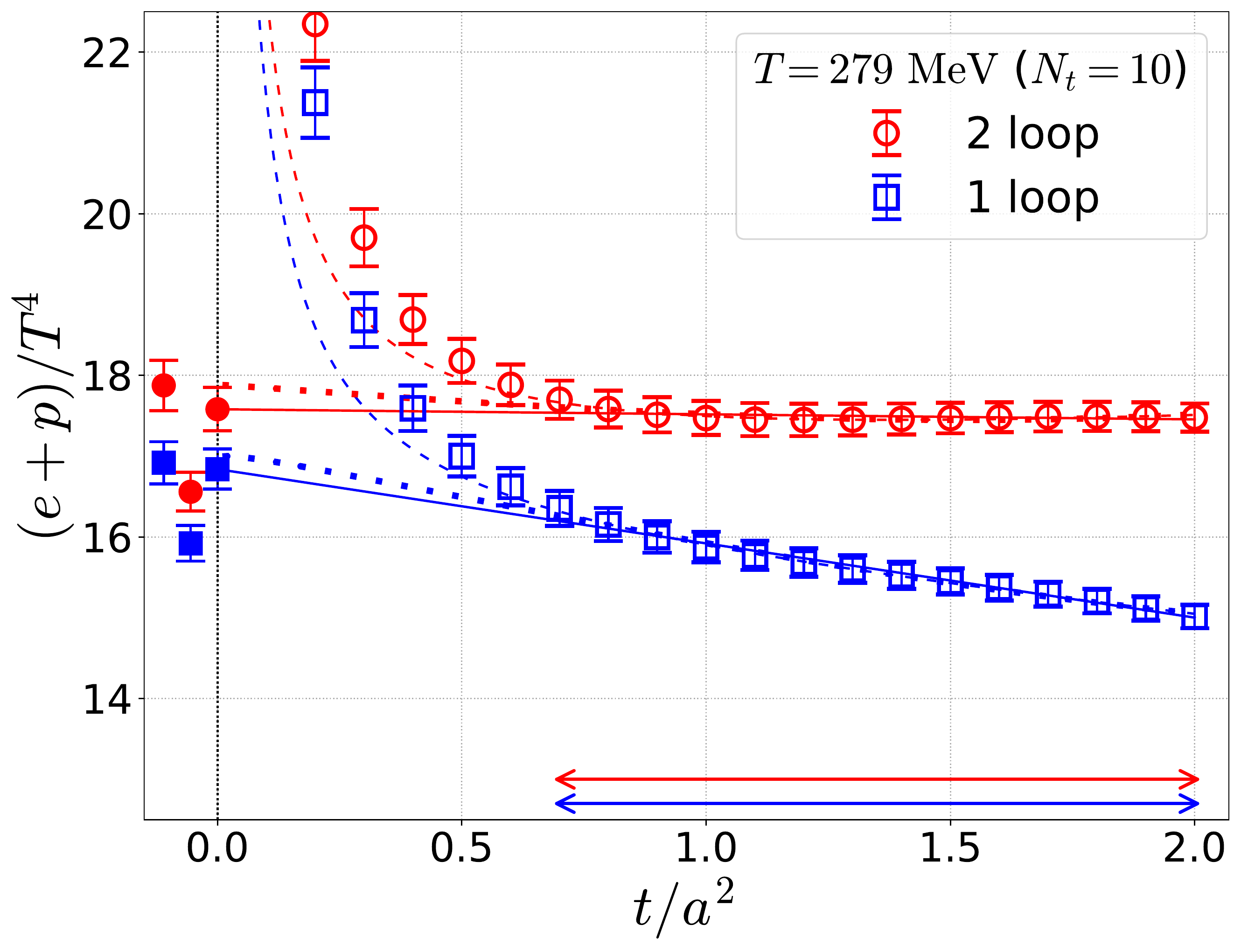}
\includegraphics[width=7cm]{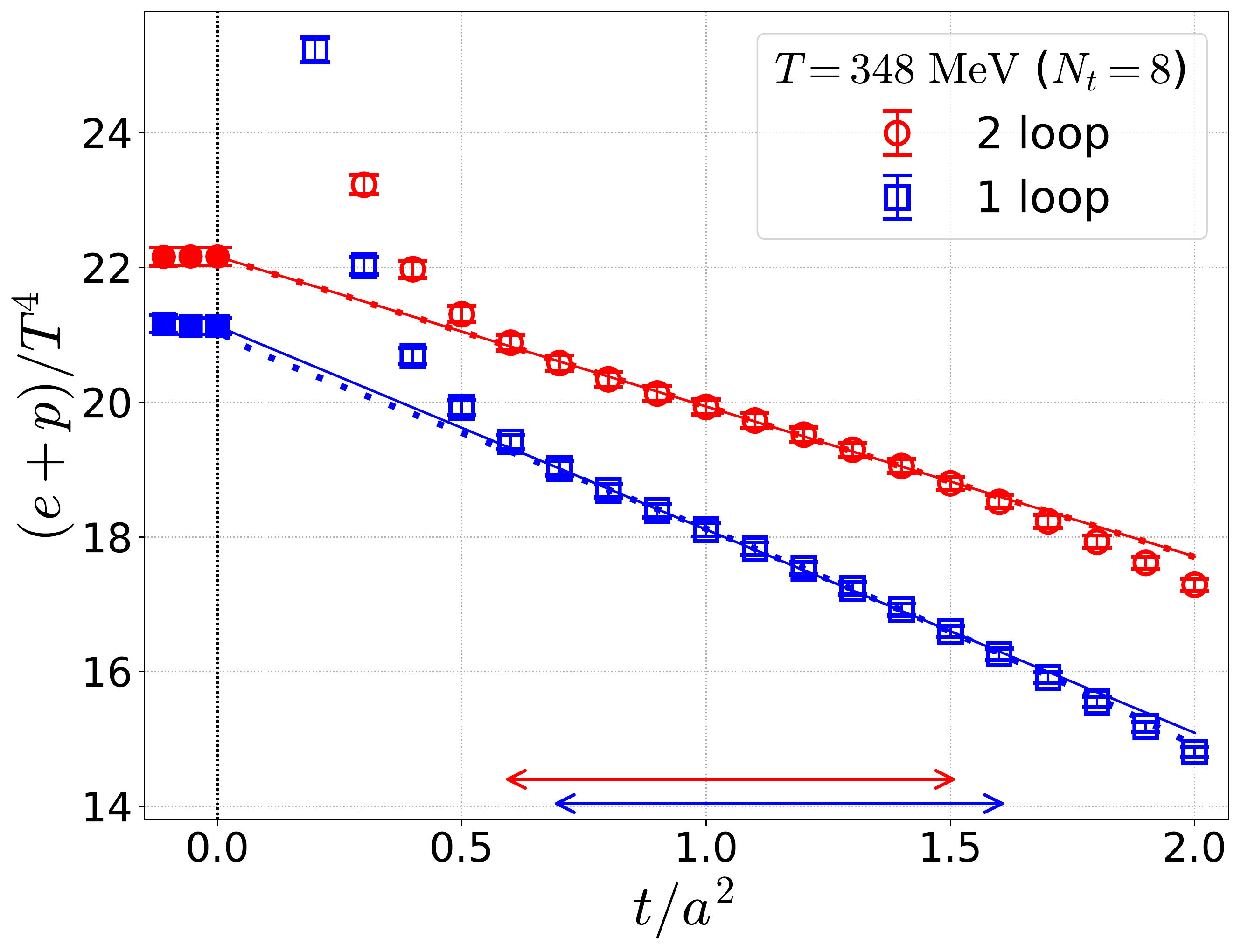}
\vspace{-5mm}
\caption{Entropy density~$(\epsilon+p)/T^4$ with the $\mu_0$-scale as function of the flow-time. 
Results using one-loop matching coefficients of Ref.~\cite{Makino:2014taa} are compared with those using two-loop matching coefficients of Ref.~\cite{Harlander:2018zpi}. 
The arrows at the bottom indicate the linear windows at each temperature.
Symbols at $t/a^2 = 0$ and solid lines with the same color are the results of the linear fits.
Fit results with the nonlinear ansatz~\eqref{eqn:5.4} and linear+log ansatz~\eqref{eqn:5.4l} or~\eqref{eqn:5.4ll} are shown by dashed and dotted curves together with the symbols at $t/a^2 < 0$ to the right and to the left, respectively.
Errors are statistical only.
}
\label{fig2:eplusp}
\end{figure}

\begin{figure}[htb]
\centering
\includegraphics[width=8cm]{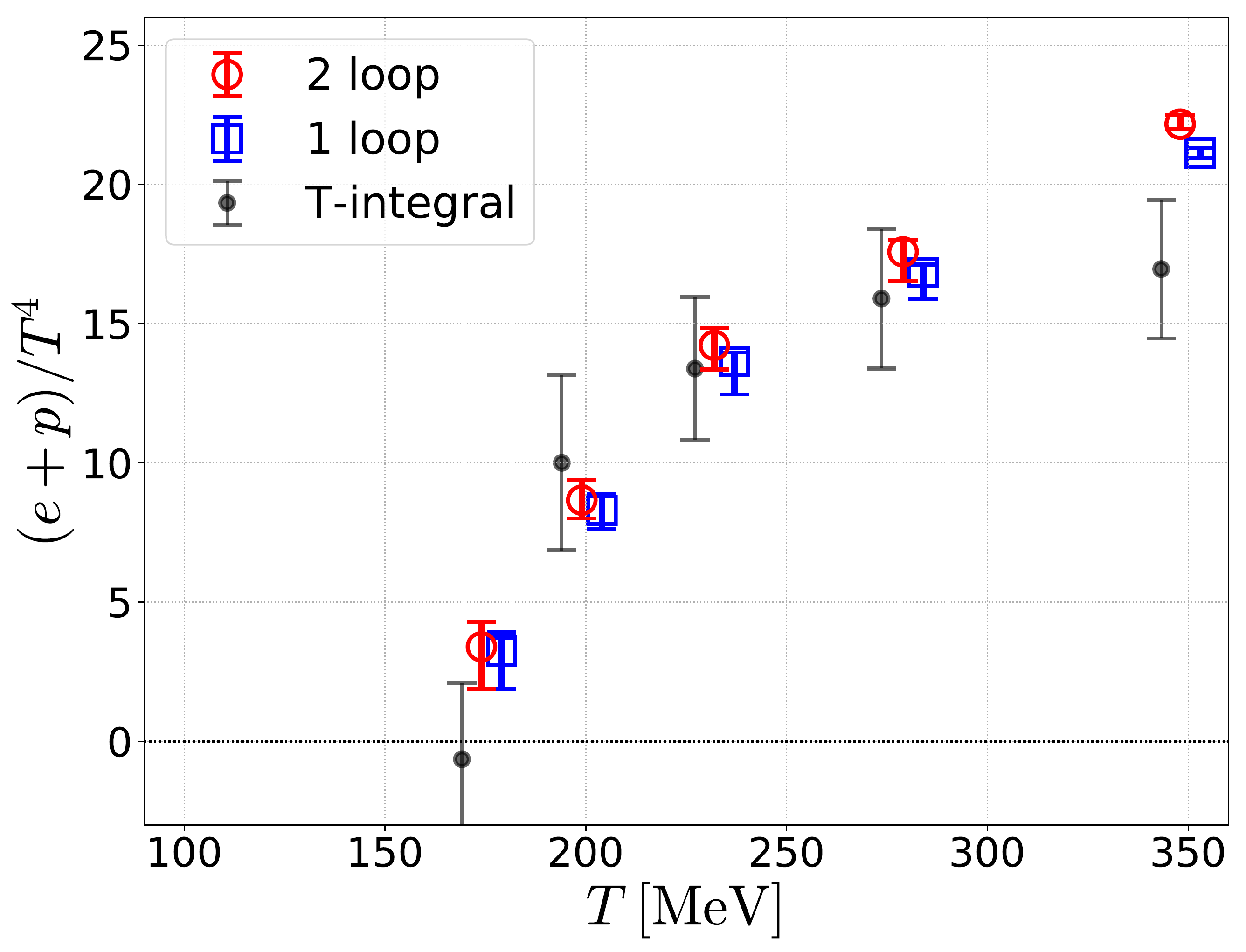}
\caption{Results of the SF\textit{t}X method for the entropy density~$(\epsilon+p)/T^4$ with the $\mu_0$-scale, as function of temperature. 
Results using one-loop matching coefficients of~Ref.~\cite{Makino:2014taa} without using the EoM are compared with results using two-loop matching coefficients of~Ref.~\cite{Harlander:2018zpi} in which the EoM is used.
Black dots are the results of the $T$-integration method~\cite{Umeda:2012er}.
Errors include systematic error due to the fit ansatz for the $t\to0$ extrapolation.
The symbols are slightly shifted horizontally to avoid overlapping.
}
\label{fig2:epluspt0}
\end{figure}

\begin{figure}[htb]
\centering
\includegraphics[width=7cm]{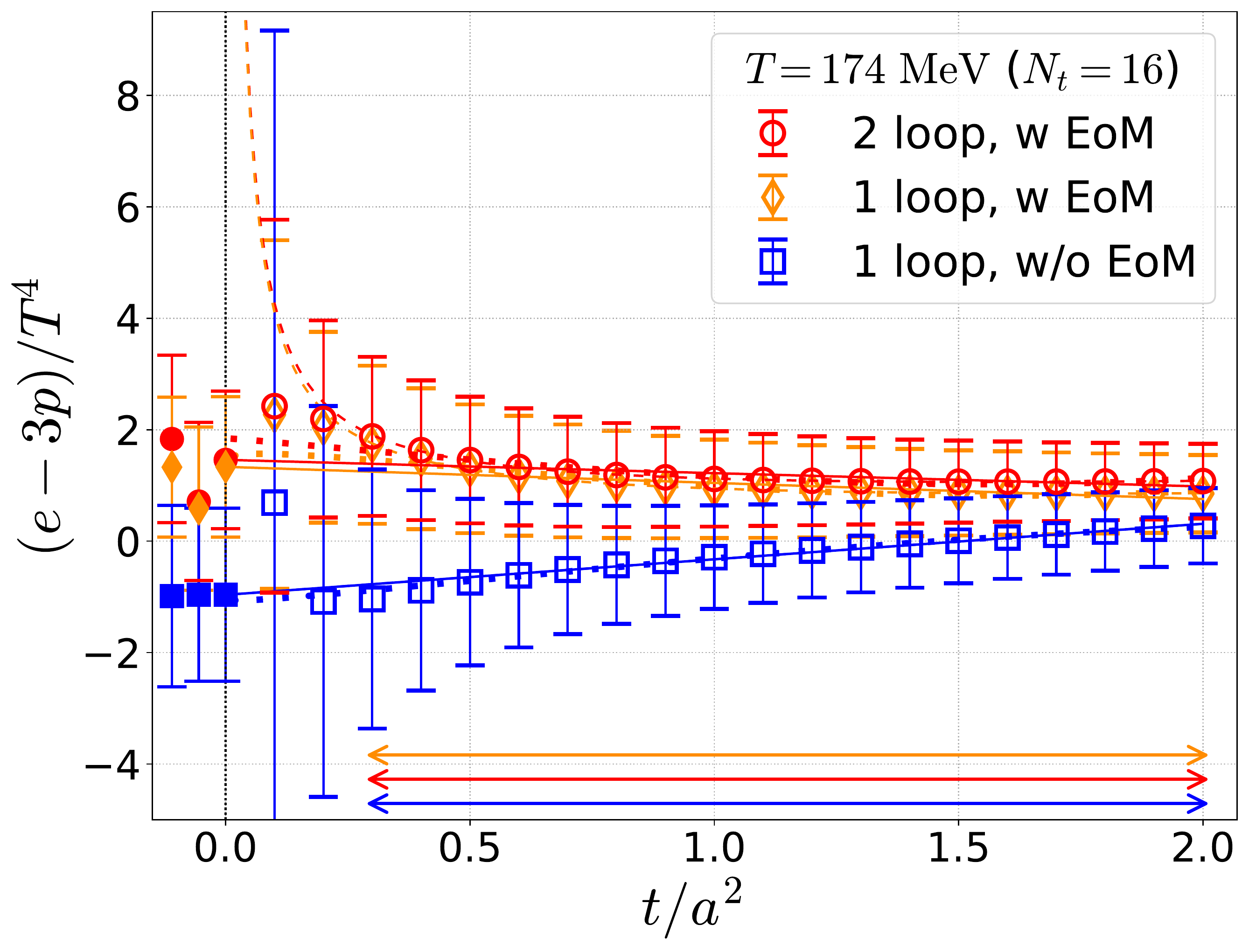}
\includegraphics[width=7cm]{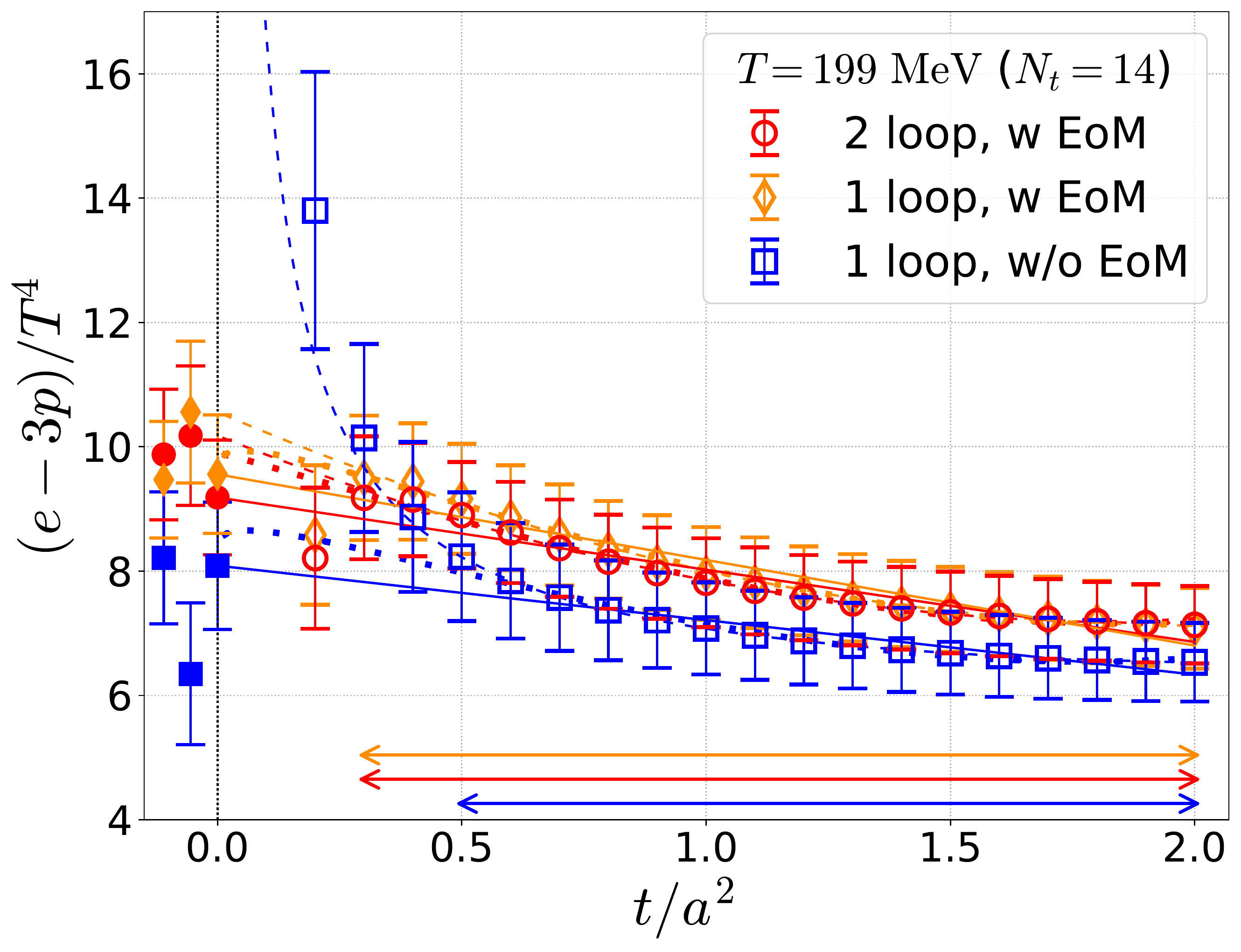}
\includegraphics[width=7cm]{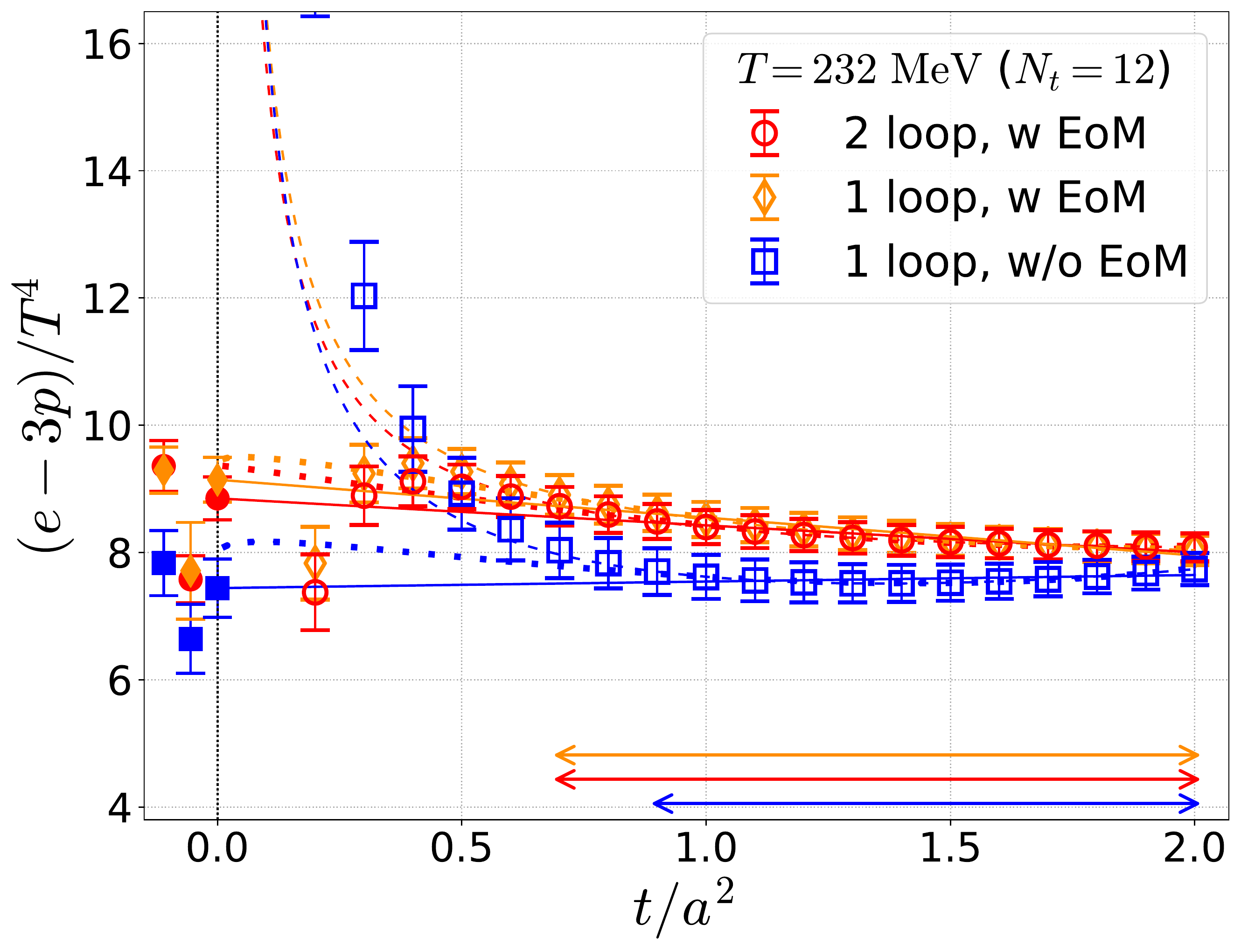}
\includegraphics[width=7cm]{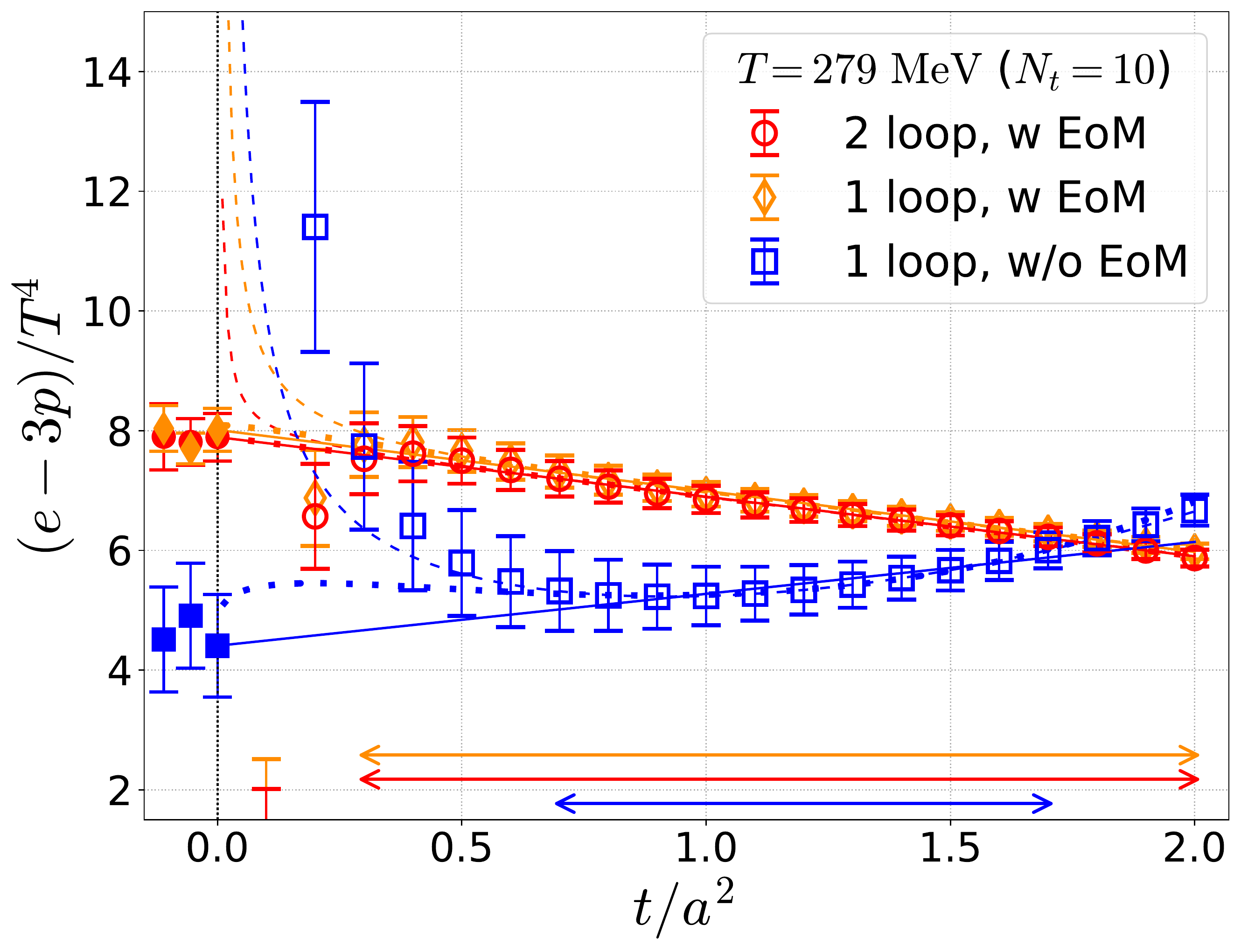}
\includegraphics[width=7cm]{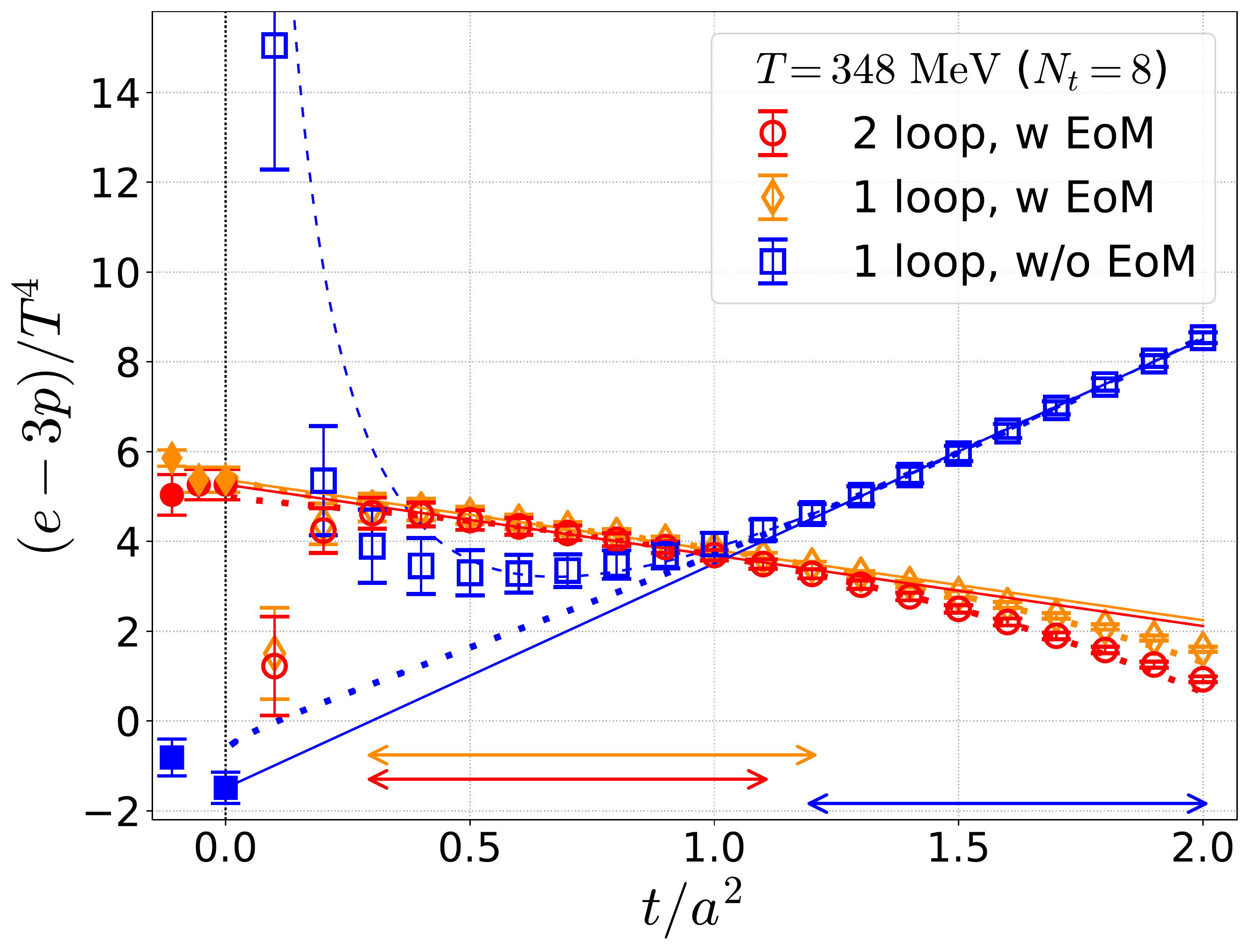}
\vspace{-5mm}
\caption{The same as Fig.~\ref{fig2:eplusp} but for the trace anomaly~$(\epsilon-3p)/T^4$.
Also shown are the results using one-loop part of the matching coefficients of~Ref.~\cite{Harlander:2018zpi} in which the EoM is used.
}
\label{fig2:eminus3p}
\end{figure}

\begin{figure}[htb]
\centering
\includegraphics[width=8cm]{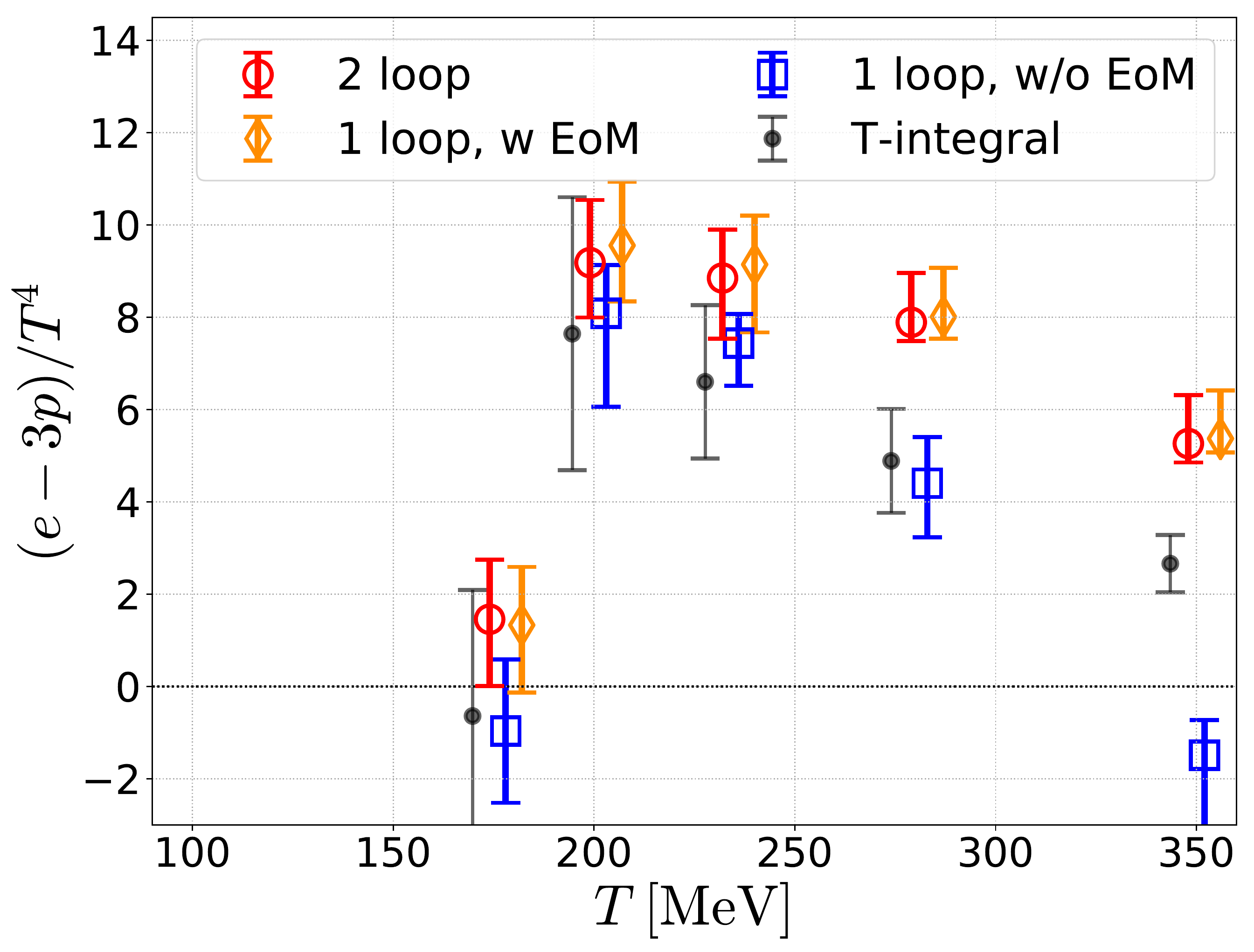}
\caption{Results of the SF\textit{t}X method for the trace anomaly~$(\epsilon-3p)/T^4$ with the $\mu_0$-scale, as function of temperature. 
Results using one-loop matching coefficients of~Ref.~\cite{Makino:2014taa}  without using the EoM are compared with results using two-loop matching coefficients of~Ref.~\cite{Harlander:2018zpi} in which the EoM is used.
Also shown are the results using one-loop part of the matching coefficients of~Ref.~\cite{Harlander:2018zpi} in which the EoM is used.
Black dots are the results of the $T$-integration method~\cite{Umeda:2012er}.
Errors include systematic error due to the fit ansatz for the $t\to0$ extrapolation.
The symbols are slightly shifted horizontally to avoid overlapping.
}
\label{fig2:eminus3pt0}
\end{figure}

%%%%%%%%%%%%%%%%%%%%%%%%%%%%%%%%%%%%%%%%%%%%%%%%%%%%
\subsection{Entropy density}

In Fig.~\ref{fig2:eplusp}, we compare the results of the entropy density at finite $t$ using the one-loop matching coefficients of Ref.~\cite{Makino:2014taa} (blue squares) and those using the two-loop coefficients of~Ref.~\cite{Harlander:2018zpi} (red circles). 
Note that, because the contribution of the EoM to the EMT given by~Eq.~\eqref{eq:(1.18)} is proportional to $\delta_{\mu\nu}$, only the trace part of the EMT is affected by the use of the EoM.
Thus, the EoM has no effects in the entropy density which is a trace-less combination of the EMT.
We find that the entropy density with the two-loop coefficients is larger than its one-loop value at finite $t$, 
but the difference becomes smaller in the~$t\to0$ limit. 

In the previous test in quenched QCD, it was reported that the use of two-loop matching coefficients generally makes the data flatter in the flow-time $t$ and thus makes the $t\to0$ extrapolation more stable~\cite{Iritani2019}.
In our study, we find in Fig.~\ref{fig2:eplusp} that, though a similar general tendency may be visible, we do not see an apparent improvement in the $t\to0$ extrapolation with the use of two-loop coefficients. 
This is caused by the fact that we have sufficiently wide linear windows for $t\to0$ extrapolation with the one-loop coefficients --- no much room was left for drastic improvement on our fine lattice.
Two-loop matching coefficients may help a study of other observables or on coarser lattices.

Physical results for the entropy density with one- and two-loop matching coefficients are shown in~Fig.~\ref{fig2:epluspt0}.
The errors include the systematic error due to the $t\to0$ extrapolation.
Here, it should be recalled that EMT data at $T\simge348$ MeV are contaminated with $O\!\left((aT)^2=1/N_t^2\right)$ lattice artifacts at $N_t \simle 8$~\cite{Taniguchi:2016ofw}. 
We find that one- and two-loop results agree well within the errors at $T<279$ MeV ($N_t>10$).

%%%%%%%%%%%%%%%%%%%%%%%%%%%%%%%%%%%%%%%%%%%%%%%%%%%%
\subsection{Trace anomaly}

In Fig.~\ref{fig2:eminus3p}, we show the results for the trace anomaly $(\epsilon-3p)/T^4$ as function of the flow-time.
The trace anomaly is just the trace part of the EMT and thus will be sensitively affected by the EoM on finite lattices.
In order to identify the effects of EoM clearly, we also compute the trace anomaly using the one-loop part of the matching  
coefficients of~Ref.~\cite{Harlander:2018zpi} in which the EoM is used. 
We find that one- and two-loop results both using the EoM are close with each other, while the one-loop results without using the EoM deviates from the results using the EoM.
We thus conclude that the deviation is mainly due to the use of the EoM.
The deviation increases with increasing $T$ (decreasing $N_t$) and becomes sizable at high temperatures, $T\simge279$ MeV ($N_t\simle10$).

Physical results of the trace anomaly extracted by the $t\to0$ extrapolation are summarized in~Fig.~\ref{fig2:eminus3pt0}.
We find that though the analyses with and without using the EoM lead to consistent results within errors at low temperatures, they show visible discrepancy at high temperatures, $T\simge279$ MeV ($N_t\simle10$).
Even with disregarding the data at $T\simge348$ MeV $(N_t=8)$ where contamination of $O\!\left((aT)^2=1/N_t^2\right)$ lattice artifacts is suggested in EMT, we see discrepancy at $T\simeq279$ MeV ($N_t=10$).
The $O\!\left((aT)^2\right)$ lattice artifacts will contaminate the EoM too.
Our results suggest that the EoM suffers from larger $O\!\left((aT)^2=1/N_t^2\right)$ discretization errors than the EMT, and has visible effect at~$N_t \simle 10$.

Final results for EoS extracted by the $t\to0$ extrapolation with the $\mu_0$-scale are summarized in~Tables~\ref{table:eos1}, \ref{table:eos2} and \ref{table:eos3}.

\begin{table}[htb]
\centering
\caption{Results for EoS by the SF\textit{t}X method with the $\mu_0$-scale
using the two-loop matching coefficients of~Ref.~\cite{Harlander:2018zpi} .
The first parenthesis is for the statistical error, and the second for the systematic error due to the fit ansatz.
}
\label{table:eos2}
\begin{tabular}{ccccc}
 $T$[MeV] & $(\epsilon+p)/T^4$ & $(\epsilon-3p)/T^4$ & $\epsilon/T^4$ & $p/T^4$ \\
\hline
174 & 3.40(75)($^{+48}_{-1.32}$) & 1.46(1.24)($^{+38}_{-75}$) & 2.90(60)($^{+44}_{-1.22}$) & 0.57(39)($^{+10}_{-00}$) \\
199 & 8.67(61)($^{+37}_{-24}$) & 9.18(93)($^{+1.00}_{-00}$) & 8.81(56)($^{+45}_{-00}$) & $-$0.12(26)($^{+00}_{-13}$) \\
232 & 14.22(29)($^{+53}_{-81}$) & 8.85(34)($^{+51}_{-1.28}$) & 12.86(24)($^{+52}_{-99}$) & 1.38(12)($^{+04}_{-00}$) \\
279 & 17.58(27)($^{+30}_{-1.03}$) & 7.89(40)($^{+01}_{-09}$) & 15.14(22)($^{+20}_{-75}$) & 2.44(12)($^{+11}_{-00}$) \\
348 & 22.16(14)($^{+00}_{-01}$) & 5.26(34)($^{+00}_{-23}$) & 18.00(13)($^{+00}_{-06}$) & 4.03(06)($^{+04}_{-73}$) \\
\hline
\end{tabular}
\end{table}

\begin{table}[htb]
\centering
\caption{Results for EoS by the SF\textit{t}X method with the $\mu_0$-scale
using the one-loop part of the matching coefficients of~Ref.~\cite{Harlander:2018zpi} .
The first parenthesis is for the statistical error, and the second for the systematic error due to the fit ansatz.
}
\label{table:eos3}
\begin{tabular}{ccccc}
 $T$[MeV] & $(\epsilon+p)/T^4$ & $(\epsilon-3p)/T^4$ & $\epsilon/T^4$ & $p/T^4$ \\
\hline
174 & 3.24(68)($^{+10}_{-1.19}$) & 1.33(1.27)($^{+00}_{-75}$) & 2.75(56)($^{+07}_{-1.13}$) & 0.56(39)($^{+01}_{-00}$) \\
199 & 8.30(57)($^{+00}_{-35}$) & 9.56(96)($^{+1.01}_{-09}$) & 8.55(52)($^{+02}_{-11}$) & $-$0.30(26)($^{+01}_{-13}$) \\
232 & 13.64(27)($^{+17}_{-1.15}$) & 9.14(35)($^{+16}_{-1.44}$) & 12.50(23)($^{+17}_{-1.26}$) & 1.16(12)($^{+01}_{-00}$) \\
279 & 16.84(25)($^{+08}_{-92}$) & 8.01(36)($^{+03}_{-32}$) & 14.61(21)($^{+07}_{-76}$) & 2.25(11)($^{+02}_{-00}$) \\
348 & 21.13(13)($^{+04}_{-00}$) & 5.37(29)($^{+49}_{-00}$) & 17.26(12)($^{+03}_{-00}$) & 3.85(05)($^{+01}_{-75}$) \\
\hline
\end{tabular}
\end{table}

\clearpage
%%%%%%%%%%%%%%%%%%%%%%%%%%%%%%%%%%%%%%%%%%%%%%%%%%%%
\section{Conclusions}
\label{sec:summary}

We presented the results of our tests of the $\mu_0=e^{-\gamma_E/2}/\sqrt{2t}$ renormalization scale and two-loop matching coefficients recently calculated by Harlander \textit{et al.}~\cite{Harlander:2018zpi}.
For this test, we revisited the case of QCD with heavy u and d quarks~\cite{Taniguchi:2016ofw}. 

We find that, comparing with the results using the conventional $\mu_d=1/\sqrt{8t}$ scale, the $\mu_0$-scale improves the quality of perturbative expressions, in particular at large $t$, and thus leads to clearer and wider linear windows so that we can carry out $t\to0$ extrapolations much confidently.
We also find that, for observables for which the linear window is clear with the conventional $\mu_d$-scale, 
the results using $\mu_0$- and $\mu_d$-scales are consistent with each other,
\textit{i.e.},  the results extrapolated to the $t\to0$ limit are insensitive to the choice of the renormalization scale, as expected. 

Concerning the test of two-loop matching coefficients, unlike the case of the one-loop matching coefficients of Ref.~\cite{Makino:2014taa}, the equation of motion for quark fields in the continuum limit is used by Harlander \textit{et al.}\ in their calculation of the two-loop matching coefficients~\cite{Harlander:2018zpi}. 
For the entropy density in which the equation of motion has no effects,
we found that the results using the two-loop coefficients are well consistent with the results using one-loop coefficients. 
On the other hand, for the trace anomaly in which the equation of motion does affect, we found discrepancies between the one- and two-loop results at high temperatures (small~$N_t$'s).
The main origin of the discrepancies was identified as the use of equation of motion by a direct comparison of the results of one-loop coefficients with and without using the equation of motion.
Our results suggest that the equation of motion suffers from large~$O\!\left((aT)^2\right)=O\!\left(1/N_t^2\right)$ discretization error at~$N_t \simle 10$. 
Therefore, one should be cautious when extracting physical quantities such as the trace anomaly which are affected by the use of the equation of motion. 
This point is not, however, the problem of the two-loop coefficients themselves, and as our results illustrate this point should be regarded as a more general problem caused by discretization errors.

We are attempting to extend applications of the SF\textit{t}X method in various directions: 
thermodynamics of $2+1$ flavor QCD at the physical point~\cite{Lat2017-kanaya,Lat2019-kanaya},
shear and bulk viscosities from two-point correlation functions of the energy-momentum tensor~\cite{Taniguchi:2017ibr},
end-point of first-order deconfining transition region in QCD near the quenched limit~\cite{Shirogane}, 
PCAC quark masses~\cite{ABaba}, etc.
The choice of the $\mu_0$-scale as well as higher order coefficients may help improving these calculations too.

%%%%%%%%%%%%%%%%%%%%%%%%%%%%%%%%%%%%%%%%%%%%%%%%%%%%
\acknowledgments
We are grateful to Prof.~R.V.~Harlander for % kindly sending us their yet unpublished results and succeeding 
valuable discussions. 
This work was in part supported by JSPS KAKENHI Grants No.~JP20H01903,
No.~JP19H05146, No.~JP19H05598, No.~JP19K03819, No.~JP18K03607, No.~JP17K05442 and No.~JP16H03982.
This research used computational resources of COMA, Oakforest-PACS, and Cygnus provided by the Interdisciplinary Computational Science Program of Center for Computational Sciences, University of Tsukuba,
K and other computers of JHPCN through the HPCI System Research Projects (Project ID:hp17208, hp190028, hp190036) and JHPCN projects (jh190003, jh190063), OCTOPUS at Cybermedia Center, Osaka University, ITO at Research Institute for Information Technology, Kyushu University, and Grand Chariot at Information Initiative Center, Hokkaido University.
The simulations were in part based on the lattice QCD code set Bridge++ \cite{bridge}.

%\clearpage
%%%%%%%%%%%%%%%%%%%%%%%%%%%%%%%%%%%%%%%%%%%%%%%%%%%%
\appendix

%%%%%%%%%%%%%%%%%%%%%%%%%%%%%%%%%%%%%%%%%%%%%%%%%%%%
\section{Group factors}
\label{sec:notations}

We normalize the gauge group generators as 
\begin{equation}
\textrm{tr}\left(T^a T^b\right) = - \frac{1}{2} \delta^{ab}.
\end{equation}
The structure constant defined by $[T^a,T^b] = f^{abc}T^c$ has quadratic Casimirs as
\begin{equation}
f^{acd}f^{bcd} = C_2(G)\, \delta^{ab},\qquad
T^aT^a = - C_2(R)\,\unitmatrix .
\end{equation}

In the perturbative expressions in Sec.~\ref{sec:EMT},
group factors are defined by
\begin{equation}
   C_{\!A}\equiv C_2(G),\qquad
   T_F\equiv T(R)\,N_f,\qquad
   C_F\equiv C_2(R).
\label{eq:(1.26)}
\end{equation}
For $G=SU(N)$ and~$R=N$,
\begin{equation}
   C_{\!A}=N,\qquad T_F=\frac{1}{2}N_f,\qquad
   C_F=\frac{N^2-1}{2N}.
\label{eq:(1.27)}
\end{equation}
In particular, for the $N_f=2+1$ QCD,
\begin{equation}
   C_{\!A}=3,\qquad T_F=\frac{3}{2},\qquad
   C_F=\frac{4}{3},
\label{eq:(1.28)}
\end{equation}
and for the quenched QCD (the $SU(3)$ pure Yang--Mills)
\begin{equation}
   C_{\!A}=3,\qquad T_F=0,\qquad C_F=0.
\label{eq:(1.29)}
\end{equation}

%%%%%%%%%%%%%%%%%%%%%%%%%%%%%%%%%%%%%%%%%%%%%%%%%%%%
\section{Consistency of matching coefficients}
\label{sec:confirmation}

In this appendix, we confirm that the matching coefficients of Refs.~\cite{Makino:2014taa} and \cite{Harlander:2018zpi} for EMT are consistent with each other.

\subsection{One-loop confirmation}

We confirm that Eqs.~\eqref{eq:(1.32)}--\eqref{eq:(1.38)} restricted to the one-loop level is consistent with the results
of~Ref.~\cite{Makino:2014taa}.

One way to see this is to construct the small flow time expansion of the
``reduced EMT'' in~Eq.~\eqref{eq:(1.25)}. Using the matrix~$\zeta(t)$ defined
by
\begin{equation}
   \Tilde{\mathcal{O}}_{i\mu\nu}(t,x)
   =\zeta_{ij}(t)\,\mathcal{O}_{j\mu\nu}(x)+O(t),
\label{eq:(1.49)}
\end{equation}
the coefficients~$\Tilde{c}_i(t)$ in
\begin{align}
   &T_{\mu\nu}(x)
\notag\\
   &=\Tilde{c}_1(t)\Tilde{\mathcal{O}}_{1,\mu\nu}(t,x)
   +\Tilde{c}_2(t)\Tilde{\mathcal{O}}_{2,\mu\nu}(t,x)
   +\Tilde{c}_3(t)\Tilde{\mathcal{O}}_{3,\mu\nu}(t,x)
   +\Tilde{c}_4(t)\Tilde{\mathcal{O}}_{4,\mu\nu}(t,x)
   +\Tilde{c}_5(t)\Tilde{\mathcal{O}}_{5,\mu\nu}(t,x)
\notag\\
   &\qquad{}+O(t),
\label{eq:(1.50)}
\end{align}
where the left-hand side is~Eq.~\eqref{eq:(1.25)}, are given by
\begin{equation}
   \Tilde{c}_i(t)
   =\frac{1}{g_0^2}\left\{
   \left(\zeta^{-1}\right)_{1i}(t)
   -\frac{1}{4}\left(\zeta^{-1}\right)_{2i}(t)\right\}
   +\frac{1}{4}\left(\zeta^{-1}\right)_{3i}(t).
\label{eq:(1.51)}
\end{equation}
Compare this to~Eq.~(4.16) of~Ref.~\cite{Makino:2014taa}. The one-loop
$\zeta(t)$ obtained in~Ref.~\cite{Makino:2014taa} then yields
\begin{align}
   \Tilde{c}_1(t)
   &=\frac{1}{g^2}
   \left\{1
   +\frac{g^2}{(4\pi)^2}
   \left[-\beta_0L(\mu,t)-\frac{7}{3}C_{\!A}+\frac{3}{2}T_F\right]
   \right\},
\label{eq:(1.52)}
\\
   \Tilde{c}_2(t)
   &=\frac{1}{4g^2}
   \left\{-1
   +\frac{g^2}{(4\pi)^2}
   \left[\beta_0L(\mu,t)+\frac{25}{6}C_{\!A}-3T_F\right]
   \right\},
\label{eq:(1.53)}
\\
   \Tilde{c}_3(t)
   &=\frac{1}{4}\left[1
   +\frac{g^2}{(4\pi)^2}C_F\left(\frac{3}{2}+\ln432\right)
   \right],
\label{eq:(1.54)}
\\
   \Tilde{c}_4(t)
   &=\frac{g^2}{(4\pi)^2}C_F\left(-\frac{1}{4}\right),
\label{eq:(1.55)}
\\
   \Tilde{c}_5(t)
   &=\frac{g^2}{(4\pi)^2}C_F\left(-\frac{3}{2}\right).
\label{eq:(1.56)}
\end{align}
Note that $\Tilde{c}_4(t)$ and~$\Tilde{c}_5(t)$ are one-loop quantities because
Eq.~\eqref{eq:(1.25)} does not contain $\mathcal{O}_{4,\mu\nu}(x)$
and~$\mathcal{O}_{5,\mu\nu}(x)$ and $\Tilde{\mathcal{O}}_{4,\mu\nu}(t,x)$
and~$\Tilde{\mathcal{O}}_{5,\mu\nu}(t,x)$ appear only through loop diagrams. To
translate these coefficients $\Tilde{c}_i(t)$ in~Eq.~\eqref{eq:(1.50)} to the
coefficients in~Eq.~\eqref{eq:(1.16)}, we have to eliminate the
operator~$\Tilde{\mathcal{O}}_{5,\mu\nu}(t,x)$ from~Eq.~\eqref{eq:(1.50)} in
favor of~$\Tilde{\mathcal{O}}_{4,\mu\nu}(t,x)$
and~$\Tilde{\mathcal{O}}_{2,\mu\nu}(t,x)$ by using the
relation~\eqref{eq:(1.19)}. In the present order of approximation, this is easy
because we can use the relation
\begin{equation}
   \frac{1}{2}\Tilde{\mathcal{O}}_{4,\mu\nu}(t,x)
   +\Tilde{\mathcal{O}}_{5.\mu\nu}(t,x)=0
\label{eq:(1.57)}
\end{equation}
that holds in the \emph{tree-level\/} in~Eq.~\eqref{eq:(1.50)} because
$\Tilde{c}_4(t)$ and~$\Tilde{c}_5(t)$ are already one-loop quantities. After
this elimination, we have
\begin{align}
   \Check{c}_1(t)&=\Tilde{c}_1(t),
\label{eq:(1.58)}
\\
   \Check{c}_2(t)&=\Tilde{c}_2(t),
\label{eq:(1.59)}
\\
   \mathring{\Check{c}}_3(t)&=\Tilde{c}_3(t),
\label{eq:(1.60)}
\\
   \mathring{\Check{c}}_4(t)&=\Tilde{c}_4(t)-\frac{1}{2}\Tilde{c}_5(t).
\label{eq:(1.61)}
\end{align}
We see that these precisely coincide with~Eqs.~\eqref{eq:(1.32)},
\eqref{eq:(1.33)}, \eqref{eq:(1.34)}, and~\eqref{eq:(1.38)} in the one-loop level.

Another way to see the consistency is the following. The one-loop result
of~Ref.~\cite{Makino:2014taa} (cf.\ Eqs.~(4.60)--(4.64)) gives, for the
coefficients in~Eq.~\eqref{eq:(1.15)},
\begin{align}
   c_1^{\text{old}}(t)
   &=\frac{1}{g^2}
   \left\{1
   +\frac{g^2}{(4\pi)^2}
   \left[-\beta_0L(\mu,t)-\frac{7}{3}C_{\!A}+\frac{3}{2}T_F\right]
   \right\},
\label{eq:(1.62)}
\\
   c_2^{\text{old}}(t)
   &=\frac{1}{4g^2}\frac{g^2}{(4\pi)^2}
   \left(\frac{11}{6}C_{\!A}+\frac{11}{6}T_F\right),
\label{eq:(1.63)}
\\
   c_3^{\text{old}}(t)
   &=\frac{1}{4}\left[1
   +\frac{g^2}{(4\pi)^2}C_F\left(\frac{3}{2}+\ln432\right)
   \right],
\label{eq:(1.64)}
\\
   c_4^{\text{old}}(t)
   &=\frac{g^2}{(4\pi)^2}\frac{3}{4}C_F,
\label{eq:(1.65)}
\\
   c_5^{\text{old}}(t)
   &=-\left\{1+\frac{g^2}{(4\pi)^2}C_F
   \left[3L(\mu,t)+\frac{7}{2}+\ln432\right]
   \right\}.
\label{eq:(1.66)}
\end{align}
They differ from the coefficients obtained from Eq.~\eqref{eq:(1.16)}:
\begin{align}
   c_1(t)
   &=\Check{c}_1(t) =\frac{1}{g^2}
   \left\{1
   +\frac{g^2}{(4\pi)^2}
   \left[-\beta_0L(\mu,t)-\frac{7}{3}C_{\!A}+\frac{3}{2}T_F\right]
   \right\},
\label{eq:(1.67)}
\\
   c_2(t)
   &=\Check{c}_2(t)+\frac{1}{4}\Check{c}_1(t) =\frac{1}{4g^2}\frac{g^2}{(4\pi)^2}
   \left(\frac{11}{6}C_{\!A}-\frac{3}{2}T_F\right),
\label{eq:(1.68)}
\\
   c_3(t)
   &=\mathring{\Check{c}}_3(t) =\frac{1}{4}\left[1
   +\frac{g^2}{(4\pi)^2}C_F\left(\frac{3}{2}+\ln432\right)
   \right],
\label{eq:(1.69)}
\\
   c_4(t)
   &=\mathring{\Check{c}}_4(t)+2\mathring{\Check{c}}_3(t) =\frac{1}{2}
   \left[
   1
   +\frac{g^2}{(4\pi)^2}C_F
   \left(\frac{5}{2}+\ln432\right)
   \right],
\label{eq:(1.70)}
\\
   c_5(t)
   &=0.
\label{eq:(1.71)}
\end{align}
These differences arise from the backreaction of the elimination of $\Tilde{\mathcal{O}}_{5\mu\nu}(t,x)$. 
Inserting $\Tilde{\mathcal{O}}_{5,\mu\nu}
=-\mathring{d}_5^{-1}d_2\Tilde{\mathcal{O}}_{2,\mu\nu}
-\mathring{d}_5^{-1}\mathring{d}_4\Tilde{\mathcal{O}}_{4,\mu\nu}$ 
[see Eq.~\eqref{eq:(1.19)}]
into~Eq.~\eqref{eq:(1.15)} with the
old coefficients~\eqref{eq:(1.62)}--\eqref{eq:(1.66)} to eliminate
$c_5^{\text{old}}\Tilde{\mathcal{O}}_{5,\mu\nu}$, we confirm that the old
coefficients~\eqref{eq:(1.62)}--\eqref{eq:(1.66)} precisely reproduce the new
coefficients~\eqref{eq:(1.67)}--\eqref{eq:(1.70)}.

\subsection{Two-loop confirmation}

As pointed out in Ref.~\cite{Suzuki:2013gza},
in the case of the pure Yang-Mills theory, a certain part of the two-loop order coefficients can be 
extracted by using the trace anomaly without any higher order calculations.
The expression in our present notation is (see Eq.~(4.69) of Ref.~\cite{Makino:2014taa}),
\begin{equation}
   c_2(t)\equiv\Check{c}_2(t)+\frac{1}{4}\Check{c}_1(t)
   =\frac{1}{4g^2}
   \left[
   \frac{g^2}{(4\pi)^2}\frac{\beta_0}{2}
   +\frac{g^4}{(4\pi)^4}
   \left(\frac{\beta_1}{2}-\frac{7}{4}C_{\!A}\beta_0\right)
   \right],
\label{eq:(1.75)}
\end{equation}
where the second term in the right-hand side contains the information in the
two-loop order. This coincides with the result obtained from Eqs.~\eqref{eq:(1.32)} and~\eqref{eq:(1.33)}
for the pure Yang-Mills theory.

We can similarly deduce three-loop $c_2(t)$ for the pure Yang-Mills theory from the two-loop coefficients. 
A concrete form is given in Ref.~\cite{Iritani2019}. 

%%%%%%%%%%%%%%%%%%%%%%%%%%%%%%%%%%%%%%%%%%%%%%%%%%%%
\section{Alternative method for flowed fermionic bilinear observables}
\label{sec:normalflow}

As discussed in Appendix~A of Ref.~\cite{Taniguchi:2016ofw}, fermionic parts of the EMT are given in terms of 
\begin{align}
   t_{\mu\nu}^f(t)
   &\equiv\frac{1}{N_\Gamma}\sum_x
   \left\langle\Bar{\chi}_f(t,x)\,\gamma_\mu
   \left(D_\nu-\overleftarrow{D}_\nu\right)\chi_f(t,x)\right\rangle
\notag\\
   &=-\frac{1}{N_\Gamma}\sum_{x,v,w}\Biggl\{
   \left\langle
   \sum_{\alpha,i}\left[\gamma_\mu D_\nu^xK(t,x;0,v)
   S_f(v,w)K(t,x;0,w)^\dagger
   \right]_{\alpha i,\alpha i}
   \right\rangle
\notag\\
   &\qquad\qquad\qquad\qquad{}
   -\left\langle
   \sum_{\alpha,i}\left[K(t,x;0,v)
   S_f(v,w)
   K(t,x;0,w)^\dagger
   \overleftarrow{D}_\nu^x\gamma_\mu
   \right]_{\alpha i,\alpha i}\right\rangle
   \Biggr\}, 
\label{eq:(A.1)}
\end{align}
and
\begin{align}
   s^f(t)
   &\equiv\frac{1}{N_\Gamma}\sum_x
   \left\langle\Bar{\chi}_f(t,x)\,\chi_f(t,x)\right\rangle
\notag\\
   &=-\frac{1}{N_\Gamma}\sum_{x,v,w}
   \left\langle
   \sum_{\alpha,i}\left\{K(t,x;0,v)
   \left[S_f(v,w)-c_{\mathrm{fl}}\,\delta_{v,w}\right]
   K(t,x;0,w)^\dagger
   \right\}_{\alpha i,\alpha i}
   \right\rangle,
\label{eq:(A.2)}
\end{align}
where the covariant derivatives in~Eq.~\eqref{eq:(A.1)} refer to the flowed gauge field~$B_\mu(t,x)$,
$N_\Gamma=\sum_x$ is the number of lattice points, $\alpha$ and $i$ denote the spinor and color indices, respectively, 
and $c_{\mathrm{fl}}$ is an improvement coefficient associated with the flowed
quark field~\cite{Luscher:2013cpa}. 
$S_f(x,y)$ is the quark propagator with the bare mass~$m_{f0}$,
and $K(t,x;s,y)$ is the fundamental solution to the flow equation defined by
\begin{equation}
   \left(\partial_t-\Delta\right)K(t,x;s,y)=0,\qquad
   K(t,x;t,y)=\delta_{x,y} .
\label{eq:(A.5)}
\end{equation}
The dagger ($\dagger$) in Eqs.~\eqref{eq:(A.1)} and~\eqref{eq:(A.2)} implies the hermitian conjugation with respect
to the gauge and spinor indices only.
Note that $K$ and~$D_\nu$ have no spinor indices.
In Ref.~\cite{Taniguchi:2016ofw}, we have computed them by introducing noise vectors to evaluate the trace over space-time points in Eqs.~\eqref{eq:(A.1)} and~\eqref{eq:(A.2)}. 
Expressions for other local fermionic bilinear operators can be written down similarly.

Here, we note that Eqs.~\eqref{eq:(A.1)} and~\eqref{eq:(A.2)} can be equally written as
\begin{align}
   t_{\mu\nu}^r(t)
   &=-\frac{1}{N_\Gamma}\sum_{x,y,v,w}\Biggl\{
   \left\langle
   \sum_{\alpha,i}\left[K(t,x;0,y)^\dagger
   \gamma_\mu D_\nu^xK(t,x;0,v)
   S_r(v,w)
   \right]_{\alpha i,\alpha i}\delta_{w,y}
   \right\rangle
\notag\\
   &\qquad\qquad\qquad\qquad{}
   +\left\langle\delta_{y,w}
   \sum_{\alpha,i}\left[
   S_r(v,w)^\dagger
   K(t,x;0,v)^\dagger
   \overleftarrow{D}_\nu^x\gamma_\mu
   K(t,x;0,y)
   \right]_{\alpha i,\alpha i}\right\rangle
   \Biggr\},
\label{eqa:(8.1)}  
\end{align}
where we have used the relation $S_r(v,w)=\gamma_5S_r(w,v)^\dagger\gamma_5$ , and
\begin{equation}
   s^r(t)
   =-\frac{1}{N_\Gamma}\sum_{x,y,v,w}
   \left\langle
   \sum_{\alpha,i}\left\{
   K(t,x;0,y)^\dagger
   K(t,x;0,v)
   \left[S_r(v,w)-c_{\text{fl}}\delta_{v,w}\right]
   \right\}_{\alpha i,\alpha i}\delta_{w,y}
   \right\rangle.
\label{eqa:(8.2)}
\end{equation}
We thus introduce a new noise field 
\begin{equation}
   \left\langle\eta_{\alpha i}(x)\right\rangle_\eta=0,\qquad
   \left\langle\eta_{\alpha i}(x)\eta_{\beta j}(y)^*\right\rangle_\eta
   =\delta_{\alpha\beta}\delta_{ij}\delta_{x,y},
\label{eqa:(8.3)}
\end{equation}
and define
\begin{eqnarray}
   \Xi(t,x)&\equiv&\sum_yK(t,x;0,y)\,\eta(y),
   \\
   Z_r(t,x)&\equiv&\sum_{v,w}K(t,x;0,v)\,S_r(v,w)\,\eta(w),
\end{eqnarray}
where contraction of spinor and color indices is understood. 
We then obtain compact expressions as
\begin{equation}
   t_{\mu\nu}^r(t)
   =-\frac{1}{N_\Gamma}2\re
   \left\langle\left\langle
   \sum_x\Xi(t,x)^\dagger\gamma_\mu D_\nu Z_r(t,x)
   \right\rangle_\eta\right\rangle,
\label{eqa:(8.6)}
\end{equation}
where the gauge field in the covariant derivative~$D_\nu$ is the flowed $B_\mu(t,x)$, and
\begin{equation}
   s^r(t)
   =-\frac{1}{N_\Gamma}
   \left\langle\left\langle
   \sum_x\Xi(t,x)^\dagger Z_r(t,x)
   \right\rangle_\eta\right\rangle
   +c_{\text{fl}}\frac{1}{N_\Gamma}
   \left\langle\left\langle
   \sum_x\Xi(t,x)^\dagger\Xi(t,x)
   \right\rangle_\eta\right\rangle.
\label{eqa:(8.7)}
\end{equation}

The building blocks obey the forward flow equations as
\begin{align}
   (\partial_t-\Delta)\Xi(t,x)&=0,&\Xi(0,x)&=\eta(x),
\\
   (\partial_t-\Delta)Z_r(t,x)&=0,&
   Z_r(0,x)&=\sum_yS_r(x,y)\eta(y),
\end{align}
which can be solved by the time-forward Runge--Kutta method
as explained in~Appendix~D.2 of~Ref.~\cite{Luscher:2013cpa}. That is, setting
$\partial_t\chi_t=\Delta(V_t)\chi_t$,
the Runge--Kutta proceeds as
\begin{align}
   \phi_0&=\chi_t,
\notag\\
   \phi_1&=\phi_0+\frac{1}{4}\Delta_0\phi_0,
\notag\\
   \phi_2&=\phi_0+\frac{8}{9}\Delta_1\phi_1-\frac{2}{9}\Delta_0\phi_0,
\notag\\
   \phi_3&=\phi_1+\frac{3}{4}\Delta_2\phi_2,
\end{align}
where
\begin{equation}
   \Delta_i=\epsilon\Delta(W_i),\qquad i=0,1,2,
\end{equation}
and
\begin{equation}
   \chi_{t+\epsilon}=\phi_3+O(\epsilon^4).
\end{equation}
These new representations are advantageous in the sense that they do not
require the Runge--Kutta steps proceeding backward in the flow time (see Appendix~B of Ref.~\cite{Taniguchi:2016ofw}) 
which is numerically demanding.

%%%%%%%%%%%%%%%%%%%%%%%%%%%%%%%%%%%%%%%%%%%%%%%%%%%%

\end{document}